\documentclass[onecolumn,superscriptaddress,nofootinbib,longbibliography,preprintnumbers]{revtex4-2}
\usepackage{amsmath,amsthm}
\usepackage{amssymb,amsfonts}
\usepackage{bm}
\usepackage[colorlinks=true, linkcolor=blue, citecolor=blue, urlcolor=blue]{hyperref}
\usepackage{graphicx}
\usepackage{dsfont}
\usepackage{float}
\usepackage{braket}
\usepackage{subfigure}
\usepackage{cases}
\usepackage{multirow}
\numberwithin{equation}{section}

\usepackage{algpseudocode}
\usepackage{fancyhdr}
\usepackage{setspace}

\usepackage{soul}

\usepackage{tikz,xcolor}
\definecolor{lime}{HTML}{A6CE39}
\DeclareRobustCommand{\orcidicon}{%
	\begin{tikzpicture}
	\draw[lime, fill=lime] (0,0) 
	circle [radius=0.16] 
	node[white] {{\fontfamily{qag}\selectfont \tiny ID}};	\draw[white, fill=white] (-0.0625,0.095) 
	circle [radius=0.007];	\end{tikzpicture}
	\hspace{-2mm}}
\foreach \x in {A,...,Z}{%
	\expandafter\xdef\csname orcid\x\endcsname{\noexpand\href{https://orcid.org/\csname orcidauthor\x\endcsname}{\noexpand\orcidicon}}
}

\newcommand{\RU}{\affiliation{School of Physics, Renmin University of China, Beijing 100872, China}}
\newcommand{\RUa}{\affiliation{Key Laboratory of Quantum State Construction and Manipulation of MoE,
\\Renmin University of China, Beijing 100872, China}}

\newcommand{\CAS}{\affiliation{Institute of Theoretical Physics, Chinese Academy of Sciences, Beijing 100190, China}}

\newcommand{\TU}{\affiliation{Center for Computational Sciences, University of Tsukuba, Tsukuba, Ibaraki 305-8577, Japan}}
\newcommand{\UTokyo}{\affiliation{Graduate School of Science, The University of Tokyo, Bunkyo-ku, Tokyo, 113-0033, Japan}}

\begin{document}

\title{Phase diagram of the single-flavor Gross--Neveu--Wilson model from the Grassmann corner transfer matrix renormalization group}

\author{Jian-Gang Kong\,}
\email{jgkong97phy@ruc.edu.cn}
\RU

\author{Shinichiro Akiyama~\orcidB{}\,}
\email{akiyama@ccs.tsukuba.ac.jp}
\TU
\UTokyo

\author{Tao Shi}
\email{tshi@itp.ac.cn}
\CAS

\author{Z. Y. Xie}
\email{qingtaoxie@ruc.edu.cn}
\RU
\RUa

\preprint{UTHEP-816, UTCCS-P-174}

\begin{abstract}
We investigate the phase structure of the (1+1)-dimensional single-flavor Gross--Neveu model with Wilson fermions using the Grassmann corner transfer matrix renormalization group (CTMRG).
The path integral is formulated as a two-dimensional Grassmann tensor network and approximately contracted by the Grassmann CTMRG algorithm. We investigate the phase diagram by varying the fermion mass and the four-fermion coupling, using the pseudoscalar condensate as an order parameter for the $\mathbb{Z}_{2}$ parity symmetry breaking phase.
The universality classes of the phase boundaries are identified through the central charge $c$ obtained via scaling analysis of the entanglement entropy.
Furthermore, we extract the quantity related to the entanglement spectrum from the converged CTMRG environments, allowing us to distinguish the topological insulator phase and the trivial phase.
The resulting phase structure suggests that the Aoki phase is separated from the other phases by critical lines characterized by $c=1/2$, while the critical lines with $c=1$ separate the topological insulating and trivial phases.
Our numerical results also indicate that the Aoki phase does not persist in the strong-coupling regime for the single-flavor theory.
\end{abstract}

\maketitle

\onecolumngrid

\section{Introduction}

Spontaneous chiral symmetry breaking describes dynamical mass generation~\cite{Nambu:1961tp,Nambu:1961fr}.
In the case of quantum chromodynamics (QCD), the pions can be interpreted as the Nambu--Goldstone bosons of the chiral symmetry breaking.
To investigate these non-perturbative phenomena, the lattice formulation of quantum field theories is indispensable~\cite{Wilson:1974sk}.
Since the lattice formulation of chiral fermions is not straightforward due to the Nielsen--Ninomiya theorem~\cite{Nielsen:1981hk}, it is practically necessary to choose a lattice fermion discretization appropriately.
Wilson fermions provide a standard lattice fermion formulation~\cite{Wilson:1975id}. 
Although the Wilson term explicitly breaks chiral symmetry, massless pions can be realized along a critical line on which chiral symmetry is expected to be restored in the continuum limit~\cite{Kawamoto:1980fd}.
Interestingly, massless pseudoscalar mesons can actually emerge as a consequence of parity-flavor symmetry breaking when $N_{f}$, the number of flavors, is even~\cite{Aoki:1983qi,Aoki:1986xr,Aoki:1987us}. In this case, the Vafa--Witten theorem forbids the spontaneous breaking of parity symmetry alone~\cite{Vafa:1984xg}.
However, when $N_{f}$ is odd, the theorem does not apply, and the flavor-singlet pseudoscalar meson can become massless.
Since parity and flavor symmetries are preserved in continuum QCD, the continuum limit should be approached without breaking symmetries.
In this sense, a precise determination of the phase boundary of the parity-flavor or parity symmetry-broken phase, referred to as the Aoki phase, in lattice QCD with Wilson fermions is vital for taking the continuum limit.
Although the parity-flavor symmetry breaking has been extensively studied within the lattice QCD simulation with the Wilson fermions~\cite{Aoki:1989rw,Aoki:1995yf,Sharpe:1998xm}, it remains challenging to investigate the Aoki phase for Wilson fermions with odd flavors. 
This difficulty originates from the sign problem in Monte Carlo simulations~\cite{Azcoiti:2012ns}.
\footnote{
It has been pointed out that the sign problem is absent in the case of the so-called central-branch Wilson fermion, since the determinant of the Wilson--Dirac operator is positive semi-definite~\cite{Misumi:2019jrt}.
}

The Gross--Neveu model~\cite{Gross:1974jv} provides a useful starting point from this viewpoint.
It is one of the well-known toy models of QCD: A renormalizable pure fermionic theory in two spacetime dimensions, describing $N_f$ flavors of massless Dirac fermions with a four-fermion interaction.
Although the Mermin--Wagner--Coleman theorem prohibits the spontaneous breaking of the continuous chiral symmetry in two dimensions~\cite{Mermin:1966fe,Coleman:1973ci}, the model shares several features with QCD: It is asymptotically free and exhibits dynamical mass generation associated with the breaking of a discrete chiral symmetry.
In particular, the existence of the parity symmetry--broken phase with Wilson fermions was originally demonstrated by Aoki in the Gross--Neveu model, based on the large-$N_{f}$ analysis~\cite{Aoki:1983qi}. The lattice Gross--Neveu model with Wilson fermions is sometimes referred to as the Gross--Neveu--Wilson (GNW) model.
One of the central motivations for studying the phase structure of the GNW model is the weak-coupling regime, in which the continuum limit is approached.
The weak-coupling expansion leads to a phase structure of the $N_{f}=1$ GNW model that is qualitatively consistent with the results of the large-$N_{f}$ analysis~\cite{Kenna:2001fs}.
Furthermore, the 't~Hooft anomaly matching condition requires the presence of the Aoki phase for the so-called central-branch Wilson fermion, even without assuming a large $N_{f}$~\cite{Misumi:2019jrt}.

Recently, the Gross--Neveu model has also attracted increasing attention from the condensed-matter perspectives~\cite{Roose:2021pba,Asaduzzaman:2022bpi}. 
This is because the model has an explicit correspondence with certain strongly correlated electronic systems, some of which can be realized in cold-atom quantum simulations~\cite{Kuno:2018pcp,Bermudez:2018eyh}.
In particular, Ref.~\cite{Bermudez:2018eyh} presents the complete phase structure of the $N_{f}=1$ GNW model within the Hamiltonian formalism, revealing the presence of the Aoki phase, a trivial band-insulating phase, and a symmetry-protected topological (SPT) phase based on matrix product states (MPS) simulations.
Since the Hamiltonian formalism does not involve a discretized time direction, no fermion doubler associated with temporal discretization appears. 
As a result, the phase structure of the lattice GNW model will differ from that obtained in the traditional Lagrangian formalism.

In this study, we provide a phase diagram of the $N_{f}=1$ GNW model based on the Lagrangian formalism employing tensor networks.
To the best of our knowledge, this work constitutes the first comprehensive numerical study of the complete phase diagram of the $N_{f}=1$ GNW model in the Lagrangian formulation.
Since tensor network methods are free from the sign problem, this approach opens a viable pathway toward future investigations of lattice QCD with Wilson fermions.
In particular, our approach is based on the Grassmann tensor network formulation of the lattice field theory~\cite{Gu:2010yh,Gu:2013gba,Shimizu:2014uva,Shimizu:2014fsa}, where the fermionic fields are directly manipulated in the numerical calculations.
Indeed, such a motivation has already led to concrete studies in the context of the $N_{f}=1$ Schwinger model~\cite{Shimizu:2017onf}.
One advantage of using the Grassmann tensor network formulation is that it fully preserves the locality inherent in the original lattice theory.
As a result, tensor network algorithms originally developed for spin systems can be straightforwardly extended to lattice fermion systems, and various applications have been made in the high-energy physics community~\cite{Takeda:2014vwa,Kadoh:2018hqq,Akiyama:2020soe,Akiyama:2020sfo,Akiyama:2021xxr,Akiyama:2021glo,Bloch:2022vqz,Asaduzzaman:2022pnw,Asaduzzaman:2023pyz,Akiyama:2023lvr,Yosprakob:2023tyr,Yosprakob:2023flr,Kanno:2024elz,Pai:2024tip,Pai:2025eia,Sugimoto:2025vui,Sugimoto:2026wnw}.
Based on the Grassmann tensor network formulation proposed in Ref.~\cite{Akiyama:2020sfo}, we develop the corner transfer matrix renormalization group (CTMRG)~\cite{Baxter:1968krk,baxter1978variational,Nishino_1996,Nishino_1997} for the Grassmann path integrals.
Although CTMRG is known to be a highly accurate algorithm for contracting two-dimensional tensor networks, it has not yet been widely adopted in the high-energy physics community.
CTMRG allows us to investigate not only the thermodynamic observables, but also the entanglement entropy and spectrum in a straightforward manner. 
Employing all these quantities, we determine the phase diagram of the $N_{f}=1$ GNW model varying the fermion mass and the four-fermion coupling constant.

This paper is organized as follows. 
In Sec.~\ref{sec:model}, we briefly review the large-$N_{f}$ phase diagram. 
We formulate the path integral of the GNW model as a Grassmann tensor network and describe the Grassmann CTMRG algorithm in Sec.~\ref{sec:GTN}. 
In Sec.~\ref{sec:results}, we first benchmark the CTMRG using the free Wilson fermion theory.
After confirming the validity and efficiency of the CTMRG, we present our main numerical results for the pseudoscalar condensate, entanglement entropy, and entanglement spectrum.
Finally, Sec.~\ref{sec:summary} is devoted to a summary and outlook.

\section{The Gross--Neveu model with Wilson fermions}
\label{sec:model}

\subsection{The model}

We consider the (1+1)-dimensional $N_{f}$-flavor GNW model, which is defined by the following action:
\begin{align}
\label{eq:GNW_action}
S&=
-\dfrac{1}{2}
\sum_{f=1}^{N_{f}}
\sum_{n\in\Lambda_{2}}
\sum_{\nu=1,2}
\left[
\bar{\psi}^{(f)}(n)
\left(
r\mathds{1} - \gamma_{\nu}    
\right)
\psi^{(f)}(n+\hat{\nu})
+
\bar{\psi}^{(f)}(n+\hat{\nu})
\left(
r\mathds{1} + \gamma_{\nu}    
\right)
\psi^{(f)}(n)
\right]
\nonumber\\
&+
M
\sum_{f,n}
\bar{\psi}^{(f)}(n)\psi^{(f)}(n)
-
\dfrac{g^{2}_{\sigma}}{2N_{f}}
\sum_{n}
\left(
\sum_{f}
\bar{\psi}^{(f)}(n)
\psi^{(f)}(n)
\right)^{2}
-
\dfrac{g^{2}_{\pi}}{2N_{f}}
\sum_{n}
\left(
\sum_{f}
\bar{\psi}^{(f)}(n)
{\rm i}\gamma_{5}
\psi^{(f)}(n)
\right)^{2}.
\end{align} 
The Wilson fermions are represented by the two-component Grassmann-valued fields $\psi^{(f)}(n) = (\psi^{(f)}_{1}(n), \psi^{(f)}_{2}(n))^{T}$ and $\bar{\psi}^{(f)}(n) = ( \bar{\psi}^{(f)}_{1}(n), \bar{\psi}^{(f)}_{2}(n))$, where $n=(n_{1},n_{2})$ denotes the lattice site on the square lattice $\Lambda_{2}$, $f$ represents the flavor index, and $\nu=1~(2)$ stands for the spatial (temporal) direction, respectively. 
The lattice extent in each direction is specified by $N_{\nu}$, such that $n_{\nu}=0,1,\cdots, N_{\nu}-1$. 
The two-dimensional Euclidean $\gamma$-matrices are denoted by $\gamma_{\nu}$
and they satisfy $[\gamma_{\mu}, \gamma_{\nu}]_{+} = 2\delta_{\mu\nu}\mathds{1}$ with the $2\times2$ unit matrix $\mathds{1}$.
In practical computations, we choose the following  representation: $\gamma_{1} = \sigma_{x}$, $\gamma_{2} = \sigma_{y}$, and $\gamma_{5} = -{\rm i}\gamma_{1}\gamma_{2} = \sigma_{z}$.
The magnitude of the four-fermion interaction is controlled by the coupling constants $g_{\sigma}^{2}$ and $g_{\pi}^{2}$. 
Hereafter, we consider $g_{\sigma}^{2} = g_{\pi}^{2} = g^{2}$.
The parameter $M$ is defined by $M=m+2r$, where $m$ denotes the fermion mass and the finite Wilson parameter $r$ is introduced to gap out the fermion doublers in the continuum limit.
Due to the presence of the finite Wilson parameter, the chiral symmetry is explicitly broken in Eq.~\eqref{eq:GNW_action} even with the massless fermion~$m=0$.
We set $r=1$ throughout this work.

\subsection{Large-$N_{f}$ phase diagram} \label{sec:largeN}

\begin{figure}[htbp]
	\centering
	\includegraphics[width=14cm]{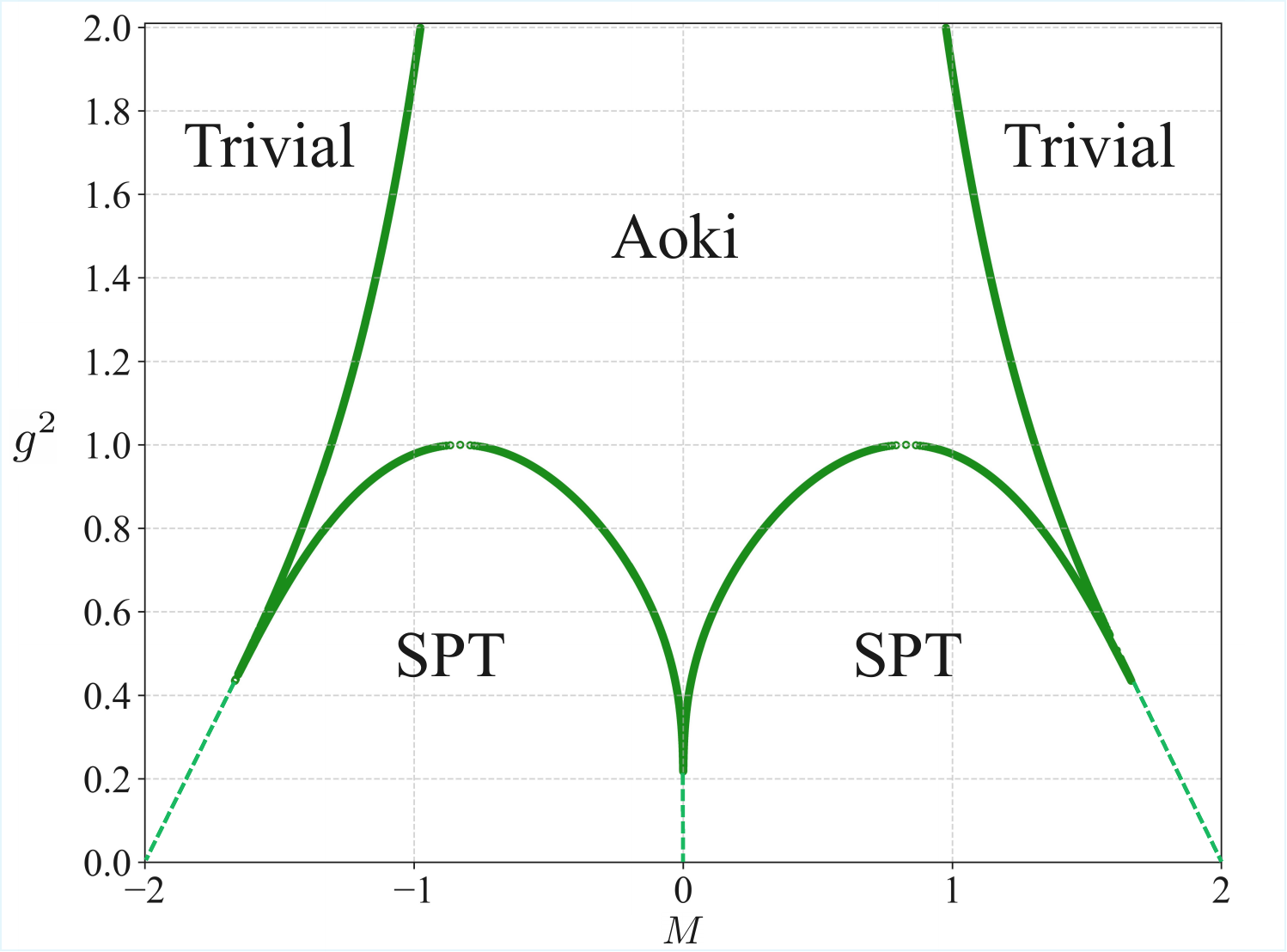}
	\caption{
        Phase diagram of the GNW model based on the large-$N_{f}$ method.
        We set $N_{1}=N_{2}=512$, which is sufficiently large to suppress finite-size effects, when solving Eqs.~\eqref{eq:gap_pi_tobesolved} and \eqref{eq:gap_sigma_tobesolved}. 
        The green lines separate the Aoki, trivial, and SPT phases, while the black dashed lines are shown as guides.
    }
	\label{fig:largeN_phase_diagram}
\end{figure} 

To make this paper self-contained, we briefly review the large-$N_{f}$ phase diagram~\cite{Aoki:1983qi}.
In the large-$N_{f}$ limit, the path integral is exactly given by the saddle point of the action.
By the Hubbard--Stratonovich transformation, we introduce the auxiliary bosonic fields $\sigma(n)$ and $\pi(n)$ into the path integral such that
\begin{align}
\label{eq:GNW_HS}
    Z
    =
    \int
    \prod_{f=1}^{N_{f}}
    \prod_{n\in\Lambda_{2}}
    {\rm d}\psi^{(f)}(n){\rm d}\bar{\psi}^{(f)}(n)~
    {\rm e}^{-S}
    =
    \int\prod_{n\in\Lambda_{2}}
    {\rm d}\sigma(n)
    {\rm d}\pi(n)
    \prod_{f=1}^{N_{f}}
    {\rm d}\psi^{(f)}(n){\rm d}\bar{\psi}^{(f)}(n)~
    {\rm e}^{-S'},
\end{align}
up to a multiplicative constant, where the original action is transformed into a new action $S'$, which reads
\begin{align}
    S'
    &=
    -\dfrac{1}{2}
    \sum_{f=1}^{N_{f}}
    \sum_{n\in\Lambda_{2}}
    \sum_{\nu=1,2}
    \left[
        \bar{\psi}^{(f)}(n)
        \left(
            \mathds{1} - \gamma_{\nu}    
        \right)
        \psi^{(f)}(n+\hat{\nu})
        +
        \bar{\psi}^{(f)}(n+\hat{\nu})
        \left(
            \mathds{1} + \gamma_{\nu}    
        \right)
        \psi^{(f)}(n)
        \right]
    \nonumber\\
    &+
    M
    \sum_{f,n}
    \bar{\psi}^{(f)}(n)\psi^{(f)}(n)
    +
    \sum_{f,n}
    \bar{\psi}^{(f)}(n)
    \left[
        \sigma(n)+{\rm i}\gamma_{5}\pi(n)
    \right]
    \psi^{(f)}(n)
    +
    \frac{N_{f}}{2g^{2}}
    \sum_{n}
    \left[
        \sigma(n)^{2}+\pi(n)^{2}
    \right].
\end{align}
The Grassmann integrals in Eq.~\eqref{eq:GNW_HS} can be carried out for $S'$.
One finds
\begin{align}
\label{eq:GNW_Z_Nf}
    Z=
    \int\prod_{n\in\Lambda_{2}}
    {\rm d}\sigma(n){\rm d}\pi(n)~
    {\rm e}^{-N_{f}|\Lambda_{2}|V_{\rm eff}},
\end{align}
where the explicit form of the effective potential $V_{\rm eff}$ can be expressed in momentum space as
\begin{align}
    V_{\rm eff}=
    \frac{1}{2g^{2}}
    \left(
        \tilde{\sigma}^{2}+\tilde{\pi}^{2}
    \right)
    -
    \frac{1}{|\Lambda_{2}|}
    \sum_{k}
    \log\left[
        \left\{
            m+\tilde{\sigma}
            +\sum_{\nu}
            \left(
                1-\cos \frac{2\pi k_{\nu}}{N_{\nu}}
            \right)
        \right\}^{2}
        +
        \tilde{\pi}^{2}
        +
        \sum_{\nu}
        \sin^{2} \frac{2\pi k_{\nu}}{N_{\nu}}
    \right],
\end{align}
where we have treated the auxiliary bosonic fields as constants, $\sigma(n)=\tilde{\sigma}$ and $\pi(n)=\tilde{\pi}$, and $k_{\nu}=-N_{\nu}/2+1,\cdots,N_{\nu}/2$ with $|\Lambda_{2}|=N_{1}N_{2}$.
In the large-$N_{f}$ limit, Eq.~\eqref{eq:GNW_Z_Nf} is exactly given by the saddle point of $V_{\rm eff}$.
The saddle-point equations are
\begin{align}
\label{eq:gap_sigma}
    \frac{\delta V_{\rm eff}}{\delta \tilde{\sigma}}
    =
    \frac{\tilde{\sigma}}{g^{2}}
    -
    \frac{1}{|\Lambda_{2}|}
    \sum_{k}
    \frac{
        2m+2\tilde{\sigma}+2\sum_{\nu}
        \left(
            1-\cos (2\pi k_{\nu}/N_{\nu})
        \right)
        }
        {
        \left[
            m+\tilde{\sigma}
            +\sum_{\nu}
            \left(
                1-\cos (2\pi k_{\nu}/N_{\nu})
            \right)
        \right]^{2}
        +
        \tilde{\pi}^{2}
        +
        \sum_{\nu}
        \sin^{2} (2\pi k_{\nu}/N_{\nu})
        }
        =
        0,
\end{align}
\begin{align}
\label{eq:gap_pi}
    \frac{\delta V_{\rm eff}}{\delta \tilde{\pi}}
    =
    \frac{\tilde{\pi}}{g^{2}}
    -
    \frac{1}{|\Lambda_{2}|}
    \sum_{k}
    \frac{
        2\tilde{\pi}
        }
        {
        \left[
            m+\tilde{\sigma}
            +\sum_{\nu}
            \left(
                1-\cos (2\pi k_{\nu}/N_{\nu})
            \right)
        \right]^{2}
        +
        \tilde{\pi}^{2}
        +
        \sum_{\nu}
        \sin^{2} (2\pi k_{\nu}/N_{\nu})
        }
        =
        0.
\end{align}
In the large-$N_{f}$ limit, Eq.~\eqref{eq:gap_sigma} suggests that $\tilde{\pi}\neq0$ can be a solution depending on the value of the fermion mass $m$.
When $\tilde{\pi}\neq0$, Eq.~\eqref{eq:gap_pi} reads
\begin{align}
\label{eq:gap_pi_tobesolved}
    \frac{1}{g^{2}}
    =
    \frac{2}{|\Lambda_{2}|}
    \sum_{k}
    \frac{
        1
        }
        {
        \left[
            m+\tilde{\sigma}
            +\sum_{\nu}
            \left(
                1-\cos (2\pi k_{\nu}/N_{\nu})
            \right)
        \right]^{2}
        +
        \tilde{\pi}^{2}
        +
        \sum_{\nu}
        \sin^{2} (2\pi k_{\nu}/N_{\nu})
        },
\end{align}
which implies both $\pm\tilde{\pi}$ are the solutions.
Using Eq.~\eqref{eq:gap_pi_tobesolved}, Eq.~\eqref{eq:gap_sigma} can be 
\begin{align}
\label{eq:gap_sigma_tobesolved}
    -\frac{m}{g^{2}}
    =
    \frac{2}{|\Lambda_{2}|}
    \sum_{k}
    \frac{
        \sum_{\nu}
        \left(
            1-\cos (2\pi k_{\nu}/N_{\nu})
        \right)
        }
        {
        \left[
            m+\tilde{\sigma}
            +\sum_{\nu}
            \left(
                1-\cos (2\pi k_{\nu}/N_{\nu})
            \right)
        \right]^{2}
        +
        \tilde{\pi}^{2}
        +
        \sum_{\nu}
        \sin^{2} (2\pi k_{\nu}/N_{\nu})
        }.
\end{align}

In Fig.~\ref{fig:largeN_phase_diagram}, the green lines separating the Aoki phase from the other phases are determined by setting $\tilde{\pi}=0$ in Eqs.~\eqref{eq:gap_pi_tobesolved} and \eqref{eq:gap_sigma_tobesolved} and solving them self-consistently.
Within the Aoki phase, the $\mathbb{Z}_{2}$ parity symmetry is spontaneously broken. 
The order parameter is a pseudoscalar condensate $\langle\sum_{f}\bar{\psi}^{(f)}(n){\rm i}\gamma_{5}\psi^{(f)}(n)\rangle$, since under the parity transformation $\psi^{(f)}(n_{1},n_{2}) \mapsto \gamma_{2} \psi^{(f)}(-n_{1},n_{2})$, it changes $\langle\sum_{f}\bar{\psi}^{(f)}(n){\rm i}\gamma_{5}\psi^{(f)}(n)\rangle \mapsto -
\langle\sum_{f}\bar{\psi}^{(f)}(n){\rm i}\gamma_{5}\psi^{(f)}(n)\rangle$.
A crucial feature of the parity-broken Aoki phase predicted in the large-$N_{f}$ limit is that it extends deeply into the strong-coupling region. 

On the other hand, several works have shown that the rest part of the phase diagram is not completely trivial~\cite{Kuno:2018pcp,Bermudez:2018eyh}, and that a topological insulator phase emerges when $N_{f}$ is odd, as shown in Fig.~\ref{fig:largeN_phase_diagram}.
A topological invariant known as Zak's phase $\phi_{Z} = 2\pi l$ can be used to distinguish the SPT phase from other two phases: $l$ is not an integer in the SPT phase, and integer for the other phases.\footnote{
As discussed in Ref.~\cite{Bermudez:2018eyh}, in the vanishing-$g^{2}$ limit of the GNW model, $\phi_{{\rm Z}}$ is explicitly evaluated through an integral of the Berry connection over the Brillouin zone, leading to a compact expression,
\begin{align}
\label{eq:Zak}
    \phi_{\rm Z} = 
    \frac{N_{f}\pi}{2}
    \left[
        {\rm sgn}(M - M_{(0,0)})
        +
        {\rm sgn}(M - M_{(0,\pi)})
        +
        {\rm sgn}(M - M_{(\pi,0)})
        +
        {\rm sgn}(M - M_{(\pi,\pi)})
    \right],
\end{align} 
where $M_{(p_{1},p_{2})}$ denotes the lattice mass such that $M_{(0, 0)} = 2, M_{(0,\pi)} = 0, M_{(\pi, 0)} = 0, M_{(\pi, \pi)} = -2$.
Note that four fermion doublers arise in the corner of Brillouin zone for the free Wilson fermions.
Therefore, we observe that $\phi_{{\rm Z}}/2\pi = \pm N_{f}/2$ for $M \in (0, 2)$ and $M \in (-2, 0)$, respectively, leading to the topological insulator phase when $N_{f}$ is odd in both cases.
In general, Eq.~\eqref{eq:Zak} remains valid even when the interaction $g^{2}$ is turned on, although an additive mass renormalization is required.
}
The topological phase in the GNW model can also be understood as a one-dimensional topological insulator with edge modes localized at the boundary of the open chain~\cite{Bermudez:2018eyh}. 
It is also referred to as an SPT phase, indicating that the topological edge states are robust against any symmetry-preserving perturbations. 
The flavor-independent single-particle Hamiltonian $h_{k}$ in momentum space satisfies $T^{\dagger} h_{-k}^{*} T = h_{k}$ and $C^{\dagger} h_{-k}^{*} C = -h_{k}$, where $T$ and $C$ denote time-reversal and charge-conjugation symmetries, respectively.
The combination of these two anti-unitary symmetries defines a sublattice symmetry $S$, which acts as $S^{\dagger} h_{k} S = -h_{k}$.
All these symmetries satisfy $T^{2} = C^{2} = S^{2} = 1$. 
The corresponding symmetry class is the so-called BDI class~\cite{Schnyder:2008tya,Ryu:2010zza}.
The interaction terms in Eq.~\eqref{eq:GNW_action} also respect these symmetries, and the topological classification of the interacting GNW model remains unchanged.
In practical numerical simulations, the topological invariant cannot be easily extracted directly. Fortunately, however, signatures of nontrivial topology can be identified in the entanglement spectrum, which is accessible within our tensor network framework, as we show in later sections.
 
\section{Grassmann tensor network approach}
\label{sec:GTN}

\subsection{Grassmann tensor network formulation}
Hereafter, we consider the $N_{f}=1$ GNW model, omitting the flavor index $f$.
The path integral generated by Eq.~\eqref{eq:GNW_action} can be represented as a two-dimensional Grassmann tensor network:
\begin{align} 
\label{eq:GNW_Z}
    Z 
    = \int \prod_{n\in\Lambda_{2}} 
    {\rm d}\psi(n){\rm d}\bar{\psi}(n)
    ~{\rm e}^{-S}
    =
    {\rm gTr}
    \left[
        \prod_{n\in\Lambda_{2}}
        \mathcal{T}_{n}
    \right],
\end{align}
where ${\rm gTr}$ represents multiple weighted Grassmann integrals over auxiliary Grassmann fields introduced on every edge of the lattice $\Lambda_{2}$ with periodic boundary conditions~\cite{Akiyama:2020sfo}.
The fundamental Grassmann tensor $\mathcal{T}_{n}$ is given by the following multi-linear combination of auxiliary Grassmann fields, $\eta_{\nu}$, $\xi_{\nu}$, $\bar{\eta}_{\nu}$, $\bar{\xi}_{\nu}$ with $\nu=1,2$,
\begin{align} 
\label{eq:Grassmann_tensor1}
	(\mathcal{T}_{n})_{
		\eta_{1}
		\xi_{1}
		\eta_{2}
		\xi_{2}
		\bar{\xi}_{1}
		\bar{\eta}_{1}
		\bar{\xi}_{2}
		\bar{\eta}_{2}
	} 
	= 
    \sum_{i_{1},j_{1},i_{2},j_{2},i'_{1},j'_{1},i'_{2},j'_{2}}
	(T_{n})_{
		i_{1}j_{1}
		i_{2}j_{2}
		i'_{1}j'_{1}
		i'_{2}j'_{2}
	}
	(\eta_{1})^{p(i_{1})}
	(\xi_{1})^{p(j_{1})}
	(\eta_{2})^{p(i_{2})}
	(\xi_{2})^{p(j_{2})}
	(\bar{\xi}_{1})^{p(j'_{1})}
	(\bar{\eta}_{1})^{p(i'_{1})}
	(\bar{\xi}_{2})^{p(j'_{2})}
	(\bar{\eta}_{2})^{p(i'_{2})}.
\end{align}
These auxiliary Grassmann variables are introduced to decompose the hopping terms in Eq.~\eqref{eq:GNW_action}, and the Grassmann tensor $\mathcal{T}_{n}$ is obtained by integrating out the original Wilson fermion fields $\psi(n)$ and $\bar{\psi}(n)$, individually at each site $n$.
Due to the nilpotency of the Grassmann variables, the integer subscripts ($i_{\nu}$, $j_{\nu}$, $i'_{\nu}$, $j'_{\nu}$ with $\nu=1,2$) can take only $0$ or $1$. 
The indices of Grassmann tensors can take larger values in general, so a parity function $p$ is introduced such that $p(i) = 0, 1$ for any given index value $i$. For example, the Grassmann tensor in Eq.~\eqref{eq:Grassmann_tensor1} can be regarded as a four-leg tensor whose bond dimension is 4, where $p(i) = 0$ for $i = 1, 2$, and $p(i) = 1$ for $i = 3, 4$.
The explicit form of the coefficient tensor $T_{n}$ for the $N_{f}=1$ GNW model is derived in Ref.~\cite{Akiyama:2023rih}.

Our primary interest is in the Aoki phase, which is characterized by a pseudoscalar field $\bar{\psi}(n){\rm i}\gamma_{5}\psi(n)$.
We denote the magnitude of the condensate by $\pi=|\langle\bar{\psi}(n){\rm i}\gamma_{5}\psi(n)\rangle|$ in the following.
In the path integral formalism, $\pi$ is defined by
\begin{align}
\label{eq:def_pi}
    \pi
    =
    \left| \lim_{h\to0}\lim_{|\Lambda_{2}|\to\infty}\frac{1}{|\Lambda_{2}|}\frac{\partial}{\partial h}\ln Z_{h}
    \right|,
\end{align}
where $Z_{h}$ is the path integral defined by
\begin{align} 
\label{eq:GNW_Zh}
    Z_{h}
    = \int \prod_{n\in\Lambda_{2}} 
    {\rm d}\psi(n){\rm d}\bar{\psi}(n)
    ~{\rm e}^{-S-S_{h}},
\end{align}
with
\begin{align}
S_{h}=h\sum_{n}\bar{\psi}(n){\rm i}\gamma_{5}\psi(n).    
\end{align}
We also derive a representation of Eq.~\eqref{eq:def_pi} in terms of a Grassmann tensor network.
This is easily achieved by introducing the following local ``impurity" Grassmann tensor:
\begin{align} 
\label{eq:Grassmann_tensor2}
	(\mathcal{I}_{n})_{
		\eta_{1}
		\xi_{1}
		\eta_{2}
		\xi_{2}
		\bar{\xi}_{1}
		\bar{\eta}_{1}
		\bar{\xi}_{2}
		\bar{\eta}_{2}
	} 
	= 
    \sum_{i_{1},j_{1},i_{2},j_{2},i'_{1},j'_{1},i'_{2},j'_{2}}
	(I_{n})_{
		i_{1}j_{1}
		i_{2}j_{2}
		i'_{1}j'_{1}
		i'_{2}j'_{2}
	}
	(\eta_{1})^{p(i_{1})}
	(\xi_{1})^{p(j_{1})}
	(\eta_{2})^{p(i_{2})}
	(\xi_{2})^{p(j_{2})}
	(\bar{\xi}_{1})^{p(j'_{1})}
	(\bar{\eta}_{1})^{p(i'_{1})}
	(\bar{\xi}_{2})^{p(j'_{2})}
	(\bar{\eta}_{2})^{p(i'_{2})},
\end{align}
with
\begin{align}
    (I_{n})_{
		i_{1}j_{1}
		i_{2}j_{2}
		i'_{1}j'_{1}
		i'_{2}j'_{2}
	} 
    &=
    \frac{
        (-1)^{
	        i_{1}
            \left(
	           i'_{2} + i'_{1} + j_{2} + j_{1}
	        \right)
	        +i_{2}
            \left(
	           i'_{2} + i'_{1} + j_{2}
	        \right)
	        +j'_{1}
            \left( 
                i'_{2} + i'_{1} 
            \right)
	        +j'_{2}i'_{2}
            +i'_{1}+i'_{2}
        }
    }{
        \sqrt{2}^{\sum_{\nu}(i_{\nu}+j_{\nu}+i'_{\nu}+j'_{\nu})}
    }\nonumber\\
    &\times
    {\rm i}
    \left[
        (+{\rm i})^{j_{2}'+i_{2}+i_{2}'+j_{2}}
        (-1)^{i_{2}+i_{1}+i_{1}'+j_{2}}
        \delta_{i_{1}+i_{2}+j_{1}'+j_{2}', 1}
        \delta_{j_{1}+j_{2}+i_{1}'+i_{2}', 1}
        -
        \delta_{i_{1}+i_{2}+j_{1}'+j_{2}', 1}
        \delta_{j_{1}+j_{2}+i_{1}'+i_{2}', 1}
    \right].
\end{align}
It then follows straightforwardly that the pseudoscalar condensate $\pi$ is evaluated as a ratio of two Grassmann tensor network contractions.
The numerator is
\begin{align}
\label{eq:GNW_impurity}
    {\rm gTr}
    \left[
        \mathcal{I}_{n'}
        \prod_{n\neq n'}
        \mathcal{T}_{n}
    \right],
\end{align}
while the denominator is given by Eq.~\eqref{eq:GNW_Z}.
The Grassmann tensor $\mathcal{I}_{n}$ can again be interpreted as a four-leg tensor with bond dimension 4.

\subsection{Grassmann corner transfer matrix renormalization group algorithm}

Hereafter, we employ a graphical representation of Grassmann tensors and their network following Refs.~\cite{Akiyama:2020sfo,Akiyama:2024ush}.
The corner transfer matrix renormalization group (CTMRG) algorithm~\cite{Nishino_1996} was originally invented by Nishino and Okunishi to investigate two-dimensional classical spin models. 
CTMRG is inspired by Baxter's corner transfer matrix formulation~\cite{Baxter:1982zz} as well as the density matrix renormalization group (DMRG) algorithm~\cite{White:1992zz,White:1993zza}. 
The CTMRG approach is further developed in the condensed-matter community~\cite{Orus:2009wuu,PhysRevB.82.245119,Corboz:2014ocg,Fishman:2018lnr}, and finds its efficiency in contracting the two-dimensional tensor network representing the norm and expectation value described by tensor network states~\cite{Nishino:2000ygc,Verstraete:2004cf,Li2022}. 
In the following, we develop the CTMRG algorithm based on Refs.~\cite{Corboz:2014ocg,Kong:2026} for the Grassmann tensor network.
Its bosonic counterpart is also known as the asymmetric CTMRG~\cite{Fishman:2018lnr}, which is suitable for local tensors with general spatial symmetries and different unit cell sizes, with improved convergence. 

\begin{figure}[htbp]
	\centering
	\includegraphics[width=14cm]{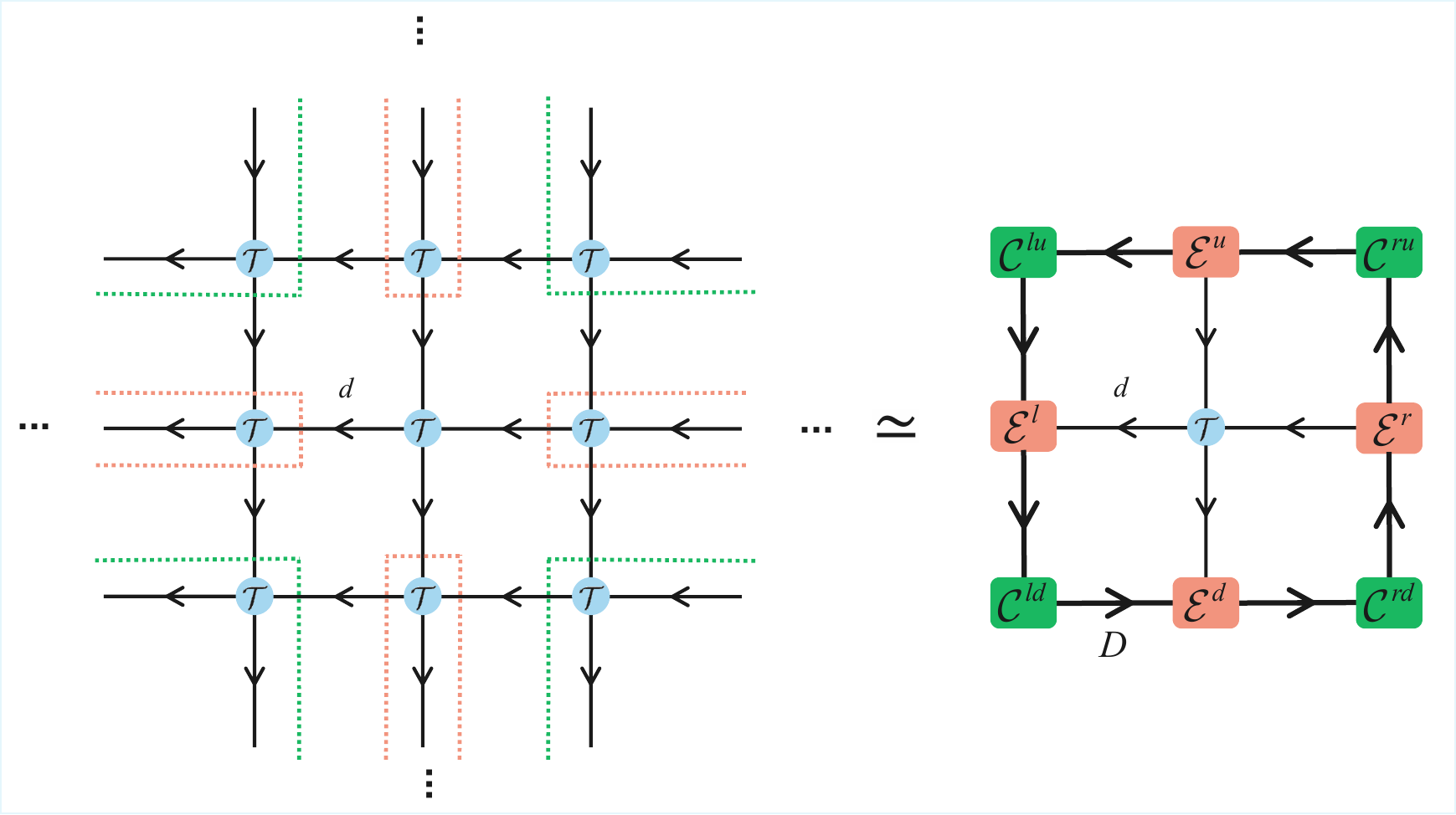}
	\caption{
        Original infinite Grassmann tensor network (left) and its effective representation by using the environment tensors (right).
        Corner matrices and three-leg edge tensors are introduced to approximate an infinite environment surrounding a local Grassmann tensor $\mathcal{T}$. 
        The local Grassmann tensor has bond dimension $d$, while the bond dimensions connecting the environment tensors are denoted by $D$.
    }
	\label{fig:ctmrg_env}
\end{figure} 

The goal of our Grassmann CTMRG is to evaluate the Grassmann path integral.
Let us consider a two-dimensional uniform Grassmann tensor network composed of an identical four-leg Grassmann tensor $\mathcal{T}_{n}$ at each lattice site $n$, with bond dimension $d$.
We introduce four types of corner matrices $\mathcal{C}^{lu}$, $\mathcal{C}^{ru}$, $\mathcal{C}^{ld}$, $\mathcal{C}^{rd}$, as well as three-leg edge tensors $\mathcal{E}^{l}$, $\mathcal{E}^{r}$, $\mathcal{E}^{u}$, $\mathcal{E}^{d}$. 
The CTMRG algorithm updates these corner and edge tensors so as to form an infinite environment surrounding the local tensor $\mathcal{T}_{n}$ as shown in Fig.~\ref{fig:ctmrg_env}. 
The arrows in Fig.~\ref{fig:ctmrg_env} are attached to all indices of the Grassmann tensors to represent their grading, which originates from the integration over the auxiliary fermions~\cite{Akiyama:2020sfo}.\footnote{
In this paper, the directions of the arrows are chosen to be opposite to those in Ref.~\cite{Akiyama:2020sfo}.
}
We denote the bond dimension of the local fundamental tensor as $d$, while that of the bonds connecting the corner matrices and edge tensors is denoted by $D$. 
This virtual bond dimension $D$ serves as a hyperparameter that controls the accuracy of approximating the infinite environments.

\begin{figure}
    \centering
    \subfigure[]{
        \includegraphics[width=15cm]{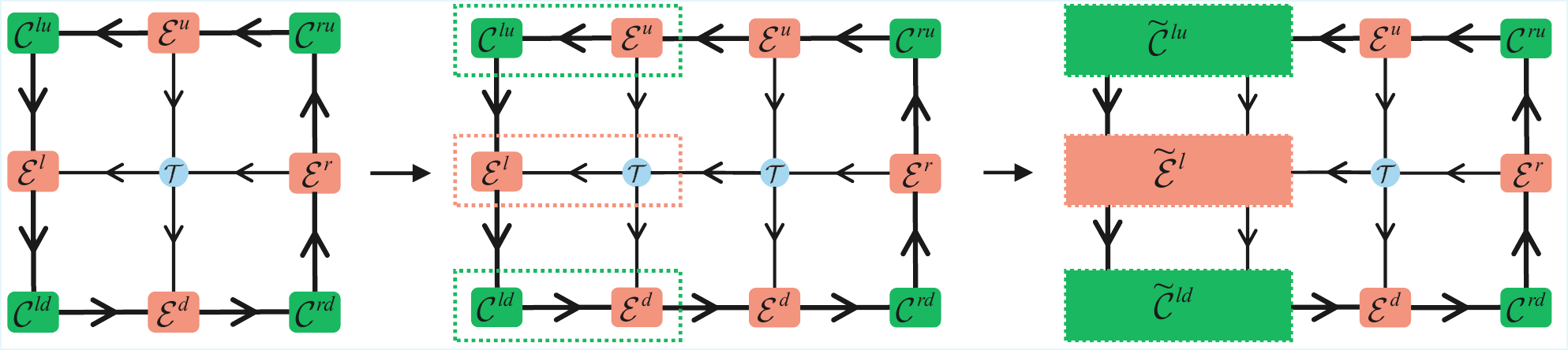}
    }\\
    \subfigure[]{
        \includegraphics[width=15cm]{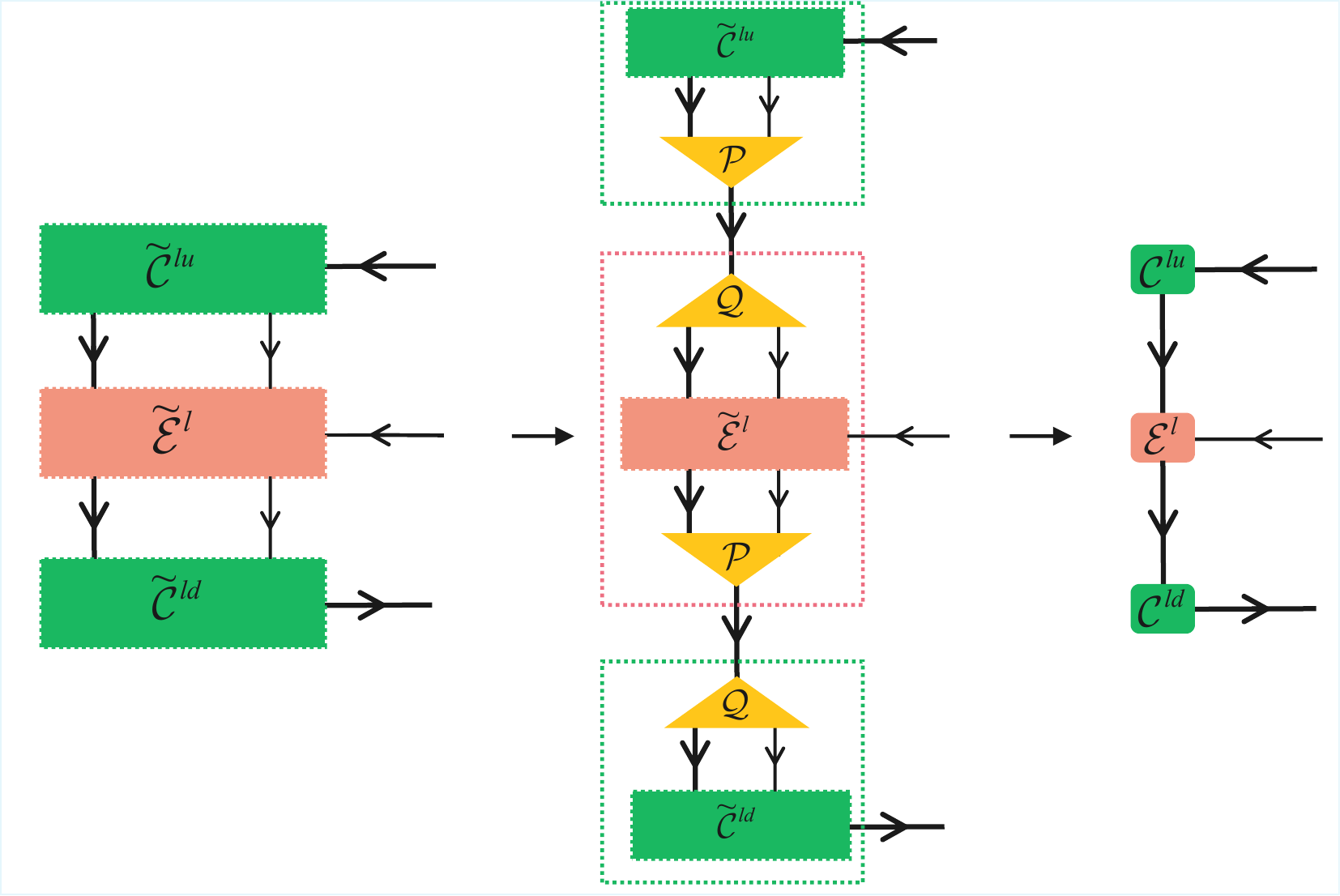}
    }
    \caption{
        The left move in the Grassmann CTMRG algorithm consists of two steps:
        (a) inserting a column of bulk tensors and absorbing it into the left environment; and
        (b) constructing the Grassmann projectors $\mathcal{P}$ and $\mathcal{Q}$ to truncate the virtual bond dimension and update the left environment tensors $\mathcal{C}^{lu}$, $\mathcal{E}^{l}$, and $\mathcal{C}^{ld}$.
    }
	\label{fig:ctmrg_renormalize}
\end{figure}

The eight environment tensors are initialized with random entries, and they are then updated iteratively until convergence is reached.
During the iterative update process, the environment tensors located in the four directions are updated sequentially, referred to as the left, right, up, and down moves.
For example, the left move is performed through a two-step procedure, as illustrated in Fig.~\ref{fig:ctmrg_renormalize}.
Firstly, a $1\times1$ unit cell of bulk tensor, together with its upward and downward environment tensors, is inserted into the Grassmann tensor network, and subsequently absorbed into the left environment, defining the enlarged environment tensors $\tilde{\mathcal{C}}^{lu}$, $\tilde{\mathcal{C}}^{ld}$, and $\tilde{\mathcal{E}}^{l}$, as shown in Fig.~\ref{fig:ctmrg_renormalize}(a). 
Next, pairs of Grassmann projectors $\mathcal{P}$ and $\mathcal{Q}$ are inserted to truncate the enlarged virtual bond dimension from $dD$ back to $D$ for $\tilde{\mathcal{C}}^{lu}$, $\tilde{\mathcal{C}}^{ld}$, and $\tilde{\mathcal{E}}^{l}$, as shown in Fig.~\ref{fig:ctmrg_renormalize}(b). 
This procedure defines a recursive update rule for $\mathcal{C}^{lu}$, $\mathcal{C}^{ld}$, and $\mathcal{E}^{l}$.
The recursive updates of the environment tensors for the remaining directions proceed in a similar manner, and all the environment tensors are updated after a single CTMRG step.
We refer to Appendix~\ref{app:construct_P_Q} for the details on how to derive $\mathcal{P}$ and $\mathcal{Q}$. 
The computational cost of the CTMRG algorithm scales as $O(d^{3}D^{3})$.

Once the converged environment tensors are obtained, the partition function $Z$ and the condensate $\langle\bar{\psi}(n){\rm i}\gamma_{5}\psi(n)\rangle$ can be easily evaluated~\cite{Liu:2022dht,Okunishi:2021but}:
\begin{align} 
\label{eq:ctmrg_Z}
    Z
    \simeq
    \raisebox{-0.48\height}{\includegraphics[width=0.38\textwidth, page=1]{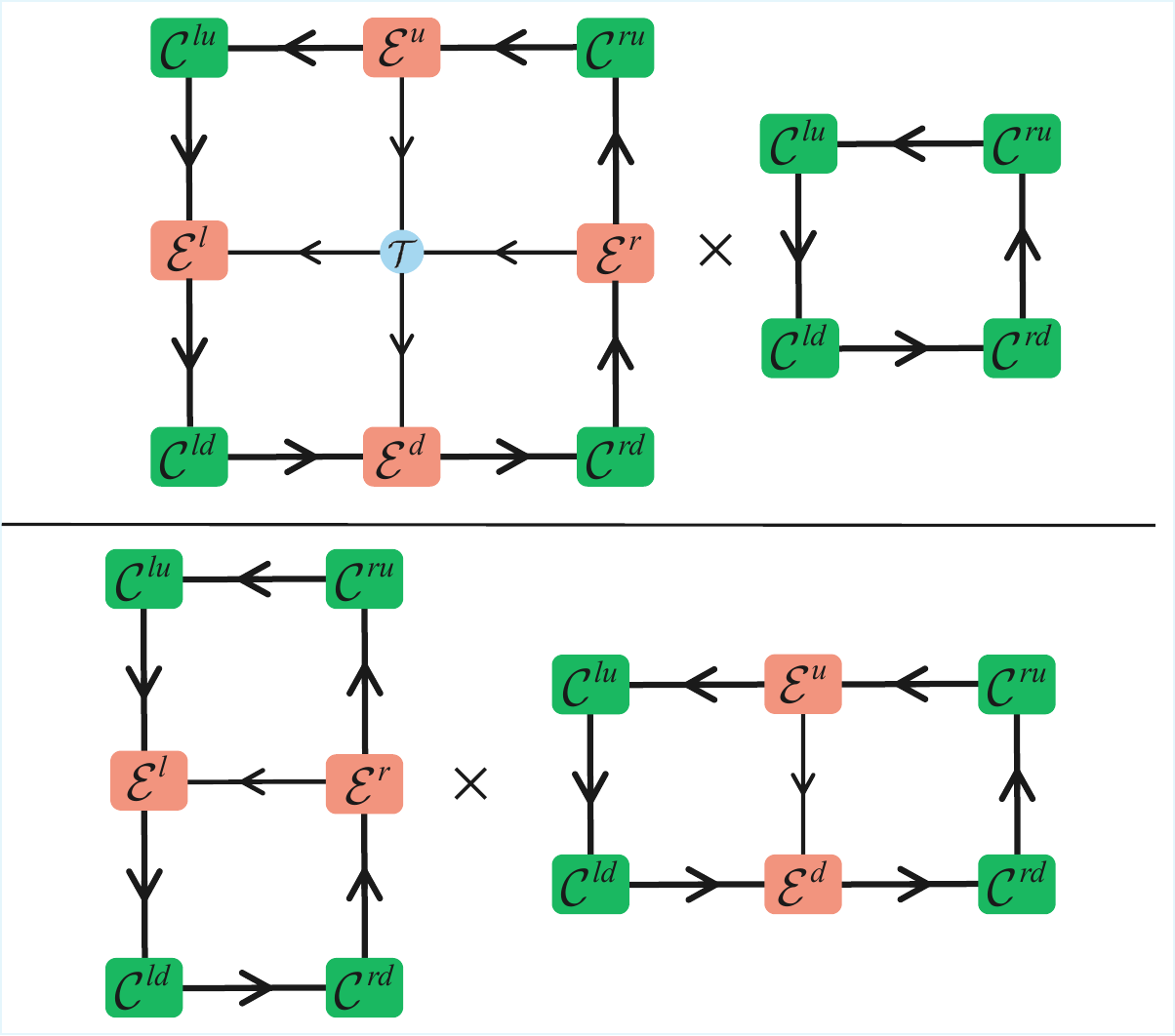}},
\end{align}
\begin{align} 
\label{eq:ctmrg_Exp}
    \langle\bar{\psi}(n){\rm i}\gamma_{5}\psi(n)\rangle
    \simeq
    \raisebox{-0.48\height}{\includegraphics[width=0.36\textwidth, page=1]{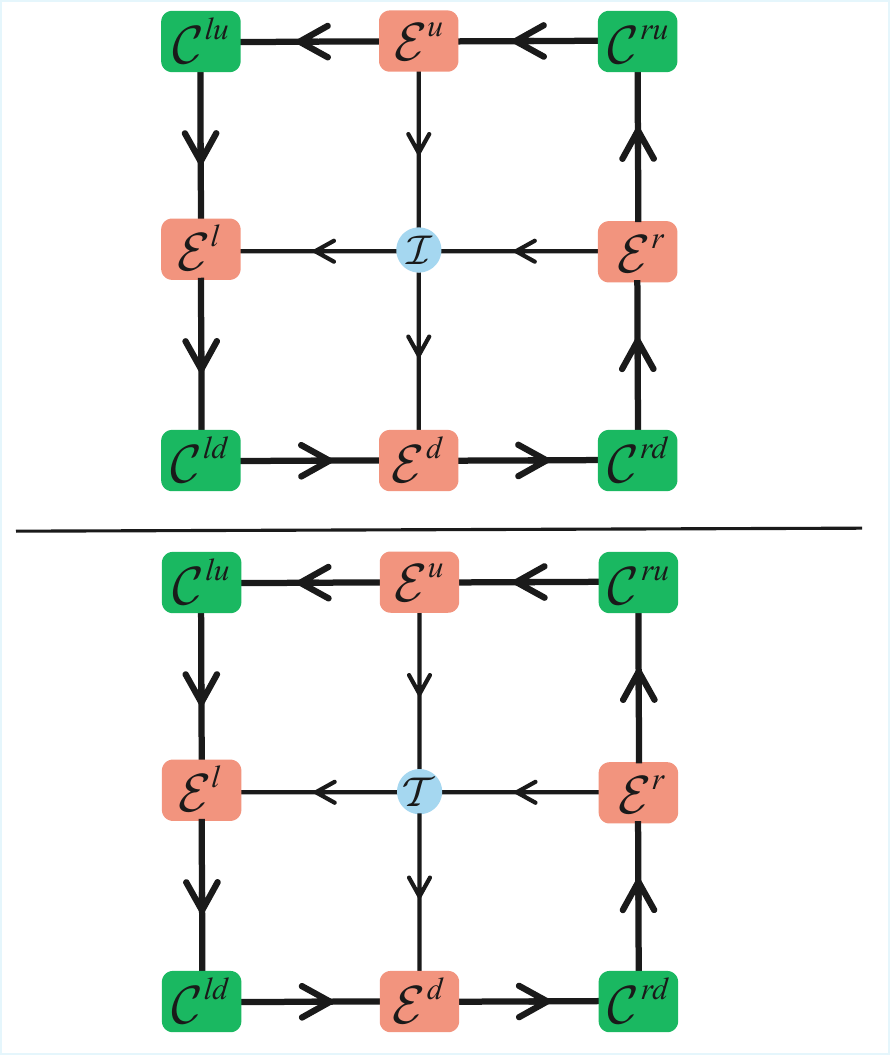}}.
\end{align}
We note that the CTMRG algorithm evaluates these quantities directly in the thermodynamic limit, assuming open boundary conditions.
However, once the environment constructed by CTMRG are sufficiently converged,
the resulting thermodynamic quantities become insensitive to the choice of boundary conditions.
Therefore, although periodic boundary conditions are assumed in the formal expressions of Eqs.~\eqref{eq:GNW_Z} and \eqref{eq:GNW_impurity}, they do not affect our final results.

\subsubsection{Correlation length, entanglement entropy, and spectrum}

The CTMRG algorithm directly deals with corner, row-to-row, and column-to-column transfer matrices. 
The corner transfer matrices are nothing but the corner matrices $\mathcal{C}^{lu}$, $\mathcal{C}^{ru}$, $\mathcal{C}^{ld}$, $\mathcal{C}^{rd}$.
On the other hand, the row-to-row transfer matrix is defined by two edge tensors $\mathcal{E}^{l}$ and $\mathcal{E}^{r}$ via
\begin{align}
\label{eq:row-to-row}
    \raisebox{-0.48\height}{\includegraphics[width=0.14\textwidth, page=1]{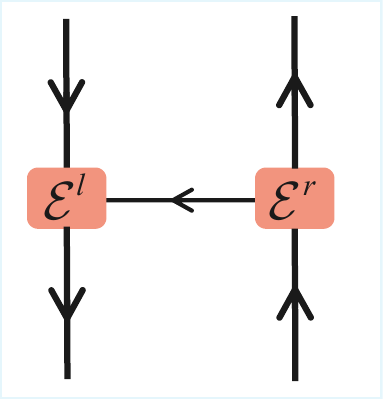}}.
\end{align}
Similarly, the column-to-column transfer matrix is defined by $\mathcal{E}^{u}$ and $\mathcal{E}^{d}$ as
\begin{align}
\label{eq:col-to-col}
    \raisebox{-0.48\height}{\includegraphics[width=0.16\textwidth, page=1]{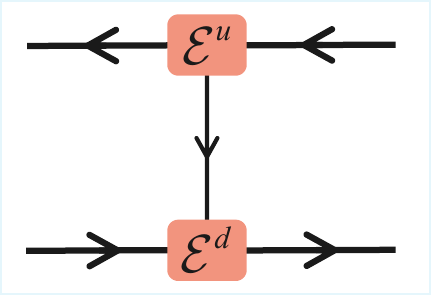}}.
\end{align}
Both Eqs.~\eqref{eq:row-to-row} and \eqref{eq:col-to-col} define $D^{2}\times D^{2}$ Grassmann matrices.
From these transfer matrices, one can evaluate the effective correlation length $\xi_{D}$ at the given bond dimension $D$~\cite{NISHINO199669}, whose definition is as follows:
\begin{align} 
\label{eq:cl_def}
    \frac{1}{\xi_D}
    = \ln \frac{\lambda_{1}}{\lambda_{2}},
\end{align}
where $\lambda_{1}$ and $\lambda_{2}$ are the leading and the sub-leading eigenvalues of row-to-row, or column-to-column, transfer matrix.

As another benefit of the CTMRG method, we can define the classical analog of entanglement entropy by
\begin{align} 
\label{eq:ee_def}
    S_{D}
    = -{\rm Tr}\left( \rho_{D}\log\rho_{D} \right) =
    -\sum_{i=1}^{D} \theta_{i} \log\theta_{i}.
\end{align}
Here, $\theta_{i}$ denote the eigenvalues of the reduced density matrix $\rho_{D}$ with bond dimension $D$, which is represented as a product of four corner matrices~\cite{Ueda:2014jxq,Ueda:2017ojh}:
\begin{align} 
\label{eq:rdm_def}
    \rho_{D} = 
    \frac{1}{Z}
    \raisebox{-0.4\height}{\includegraphics[width=0.1\textwidth, page=1]{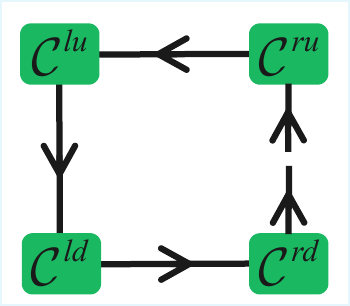}}
    .
\end{align}
When the system is at criticality, the central charge $c$ of the corresponding conformal field theory (CFT) can be estimated using the effective correlation length $\xi_{D}$ and entanglement entropy $S_{D}$ via
\begin{align} 
\label{eq:CK_formula}
    S_{D}
    \simeq
    \dfrac{c}{6} \log\xi_{D} + {\rm const}.
\end{align}
Eq.~\eqref{eq:CK_formula} is analogous to the Calabrese--Cardy formula~\cite{Calabrese:2004eu} and describes the scaling of the entanglement entropy induced by a finite bond dimension $D$~\cite{Tagliacozzo:2007rda,Pollmann:2009lnv}.
Although Refs.~\cite{Tagliacozzo:2007rda,Pollmann:2009lnv} are based on the MPS with bond dimension $D$ for one-dimensional quantum systems, the finite-$D$ scaling also holds in the case of CTMRG for two-dimensional classical systems~\cite{NISHINO199669,Ueda:2017ojh}.

We also introduce the quantity related to the entanglement spectrum by
\begin{align}
\label{eq:ES}
    \alpha_{i}=-2\log\theta_{i}.
\end{align}
Note that the entanglement spectrum serves as a useful quantity to reveal the topological phases of matter, as originally pointed out by Li and Haldane in the context of the fractional quantum Hall states~\cite{Li:2008kda}. 
Later, Ref.~\cite{Pollmann:2009ryx} demonstrates that the Haldane phase, a typical SPT phase in the $S=1$ spin chain, exhibits double degeneracies in the entire entanglement spectrum. 
These degeneracies are quite robust and generated by the same set of symmetries that protect the Haldane phase. 

\section{Numerical results}
\label{sec:results}

\subsection{Benchmarking with free Wilson fermions}

\begin{figure}
    \centering
    \subfigure[$M=1$]{
        \includegraphics[width=8.6cm]{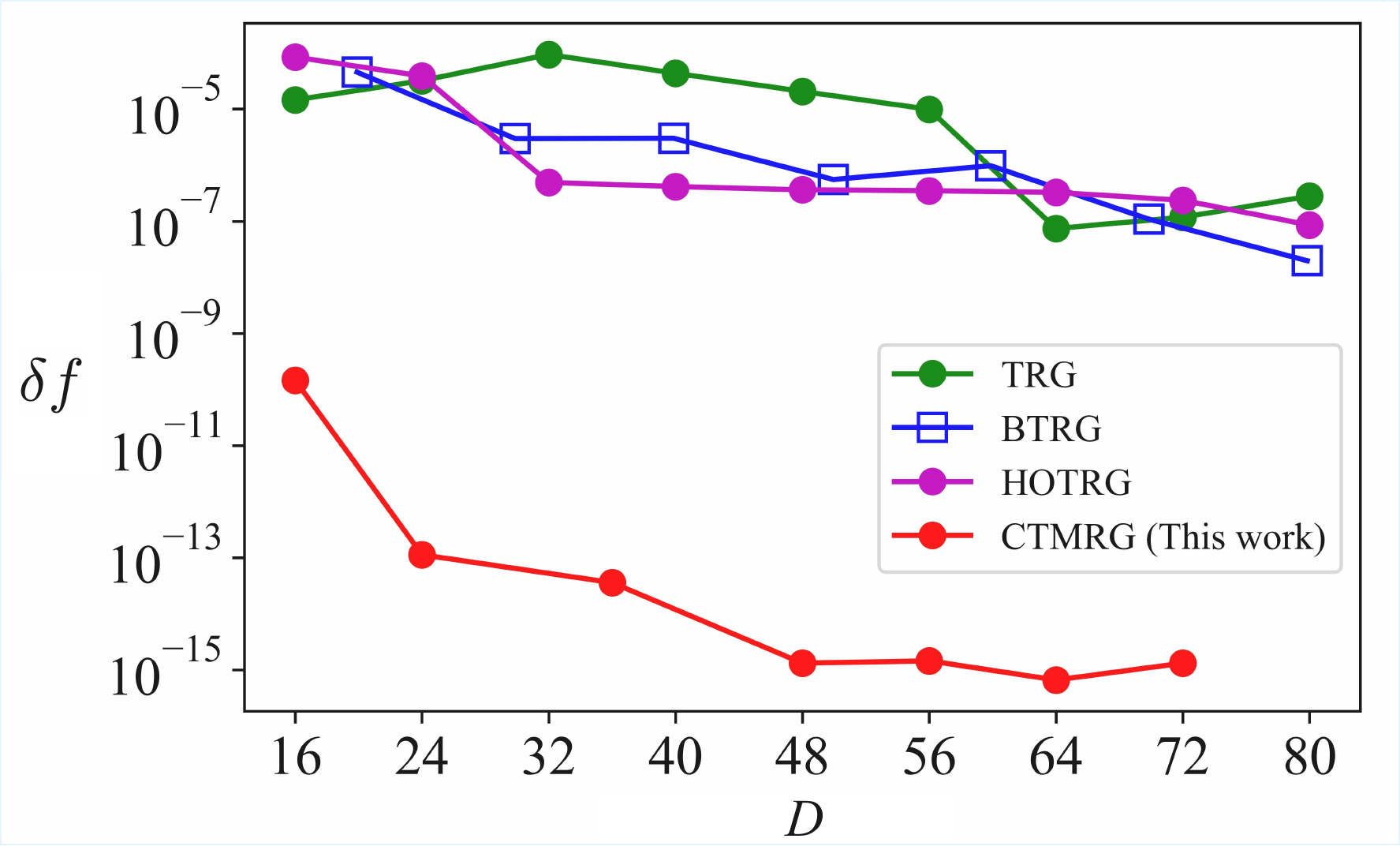}
    }
    \subfigure[$M=0$]{
        \includegraphics[width=8.6cm]{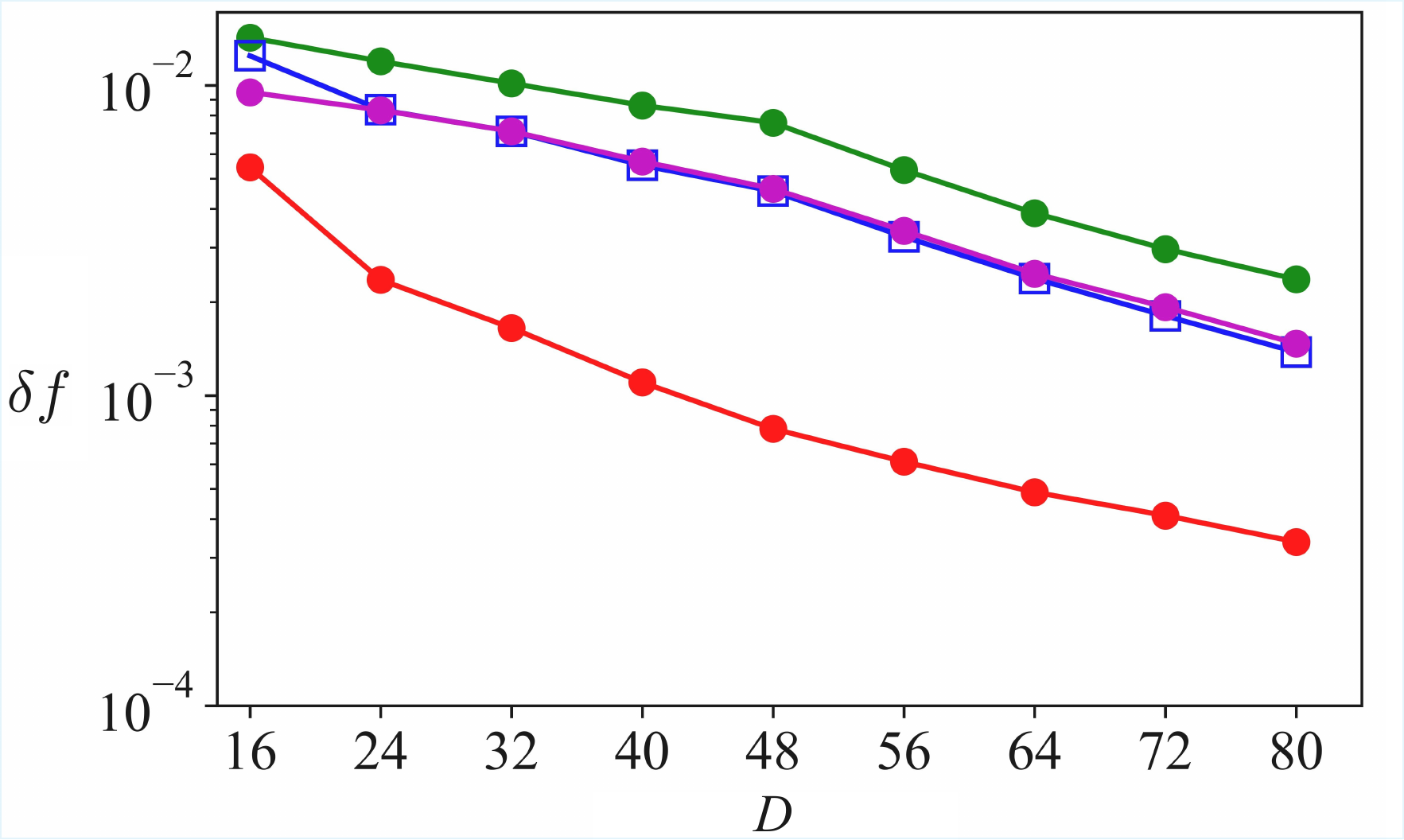}
    }
    \caption{
        The relative error $\delta f$ against the bond dimension $D$ for different Grassmann tensor network algorithms.
        When the system is away from criticality, the CTMRG outperforms other algorithms (a).
        Although the finite-$D$ effect is enhanced in all the algorithms near criticality, the CTMRG still shows the highest accuracy among these algorithms (b).
    }
    \label{fig:GNW_g2_0}
\end{figure}

We first validate the efficiency of the Grassmann CTMRG algorithm by benchmarking with the single-flavor free Wilson fermion theory, which is defined by the action in Eq.~\eqref{eq:GNW_action} with $N_{f}=1$ and $g^{2}_{\sigma}=g^{2}_{\pi}=0$.
We compute the free energy by varying the bond dimension $D$ and observe the relative error $\delta f$ defined by
\begin{align}
    \delta f
    =
    \left|
        \frac{\ln Z_{D}-\ln Z_{\rm{exact}}}{\ln Z_{\rm{exact}}}
    \right|,
\end{align}
where the resulting partition function at bond dimension $D$ is denoted by $Z_{D}$, while $Z_{\rm{exact}}$ denotes the exact solution.

Fig.~\ref{fig:GNW_g2_0}(a) shows that the Grassmann CTMRG reproduces the analytic solution almost exactly within double-precision accuracy at $M=1$, where the system is away from criticality, as discussed below.
For comparison, we also present results obtained using other conventional Grassmann tensor network algorithms, involving the Grassmann TRG~\cite{Levin:2006jai,Shimizu:2014uva}, Grassmann bond-weighted TRG (BTRG)~\cite{PhysRevB.105.L060402,Akiyama:2022pse}, and Grassmann higher-order TRG (HOTRG)~\cite{PhysRevB.86.045139,Sakai:2017jwp}.
Fig.~\ref{fig:GNW_g2_0}(a) clearly shows that the CTMRG outperforms conventional algorithms at the same bond dimension.
We also provide a similar benchmark at $M=0$, where the theory acquires an additional $U(1)$ symmetry beyond the normal $U(1)_{V}$ symmetry~\cite{Creutz:2011cd,Kimura:2011ik}. 
As shown in Fig.~\ref{fig:GNW_g2_0}(b), although the accuracy of all Grassmann tensor network algorithms deteriorates in this case, the CTMRG still achieves the highest accuracy among them.

\subsection{Phase diagram}

\begin{figure}[htbp]
	\centering
	\includegraphics[width=15cm]{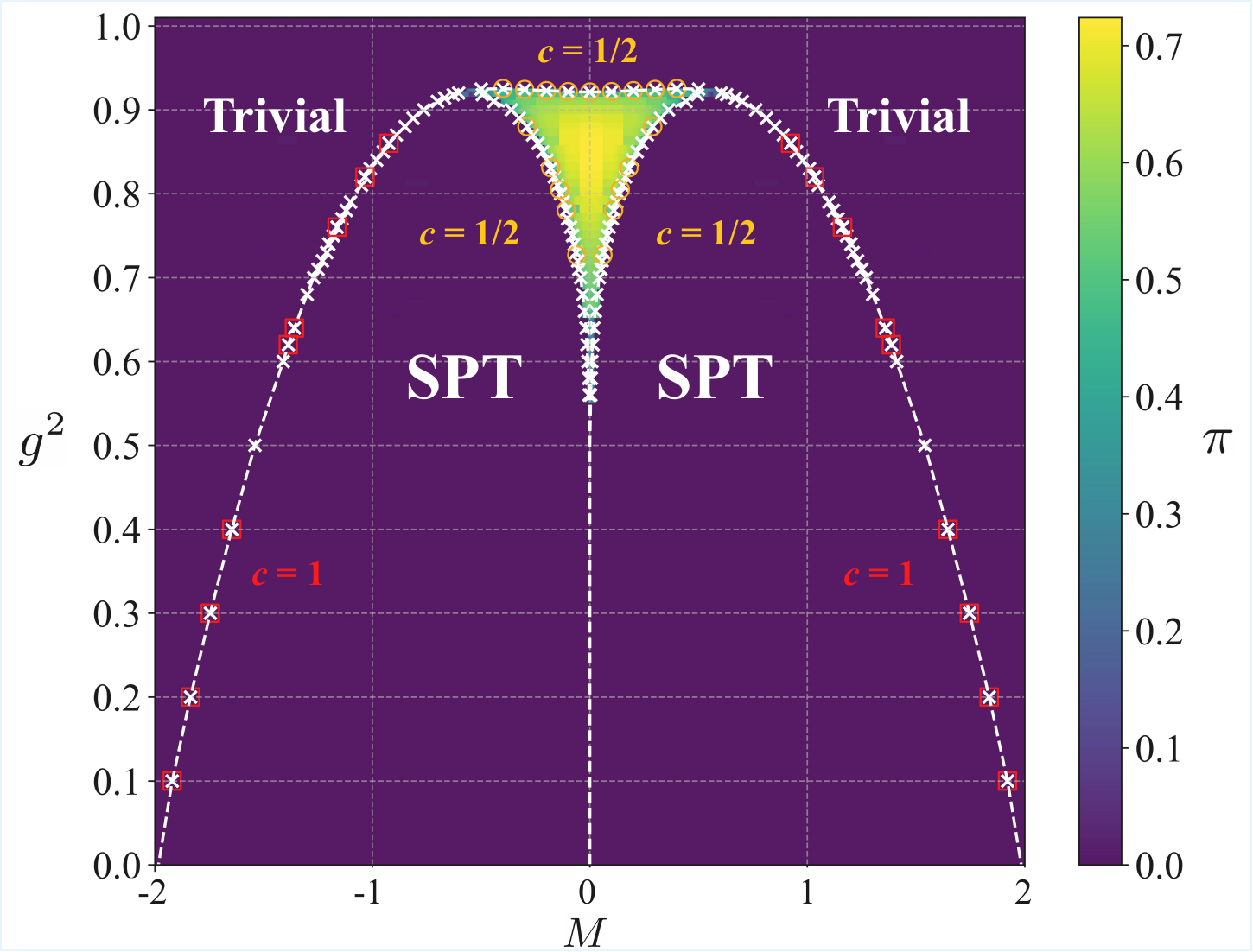}
	\caption{
        Phase diagram of the $N_{f}=1$ GNW model on the $(M, g^{2})$ plane from the Grassmann CTMRG with $D = 128$.
        The model exhibits three distinct phases: the Aoki phase, the topological insulator phase, and the trivial phase.
        The heat map represents the magnitude of the pseudoscalar condensate, which serves as the order parameter of the Aoki phase.
        White crosses indicate the parameter points at which Grassmann CTMRG detects critical behavior via peaks in the correlation length and entanglement entropy.
        Yellow circles denote the points where the extracted central charge is consistent with $c=1/2$, while red squares indicate the points consistent with $c=1$.
        The Aoki phase is separated from the other phases by critical lines characterized by $c=1/2$.
        The topological insulator and trivial phases are separated by critical lines with $c=1$.
    }
	\label{fig:phase_diagram}
\end{figure}

We now present the overall phase structure obtained from our Grassmann CTMRG computations in Fig.~\ref{fig:phase_diagram}.
The CTMRG identifies three distinct phases: the Aoki phase, the topological insulator phase, and the trivial phase.
The Aoki phase is separated from the other phases by critical lines characterized by $c=1/2$.
In contrast, the topological insulator and trivial phases are separated by critical lines with $c=1$.
The detailed characterization of each phase is presented below.
We find several consistencies with the large-$N_{f}$ phase diagram in Fig.~\ref{fig:largeN_phase_diagram}.
First, the phase diagram is mirror-symmetric with respect to $M=0$.
Second, the critical lines separating the Aoki phase and the topological insulating phase exhibit a characteristic two-lobe structure.
Although the trident-like shape of the Aoki phase shown in Fig.~\ref{fig:largeN_phase_diagram} is not clearly visible in Fig.~\ref{fig:phase_diagram}, our numerical results presented below suggest that such a trident-like structure indeed exists. 
The main difference from the large-$N_{f}$ phase diagram is that the Aoki phase is entirely surrounded by critical lines, and our numerical results indicate that the Aoki phase does not persist in the strong-coupling regime when $N_{f}=1$.

\subsubsection{Aoki phase at $M=0$}

\begin{figure}[htbp]
	\centering
	\includegraphics[width=14cm]{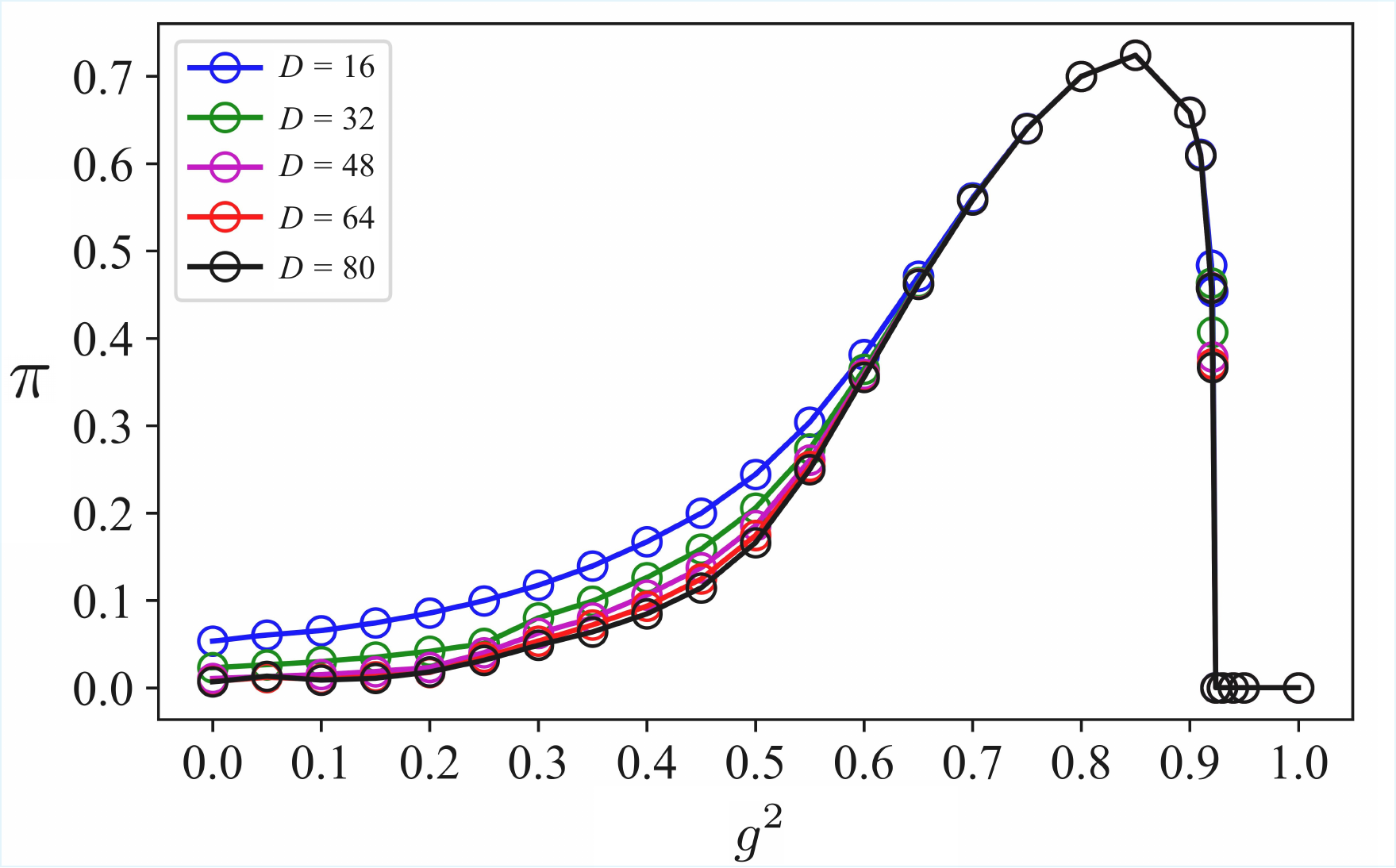}
	\caption{
        Pseudoscalar condensate at $M=0$ as a function of $g^{2}$ varying the bond dimension $D$.
    }
	\label{fig:GNW_m_-2_scan}
\end{figure}

\begin{figure}[htbp]
	\centering
    \subfigure[$g^{2}=0.2$]{
    \includegraphics[width=8.6cm]{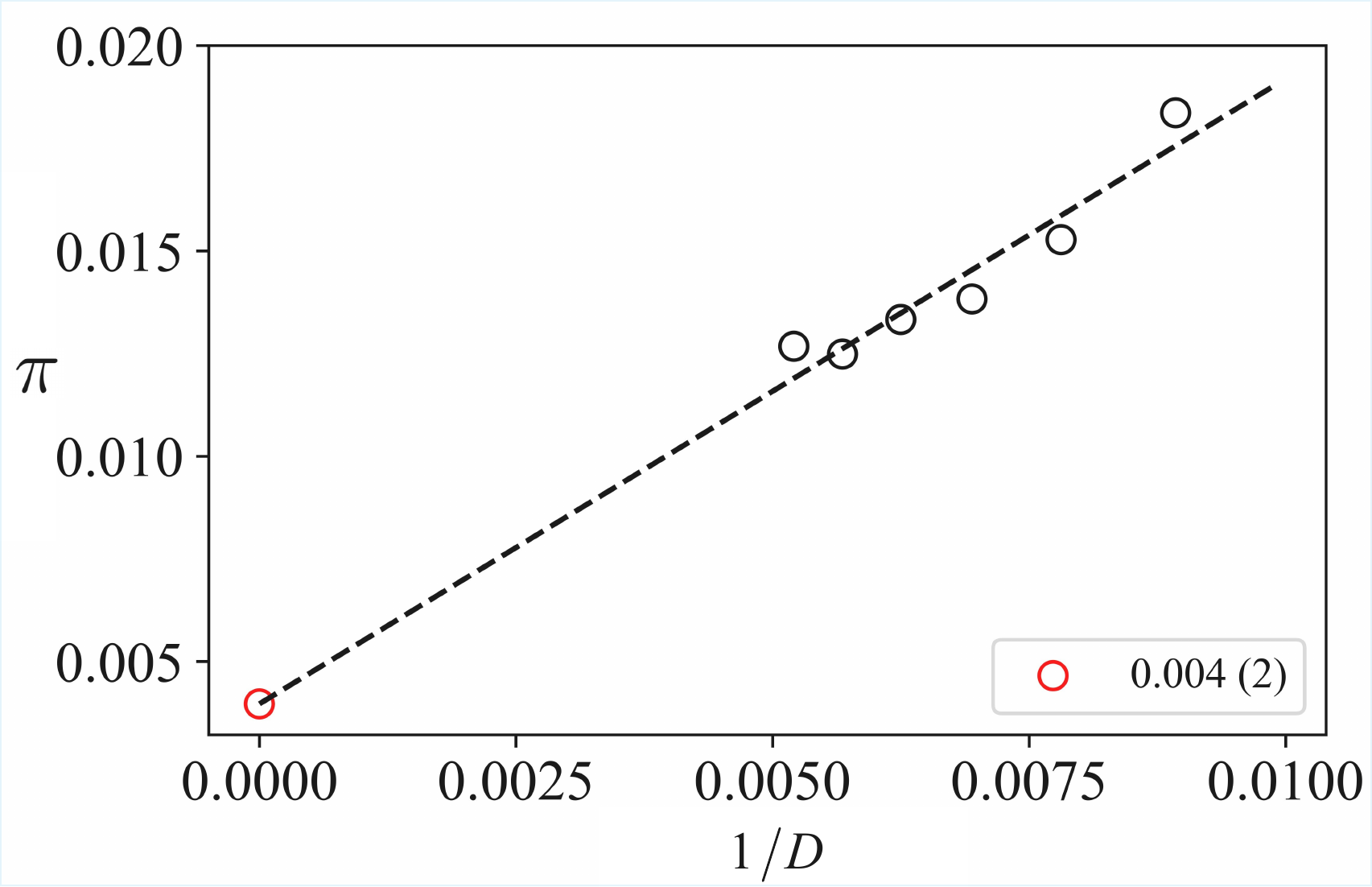}
    }
    \subfigure[$g^{2}=0.3$]{
    \includegraphics[width=8.6cm]{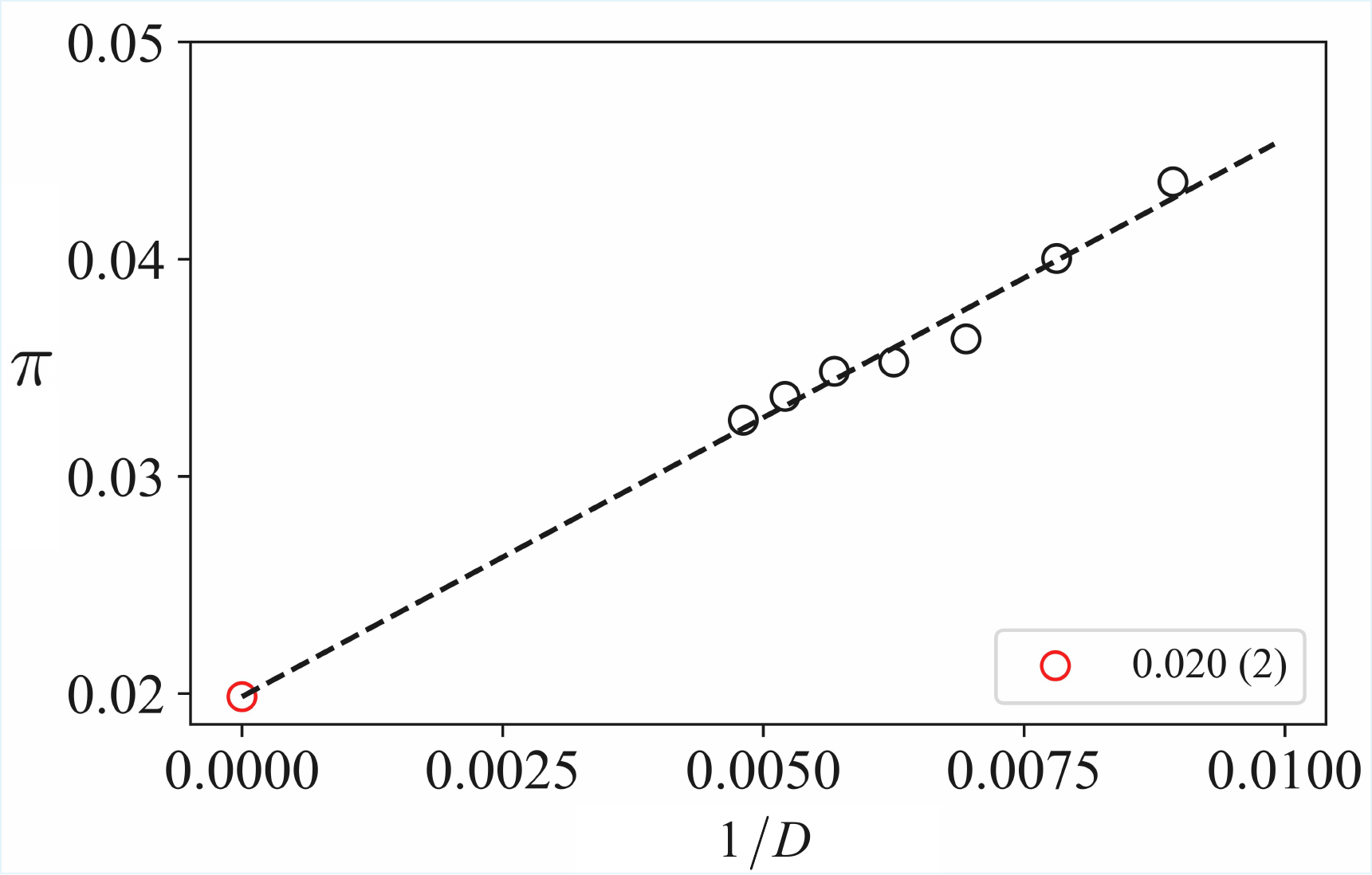}
    }
   \\
    \subfigure[$g^{2}=0.4$]{
    \includegraphics[width=8.6cm]{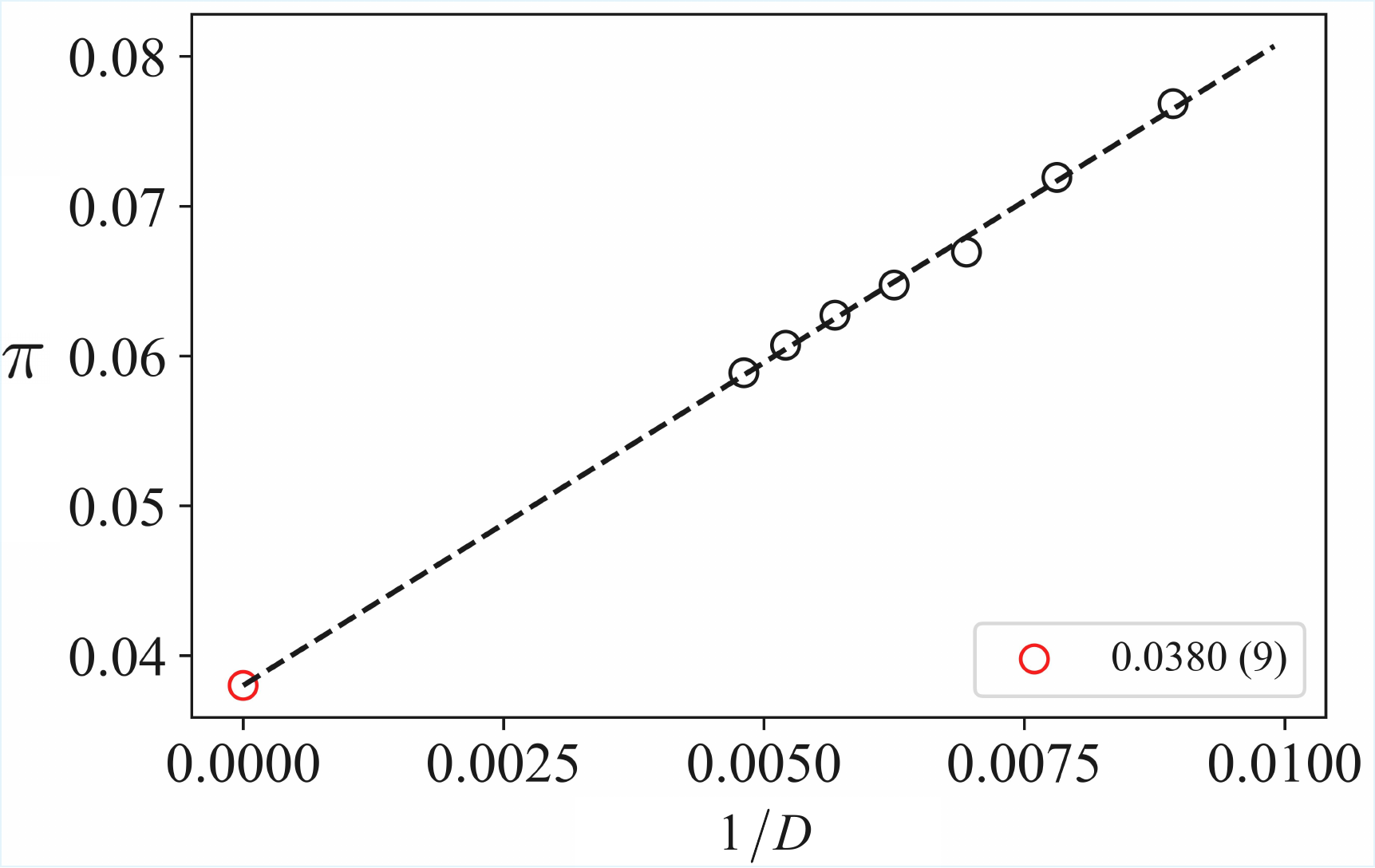}
    }
    \subfigure[$g^{2}=0.5$]{
    \includegraphics[width=8.6cm]{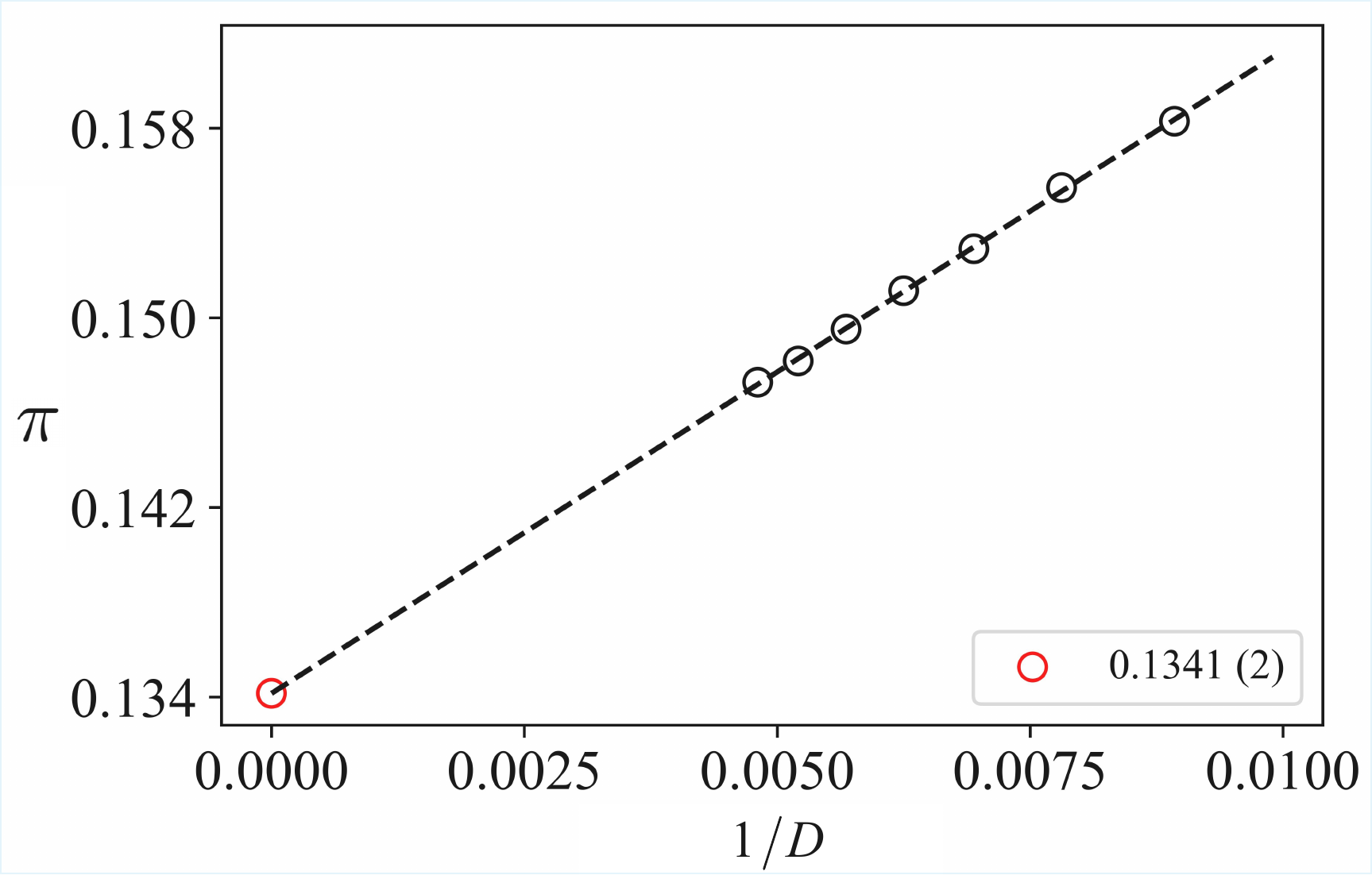}
    }
	\caption{
        Extrapolation of pseudoscalar condensate to the $D\to\infty$ limit at weak coupling.
        Dark circles show the results at finite $D$, while red ones denote the extrapolated values.
        Dashed lines denote our fitting form $\pi=a/D+C$.
    }
	\label{fig:extrap_m_-2}
\end{figure}

We first investigate the parity symmetry--broken phase at $M=0$.
Fig.~\ref{fig:GNW_m_-2_scan} shows the resulting pseudoscalar condensate as a function of $g^{2}$.
The finite-$D$ effects appear to be well suppressed for $g^{2}\gtrsim0.7$, where a clear signal of spontaneous parity symmetry breaking is observed up to $g^{2}\sim0.9$.
In contrast, the behavior of the pseudoscalar condensate strongly suggests that parity symmetry remains unbroken at $g^{2}\gtrsim0.9$.
We also observe that the finite-$D$ effect is enhanced at smaller values of $g^{2}$. 
In particular, the magnitude of the pseudoscalar condensate decreases as the bond dimension is increased.
As already suggested by Fig.~\ref{fig:GNW_g2_0}(b), the finite-$D$ effects are expected to become significant in the weak-coupling regime at $M=0$.
Therefore, we further push the bond dimension of the CTMRG up to $D=208$.
Fig.~\ref{fig:extrap_m_-2} shows the pseudoscalar condensate in the weaker coupling regime extrapolated to the $D\to\infty$ limit.
We fit the results assuming $\pi=a/D+C$, where $a$ and $C$ are free parameters.
We find a non-zero pseudoscalar condensate at least for $g^2\ge0.2$.

The vanishing pseudoscalar condensate in the strong-coupling region is highly contrasted with the large-$N_{f}$ prediction, where the Aoki phase persists in the strong coupling limit. 
However, we expect that when the four-fermion interaction term dominates the action in Eq.~\eqref{eq:GNW_action}, the contribution from the kinetic term becomes negligible and the theory is trivially gapped.
Therefore, the CTMRG result and the large-$N_{f}$ phase diagram suggest that the critical $g^{2}$ will be pushed further into the strong-coupling regime as $N_{f}$ is increased.
We further note that this observation may not be in contradiction with a recent prediction from the ’t~Hooft anomaly matching at $M=0$, which implies that parity symmetry is broken for any value of $g^{2}$ in the $N_{f}=1$ GNW model~\cite{Misumi:2019jrt}.
This can be because the discussion in Ref.~\cite{Misumi:2019jrt} is based on a continuum approximation, which is expected to be valid near the continuum limit. 
In contrast, our numerical analysis treats the system purely as a lattice model, and the large-$g^{2}$ regime lies far outside the vicinity of the continuum limit.
At the same time, since the prediction in Ref.~\cite{Misumi:2019jrt} should hold at the weak-coupling regime, we anticipate that a much larger bond dimension is required to numerically resolve a finite pseudoscalar condensate for $g^{2}\le0.2$.
\footnote{
Since the CTMRG assumes the open boundary condition, the ’t~Hooft anomaly matching argument discussed in Ref.~\cite{Misumi:2019jrt} cannot be directly applied to our results.
However, we further remark that the resulting pseudoscalar condensate in Fig.~\ref{fig:GNW_m_-2_scan} is not sensitive to the choice of boundary conditions.
As demonstrated in Appendix~\ref{app:GTRG}, the pseudoscalar condensate computed by the HOTRG algorithm, which assumes periodic boundary conditions, also exhibits a critical coupling at which the Aoki phase terminates.
}

\subsubsection{Phase boundaries of the Aoki phase}
\label{subsubsec:phase_boundaries_Aoki}

\begin{figure}[htbp]
	\centering
	\includegraphics[width=14cm]{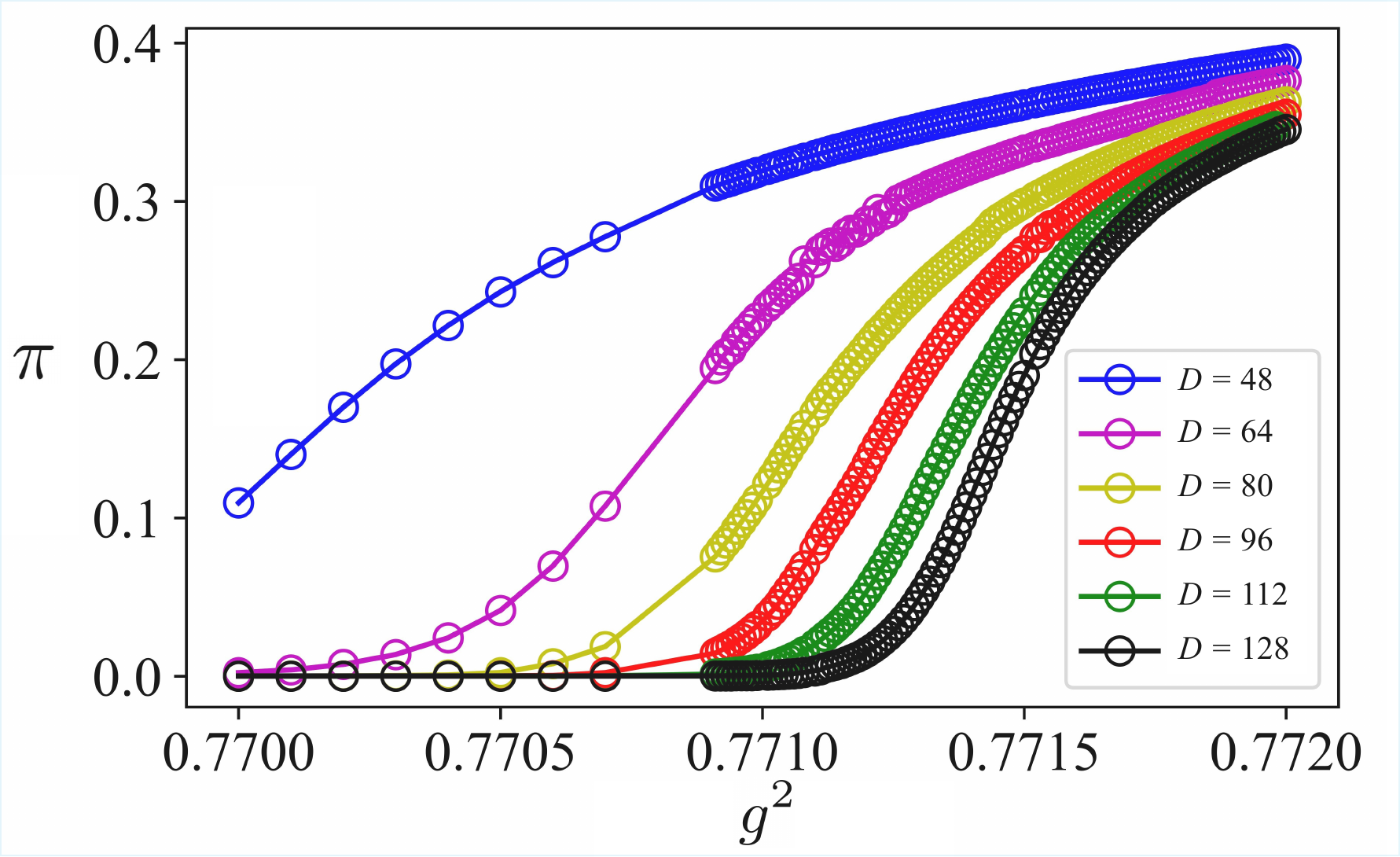}
	\caption{
        Pseudoscalar condensate at $M=0.1$ as a function of $g^{2}$ varying the bond dimension $D$.
    }
	\label{fig:PS_m_1p9_lobe}
\end{figure}

\begin{figure}
    \centering
    \subfigure[]{
        \includegraphics[width=8.7cm]{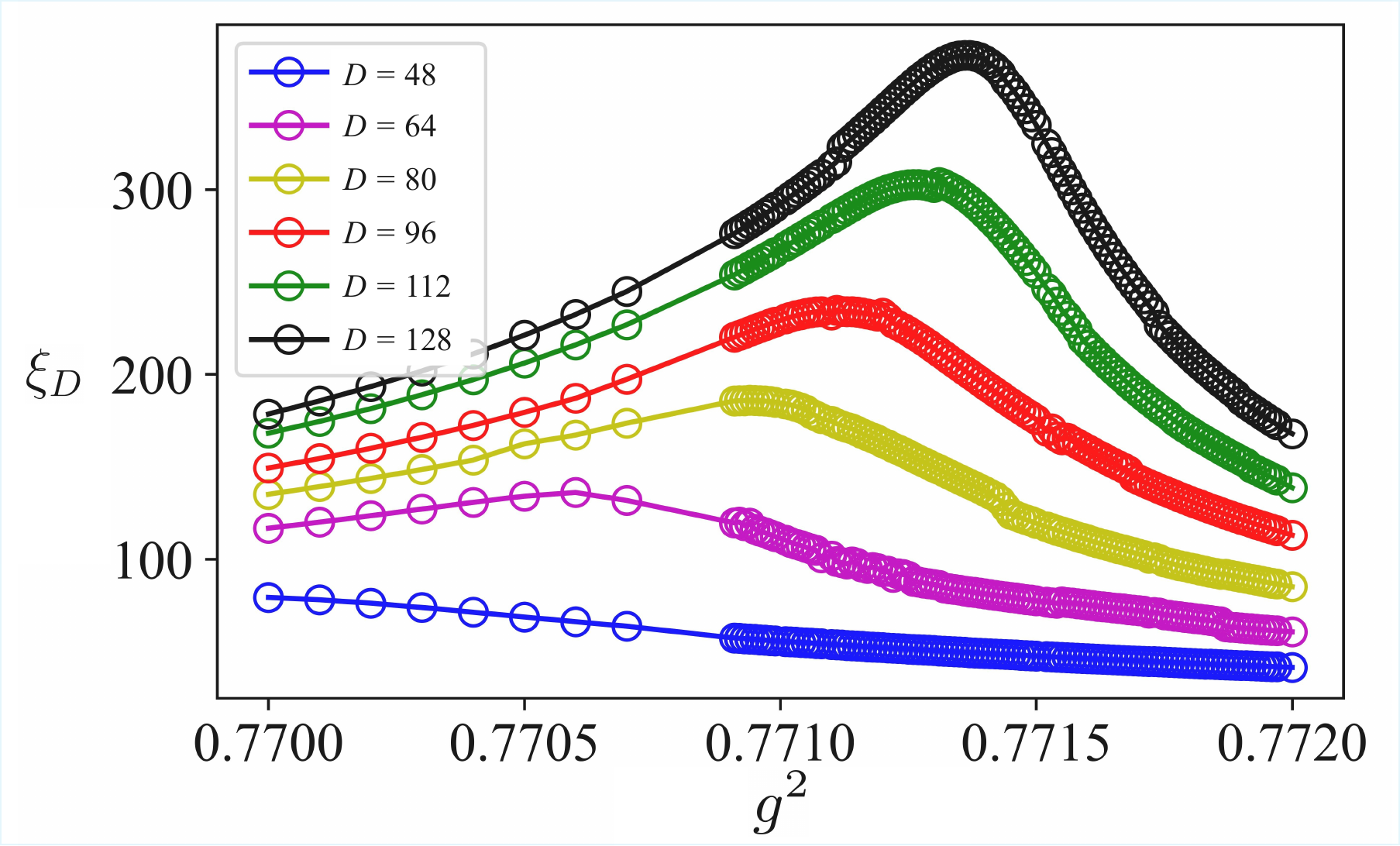}
    }
    \subfigure[]{
        \includegraphics[width=8.6cm]{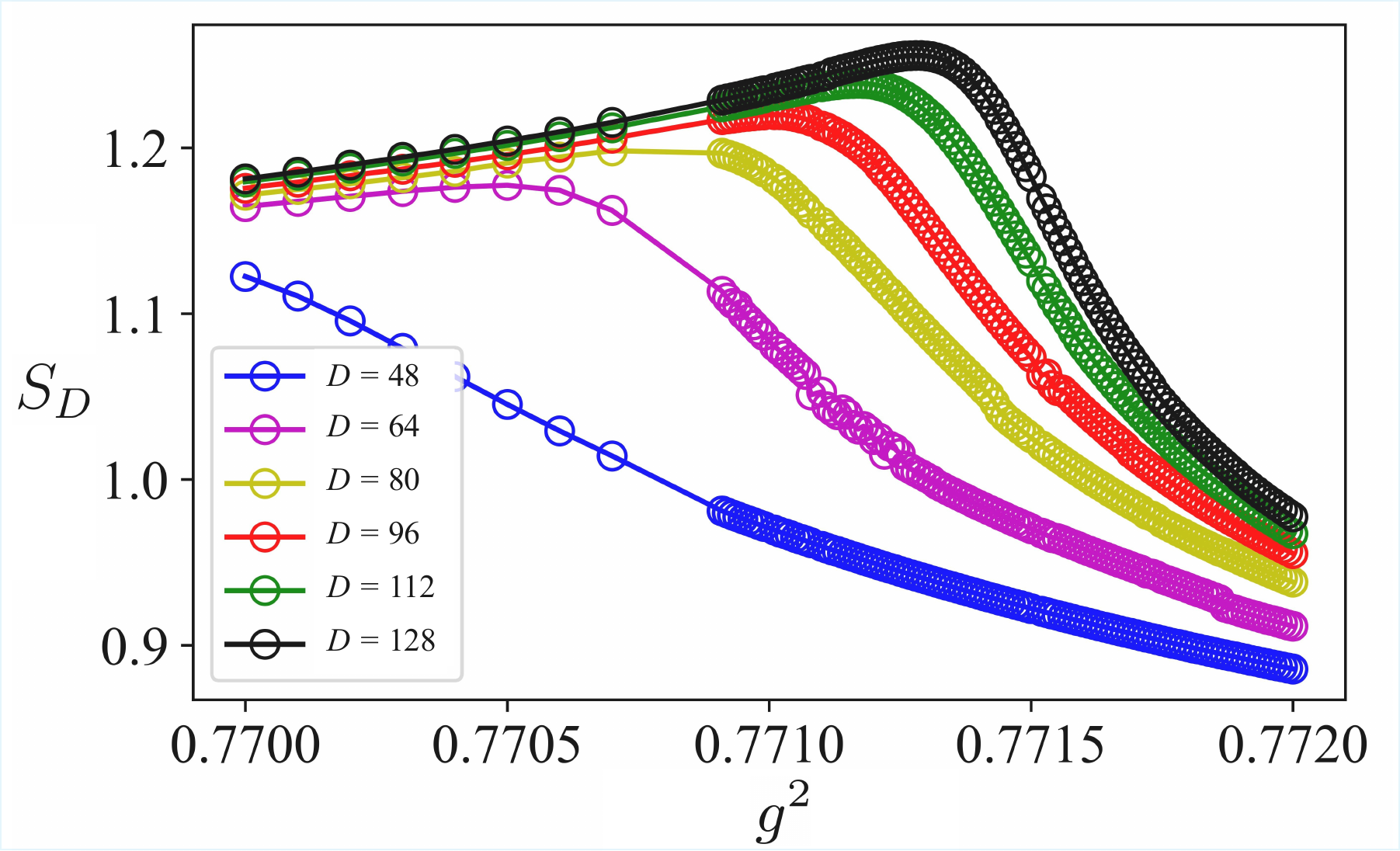}
    }
    \caption{
        Correlation length (a) and entanglement entropy (b) at $M=0.1$ as a function of $g^{2}$ varying the bond dimension $D$.
    }
    \label{fig:EE-Xi_m_1p9_lobe}
\end{figure}

\begin{figure}
    \centering
    \subfigure[$g^{2}=0.77181$]{
    \includegraphics[width=8.6cm]{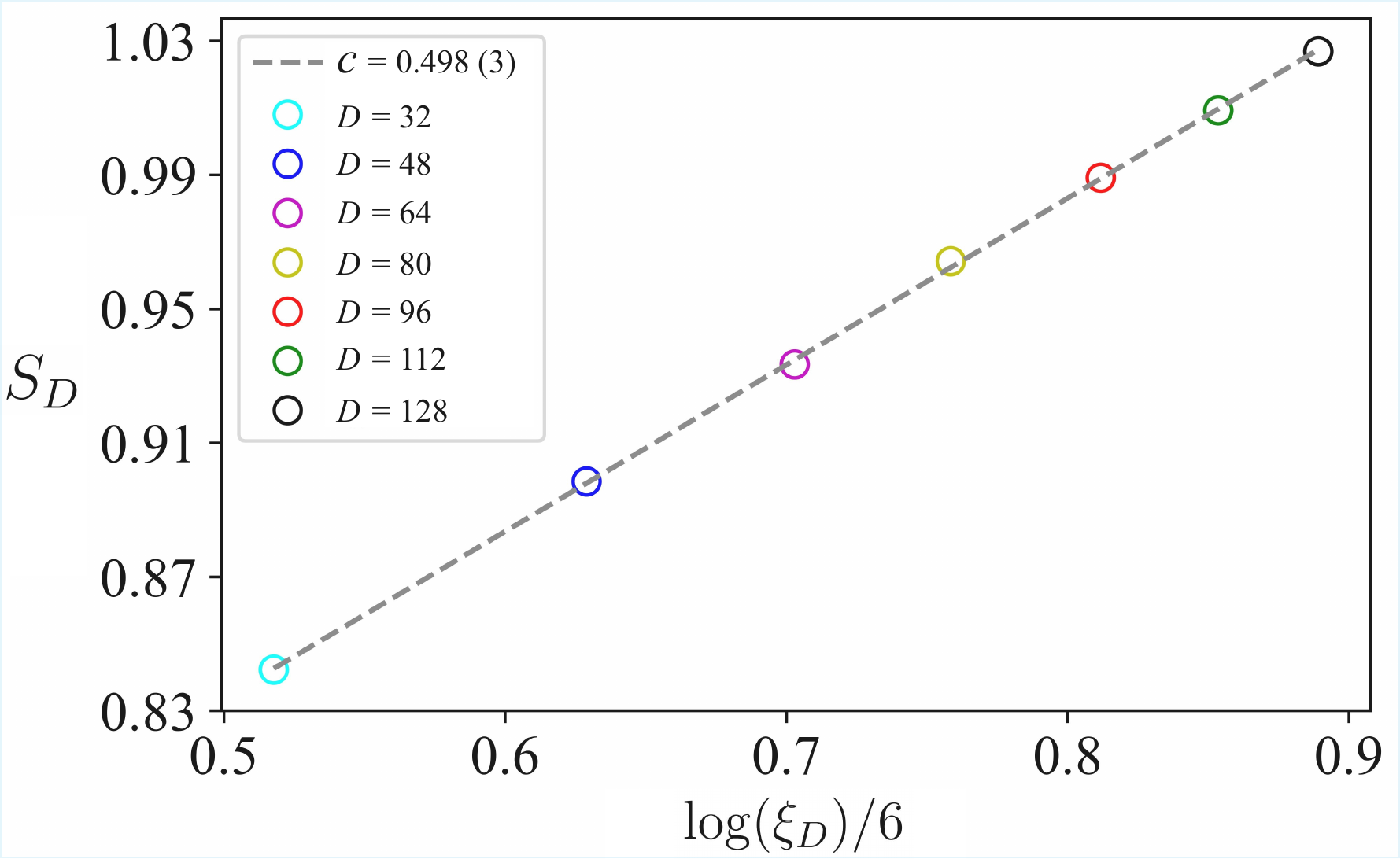}
    }
    \subfigure[$g^{2}=0.92161$]{
    \includegraphics[width=8.6cm]{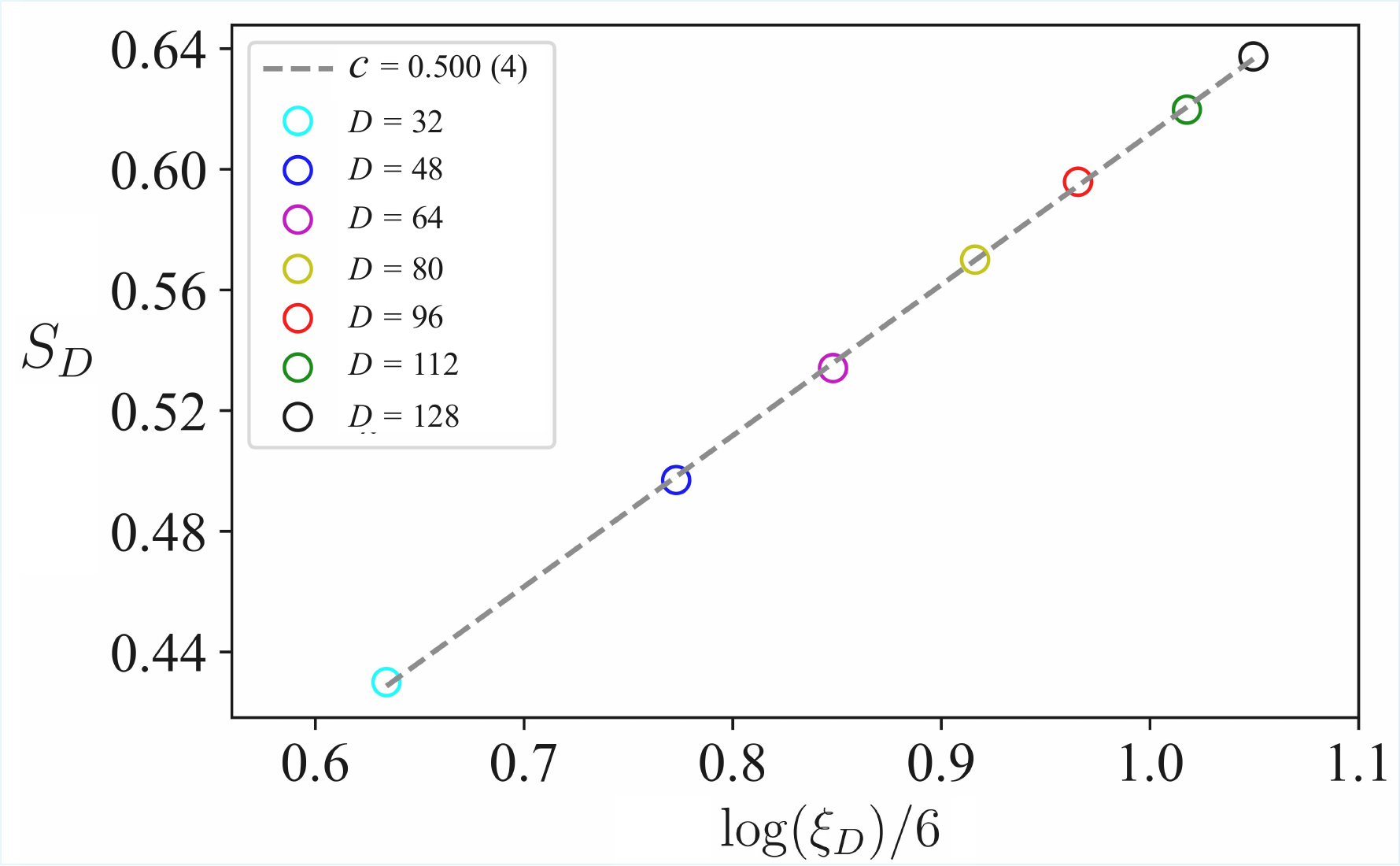}
    }
    \caption{
        Entanglement entropy as a function of the effective correlation length at $M=0.1$.
    }
    \label{fig:ising_CFT}
\end{figure}

According to the large-$N_{f}$ saddle-point approximation, the phase diagram exhibits two lobes within which the parity symmetry remains unbroken in the weak-coupling regime.
We now study the phase boundaries of the Aoki phase by employing the pseudoscalar condensate, correlation length, and entanglement entropy.
Fig.~\ref{fig:PS_m_1p9_lobe} shows the pseudoscalar condensate at $M=0.1$ obtained at various bond dimensions.
As $g^2$ increases, the system undergoes a phase transition from the symmetric phase to the Aoki phase.
We find that the extent of the Aoki phase is typically overestimated at small $D$.
Fig.~\ref{fig:EE-Xi_m_1p9_lobe} shows the correlation length and entanglement entropy in the same parameter region as Fig.~\ref{fig:PS_m_1p9_lobe}.
Although it is not straightforward to locate the critical $g^{2}$ based on the pseudoscalar condensate alone, we instead use the correlation length and the entanglement entropy via Eq.~\eqref{eq:CK_formula}. 
\footnote{
We also note that the data collapse of the entanglement entropy can provide an alternative way to identify both the critical point and the universality class, as discussed in Appendix~\ref{app:FES}.
}
A linear fit of $S_{D}$ as a function of $\log\xi_{D}$ yields an estimate of the central charge $c$.
We find that Eq.~\eqref{eq:CK_formula} indeed holds at $(M, g^{2})=(0.1, 0.77181)$ with $c=0.498(3)$, which is in agreement with the two-dimensional Ising universality class, as shown in Fig.~\ref{fig:ising_CFT}(a).
We note that by assuming the critical coupling $g^{2}_{c}=0.77181$ at $M=0.1$, the entanglement entropy $S_{D}$ for different $D$ collapse onto a single curve, as shown in Appendix~\ref{app:FES}.
This further supports that the system is critical at $(M, g^{2})=(0.1, 0.77181)$.

Based on the previous observation of the pseudoscalar condensate at $M=0$ in the large-$g^{2}$ region, as shown in Fig.~\ref{fig:GNW_m_-2_scan}, it is natural to expect the existence of another phase boundary of the Aoki phase in the strong-coupling regime.
We indeed find an additional criticality in the strong-coupling region. 
Eq.~\eqref{eq:CK_formula} again helps us estimate another critical coupling as shown in Fig.~\ref{fig:ising_CFT}(b).
We observe that Eq.~\eqref{eq:CK_formula} holds at $(M, g^{2})=(0.1, 0.92161)$ with $c=0.500(4)$.
Therefore, both of the critical points are described by the two-dimensional Ising CFT.
By repeating the above analysis for various values of $M\in[-2,2]$, we find that the $c=1/2$ critical lines always separate the Aoki phase.
We also find that the structure of the Aoki phase is symmetric with respect to $M=0$, as shown in Fig.~\ref{fig:phase_diagram}.
Before further investigating the fate of the two critical lines found by the CTMRG, we examine in more detail the two-lobe structure predicted by the large-$N_{f}$ phase diagram.

\subsubsection{Topological insulating phase}

So far, our CTMRG simulations indicate that the Aoki phase is separated by critical lines characterized by $c=1/2$. 
In the weak-coupling regime, these $c=1/2$ critical lines partially form the two-lobe structure.
Away from the $M=0$ line in the weak-coupling region, the CTMRG finds that the Aoki phase rapidly disappears.
Nevertheless, signatures of criticality are still captured through the behavior of the correlation length and the entanglement entropy.

\begin{figure}
    \centering
    \subfigure[]{
    \includegraphics[width=8.6cm]{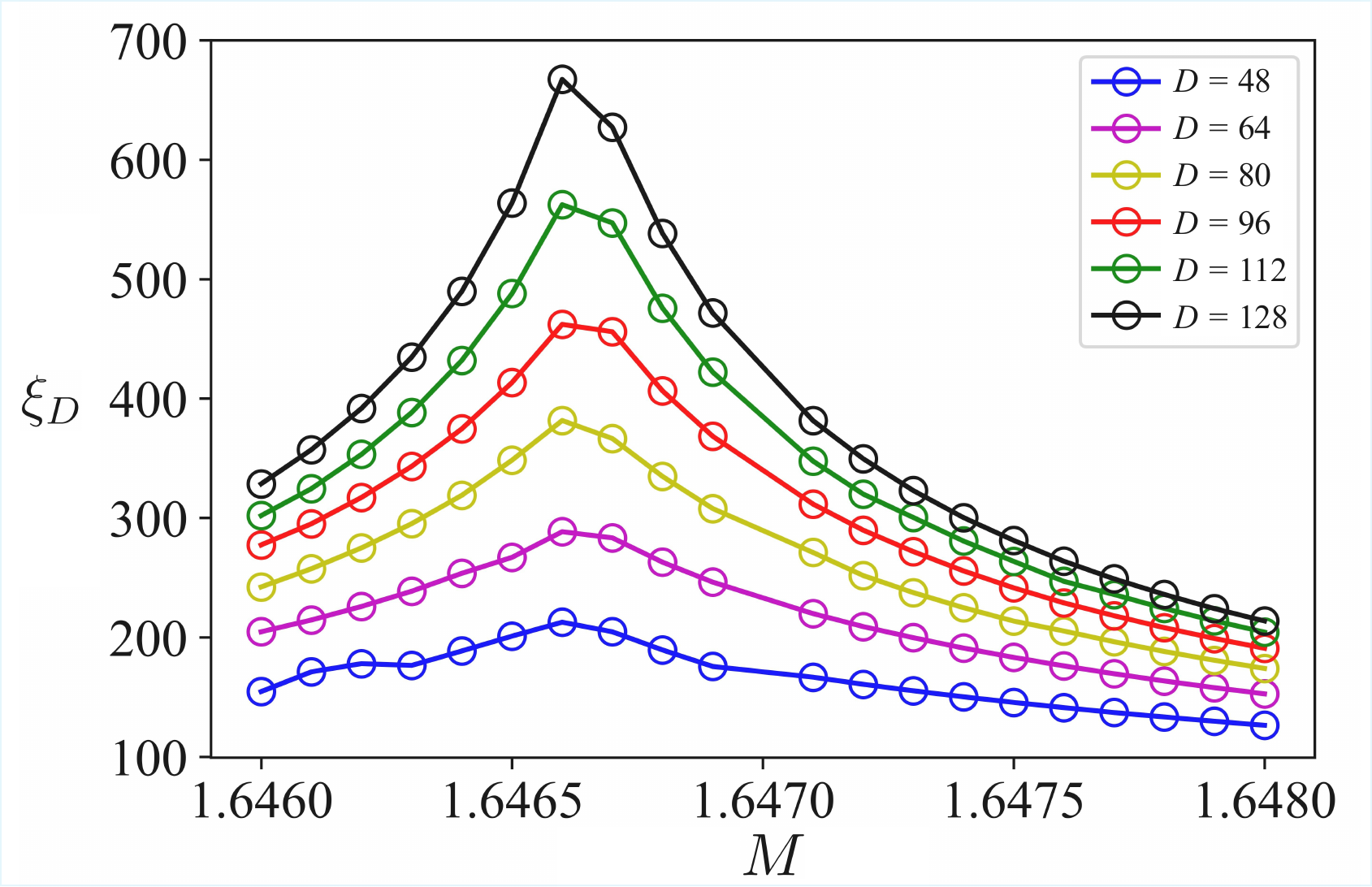}
    }
    \subfigure[]{
    \includegraphics[width=8.6cm]{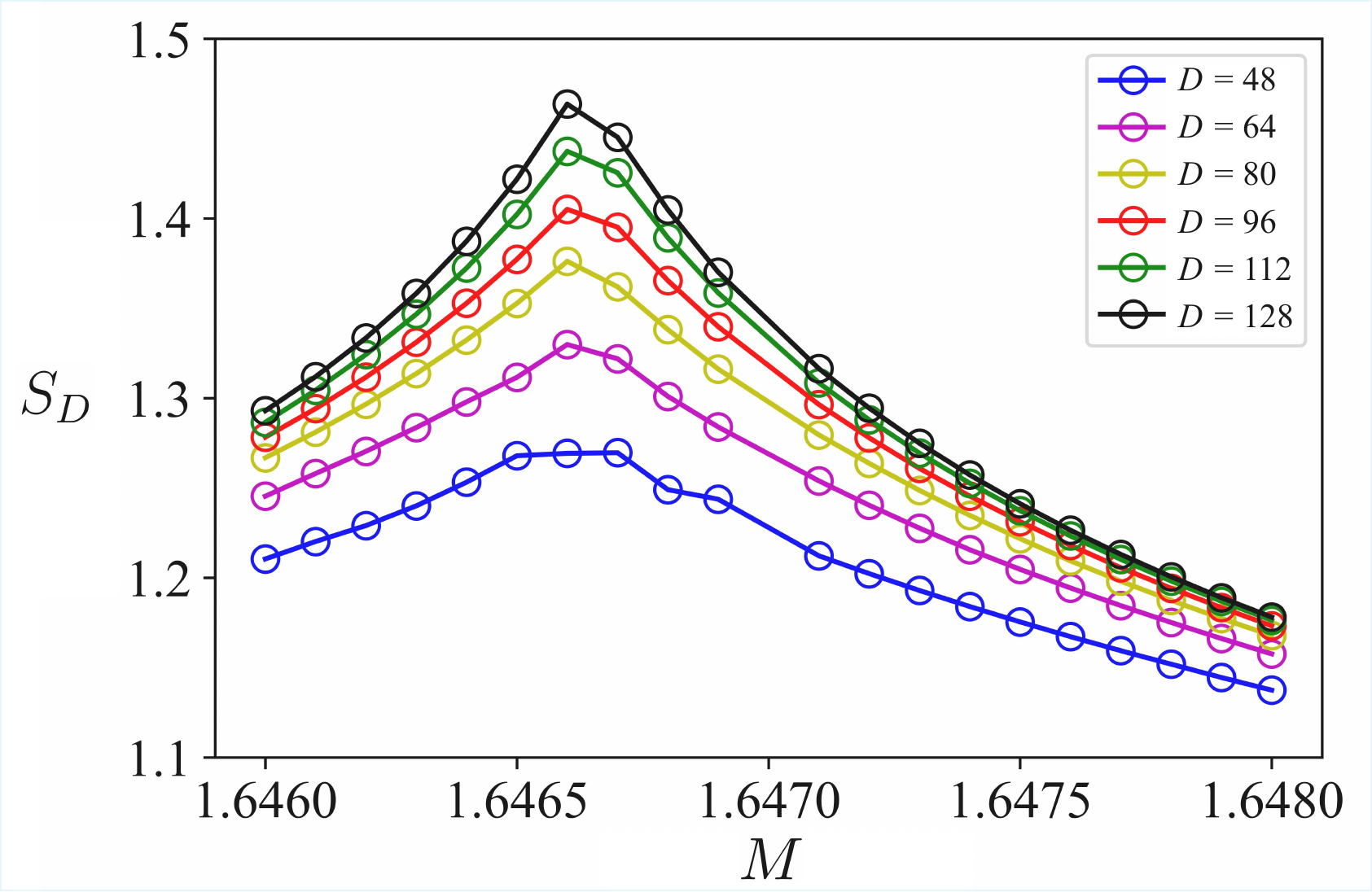}
    }
    \caption{
        Correlation length (a) and entanglement entropy (b) as a function of $M$ at $g^{2}=0.4$ varying the bond dimension $D$.
    }
    \label{fig:c1_criticality}
\end{figure}

\begin{figure}[htbp]
	\centering
	\includegraphics[width=14cm]{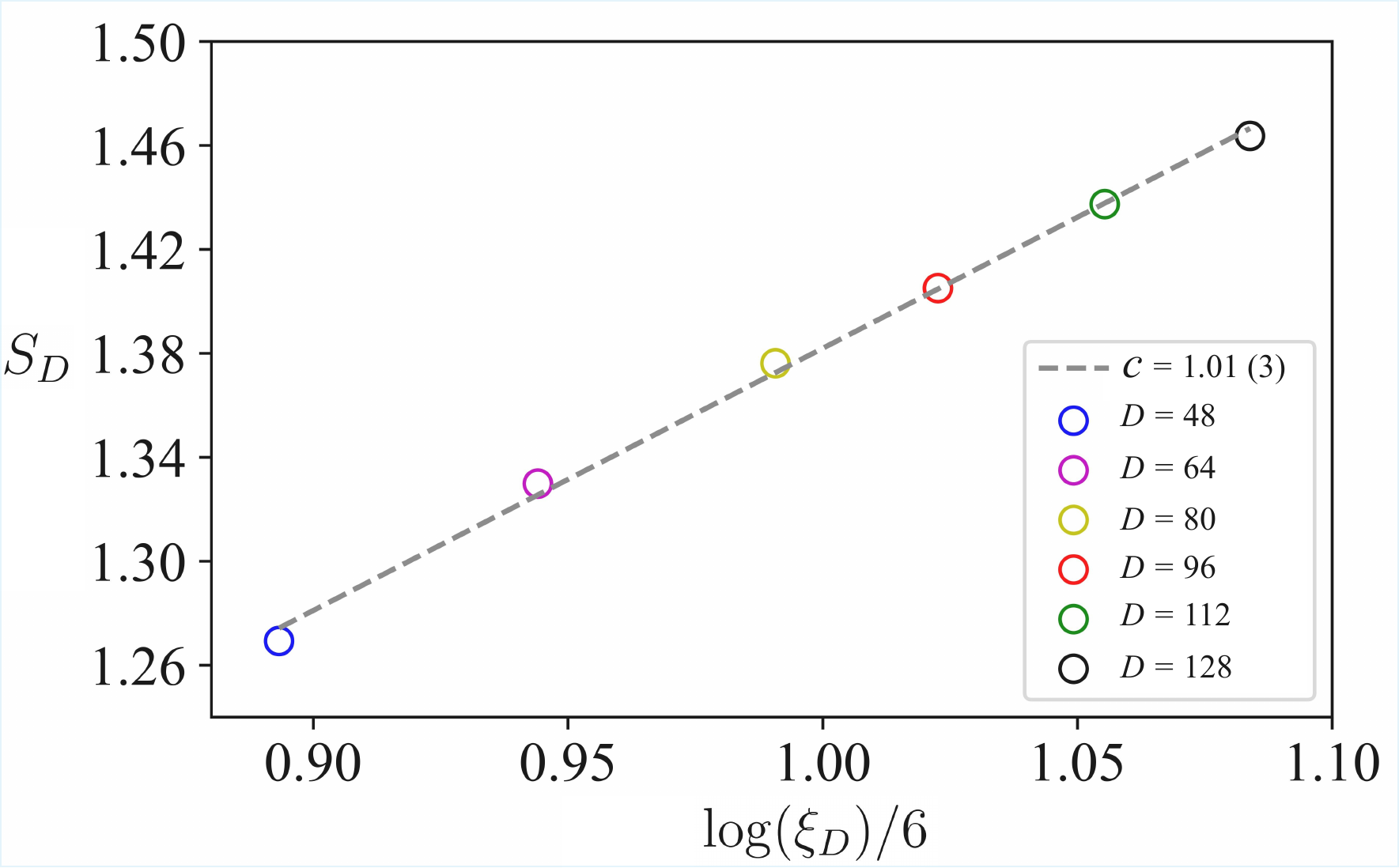}
	\caption{
        Entanglement entropy as a function of the effective correlation length at $(M,g^{2})=(1.6466,0.4)$.
    }
	\label{fig:c1_CFT}
\end{figure}

Fig.~\ref{fig:c1_criticality} shows the correlation length and entanglement entropy as a function of $M$ at $g^{2}=0.4$. 
Although no finite pseudoscalar condensate is observed in this parameter region, both quantities exhibit pronounced peaks around $M\sim1.6466$.
These observations indicate the presence of a critical point, which is further confirmed by the finite-entanglement scaling analysis based on Eq.~\eqref{eq:CK_formula}, as shown in Fig.~\ref{fig:c1_CFT}.
From the linear scaling behavior in terms of $\log\xi_{D}$, we extract a central charge $c=1.01(3)$ at $(M, g^{2})=(1.6466,0.4)$.
The resulting $c=1.01(3)$ seems natural, as the model is expected to be governed by a massless Dirac fermion, which is characterized by $c=1$, in the weak-coupling limit near $M=\pm2$.
By repeating this analysis for various $g^{2}$ in the weak-coupling regime, we indeed find that the two-lobe structure is characterized by the $c=1/2$ critical line around $M=0$ and the $c=1$ critical lines away from $M=0$.
We also find two identical $c=1$ critical lines, the one seemingly originating near $M=2$ and the other near $M=-2$, which are symmetric with respect to $M=0$, as shown in Fig.~\ref{fig:phase_diagram}.
This result is consistent with recent MPS simulations based on the Hamiltonian formalism~\cite{Bermudez:2018eyh}, in which the single-lobe structure is characterized by a $c=1/2$ critical line at large $g^{2}$ and by a $c=1$ line at weak couplings. 
The difference in the number of lobes arises from the absence of doubler modes associated with the continuous temporal direction in the Hamiltonian formalism.

\begin{figure}[htbp]
	\centering
    \subfigure[$M=2.8$]{
        \includegraphics[width=5.2cm]{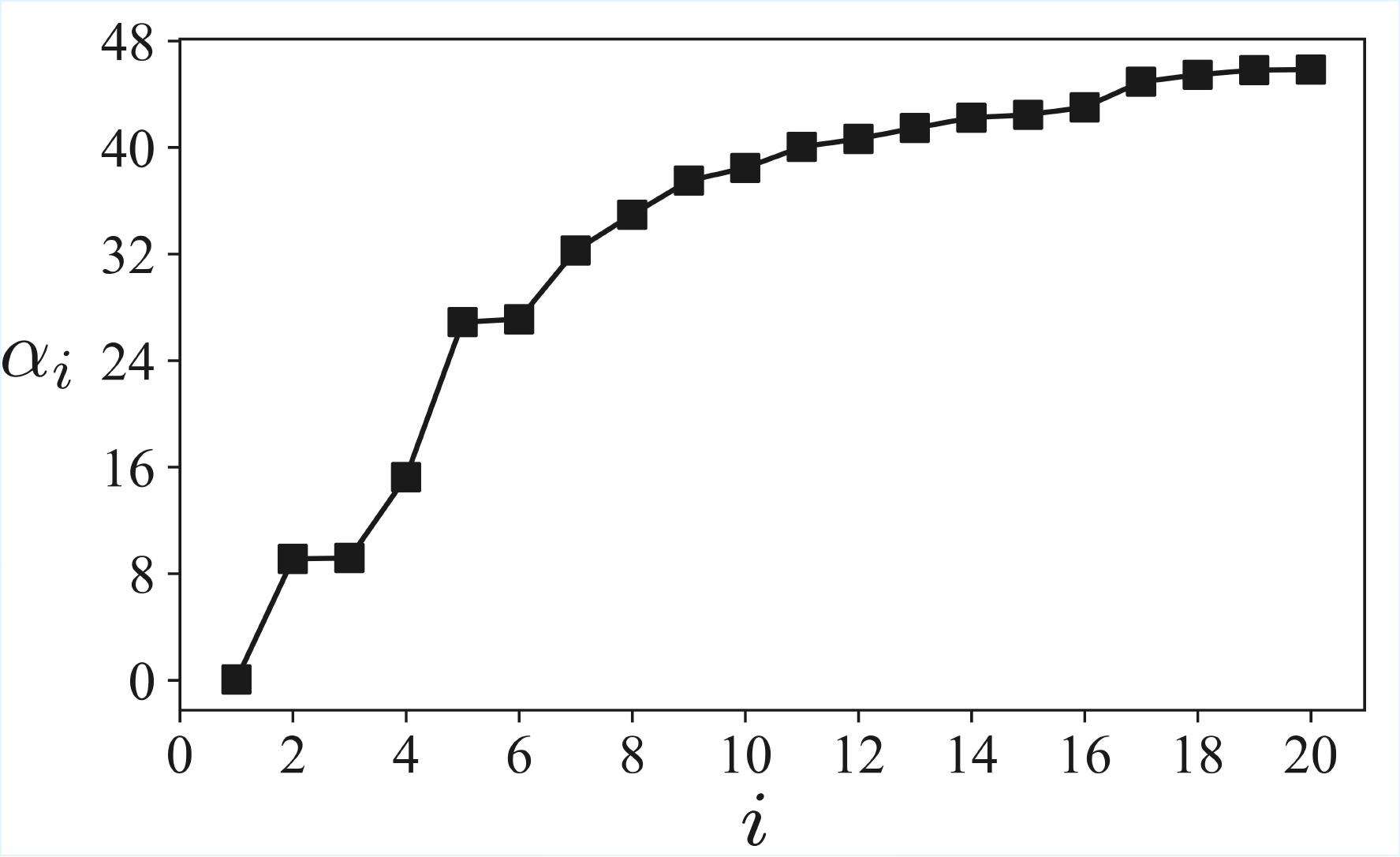}
    }
    \subfigure[$M=0.1$]{
        \includegraphics[width=5.2cm]{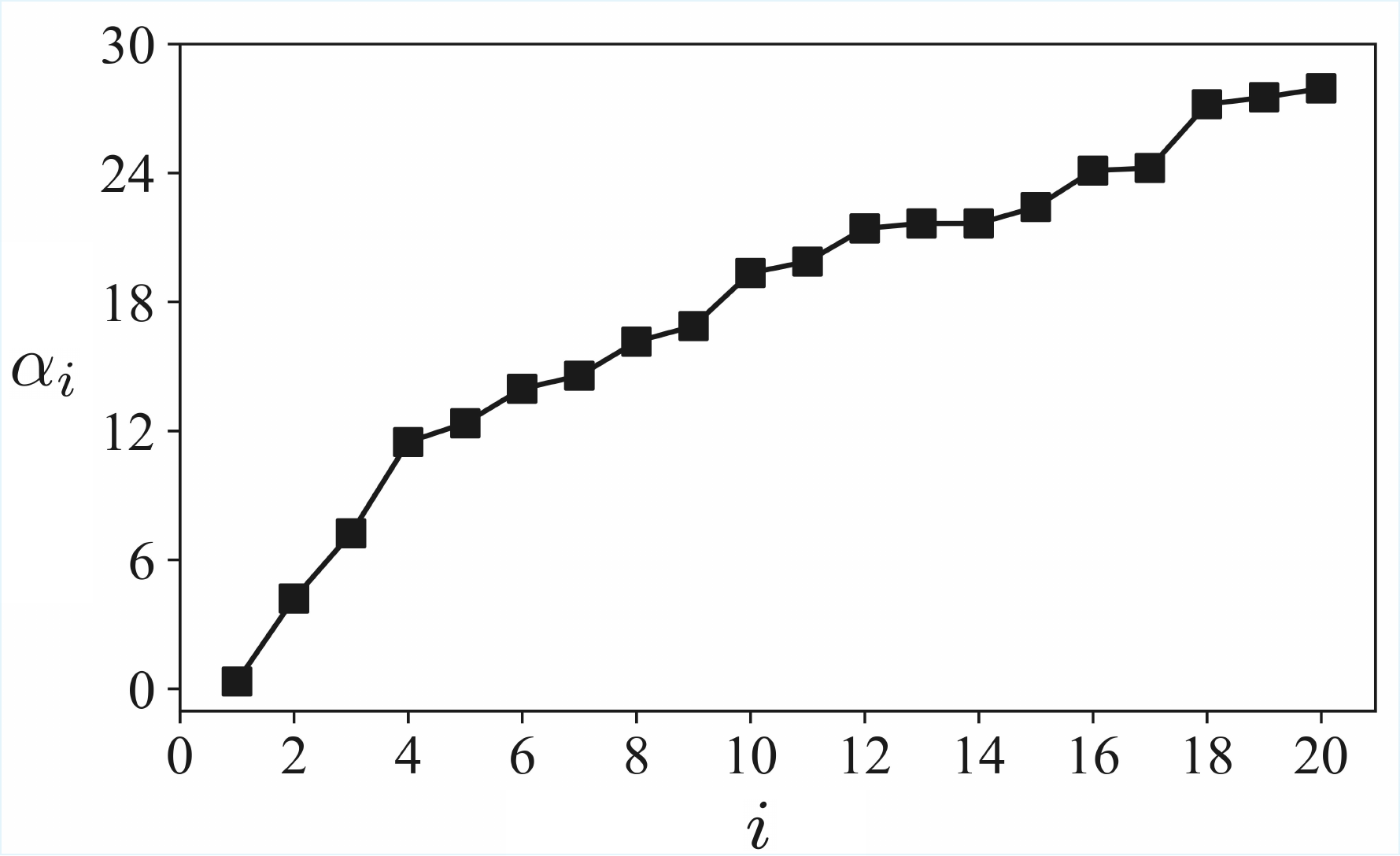}
    }
    \subfigure[$M=0.6$]{
        \includegraphics[width=5.2cm]{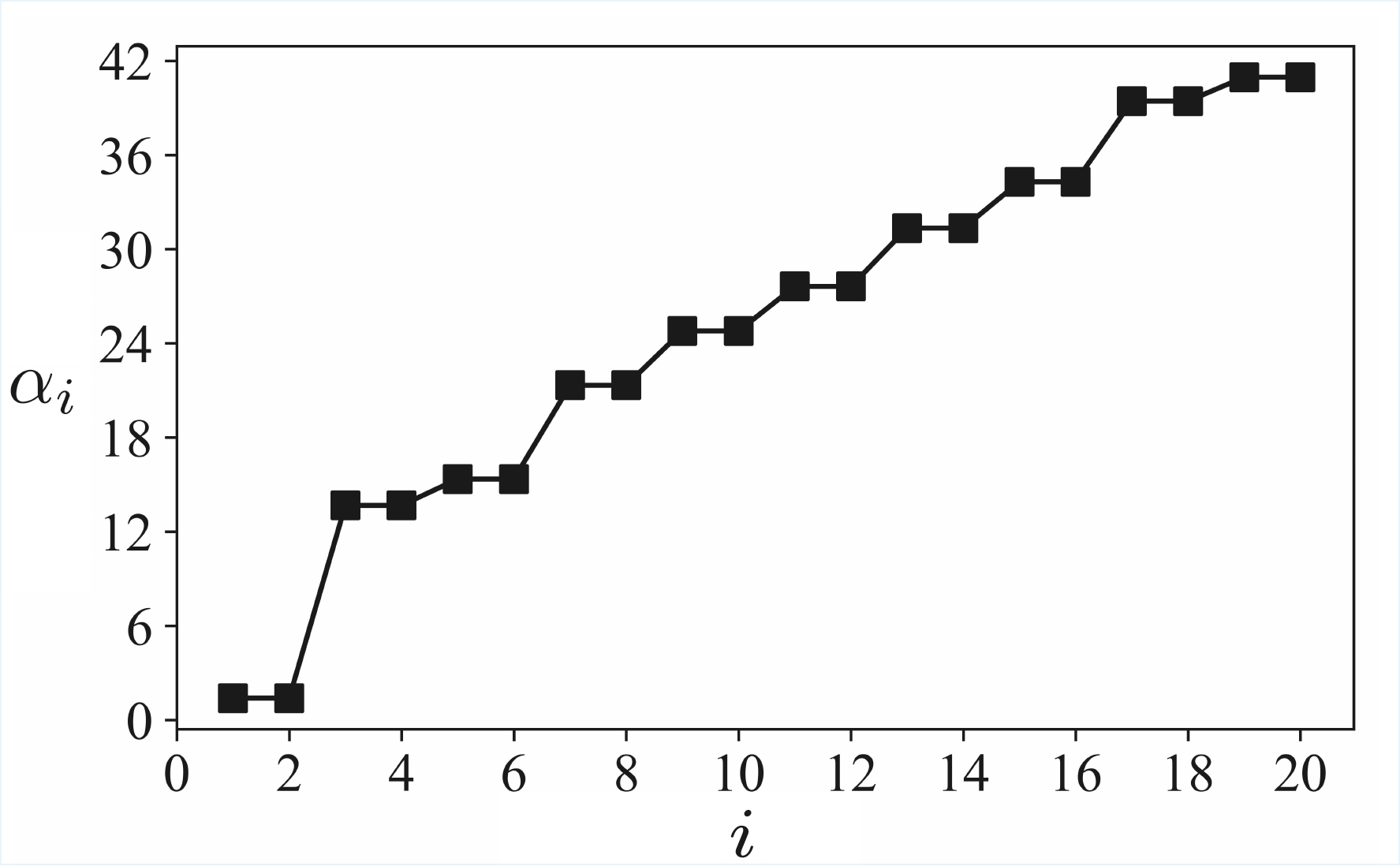}
    }
	\caption{
        Entanglement spectrum at $g^{2}=0.8$.
        These three choices of $M$ are representative of the trivial phase (a), the Aoki phase (b), and the SPT phase (c).
        Pairs of doubly degenerate spectra appear only inside the lobe structure.
    }
	\label{fig:ES}
\end{figure}

Here, we further investigate the phases realized inside these two lobes by analyzing the entanglement spectrum introduced in Eq.~\eqref{eq:ES}. 
Fig.~\ref{fig:ES} shows the lowest twenty entanglement spectra obtained at $g^{2}=0.8$ for three representative values of $M$.
Figs.~\ref{fig:ES}(a) and \ref{fig:ES}(b) correspond to $M=2.8$ and $M=0.1$, respectively, both of which lie outside the lobes. 
In contrast, Fig.~\ref{fig:ES}(c) is obtained inside the lobe, where the CTMRG consistently yields a doubly degenerate entanglement spectrum.
We find the same behavior throughout the entire phase diagram.
We expect that these doubly degenerate entanglement spectra provide strong evidence for a topological insulating phase.
In fact, Ref.~\cite{Bermudez:2018eyh} presents an analytical argument based on large $N_{f}$, according to which a topological insulator is realized within these two lobes for odd $N_{f}$.
We further note that a similar phase identification based on entanglement spectra has been reported for the Creutz–Hubbard model, in which the topological insulating phase is indeed characterized by a twofold-degenerate entanglement spectrum~\cite{Junemann:2016fxu}.

\subsubsection{Triple point}

\begin{figure}[htbp]
	\centering
	\includegraphics[width=9cm]{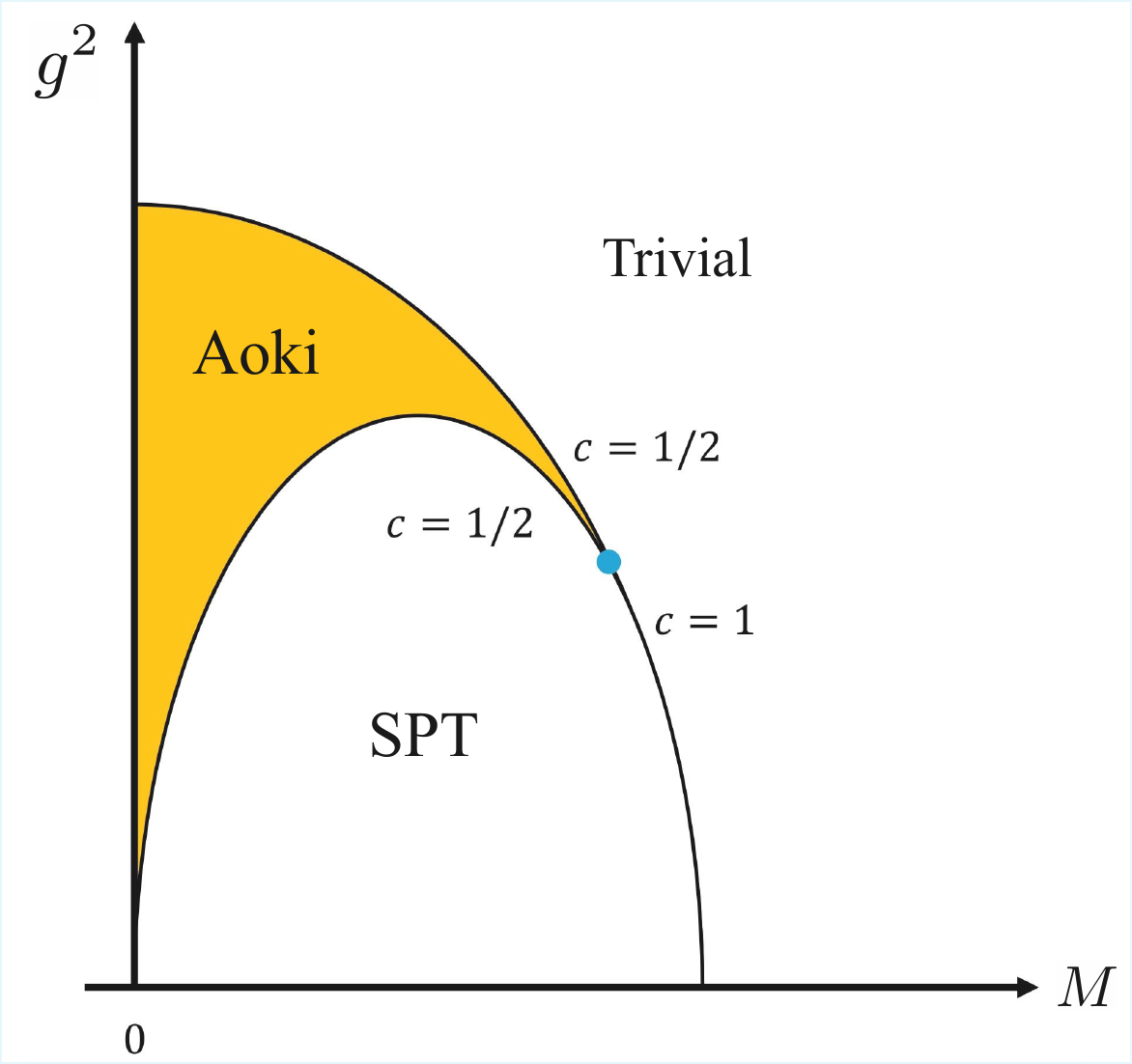}
	\caption{
        Schematic phase diagram near a triple point, denoted by a blue point, at which the Aoki phase terminates, and two $c=1/2$ critical lines merge into a single $c=1$ critical line.
    }
	\label{fig:schematic_PD_GNW}
\end{figure}

We finally address the fate of two critical lines characterized by $c=1/2$.
Since we have already found the $c=1$ critical line, separating the topological insulating phase from the trivial phase, we expect the phase structure schematically shown in Fig.~\ref{fig:schematic_PD_GNW}: two $c=1/2$ critical lines merge at a certain point, while a single critical line with $c=1$ remains.
We note that a similar scenario has also been confirmed by the recent MPS simulation~\cite{Bermudez:2018eyh}.
The phase structure shown in Fig.~\ref{fig:schematic_PD_GNW} suggests that the Aoki phase indeed exhibits a trident-like shape, as predicted in the large-$N_{f}$ phase diagram shown in Fig.~\ref{fig:largeN_phase_diagram}.

\begin{figure}[htbp]
	\centering
	\includegraphics[width=10cm]{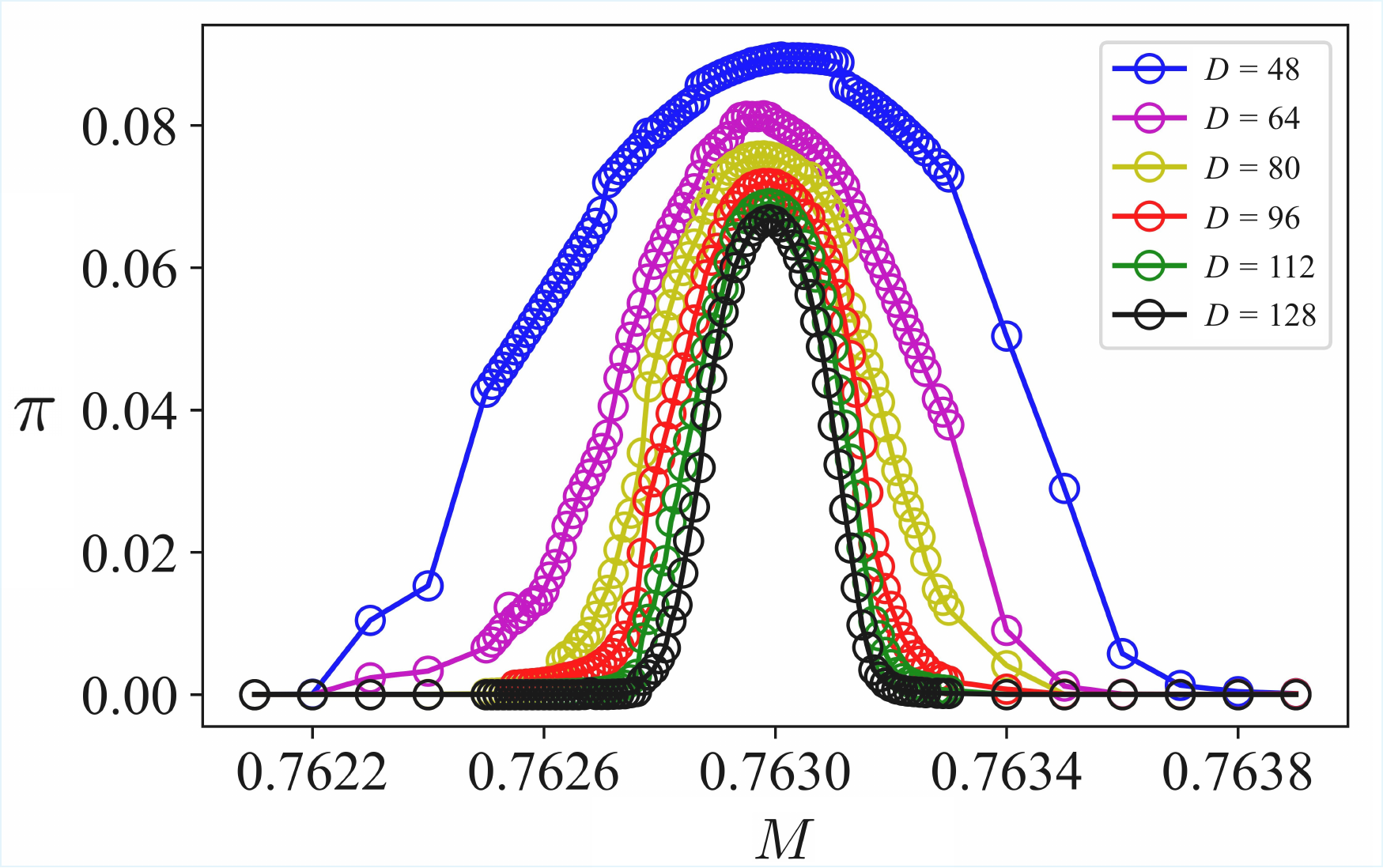}
	\caption{
        Pseudoscalar condensate at $g^{2}=0.9$ as a function of $M$ varying the bond dimension $D$.
    }
	\label{fig:GNW_ps_g209}
\end{figure}

Fig.~\ref{fig:GNW_ps_g209} shows the pseudoscalar condensate at $g^{2}=0.9$ as a function of $M$, indicating the existence of two independent transition points.
We note that the CTMRG suggests the presence of three transition points in the region with $M>0$ at $g^{2}=0.9$: two around $M\sim0.763$ as shown in Fig.~\ref{fig:GNW_ps_g209}, and another around $M\sim0.363$, as suggested in Fig.~\ref{fig:phase_diagram}.
This result is consistent with the phase structure schematically illustrated in Fig.~\ref{fig:schematic_PD_GNW}.

\begin{figure}
    \centering
    \subfigure[]{
    \includegraphics[width=8.6cm]{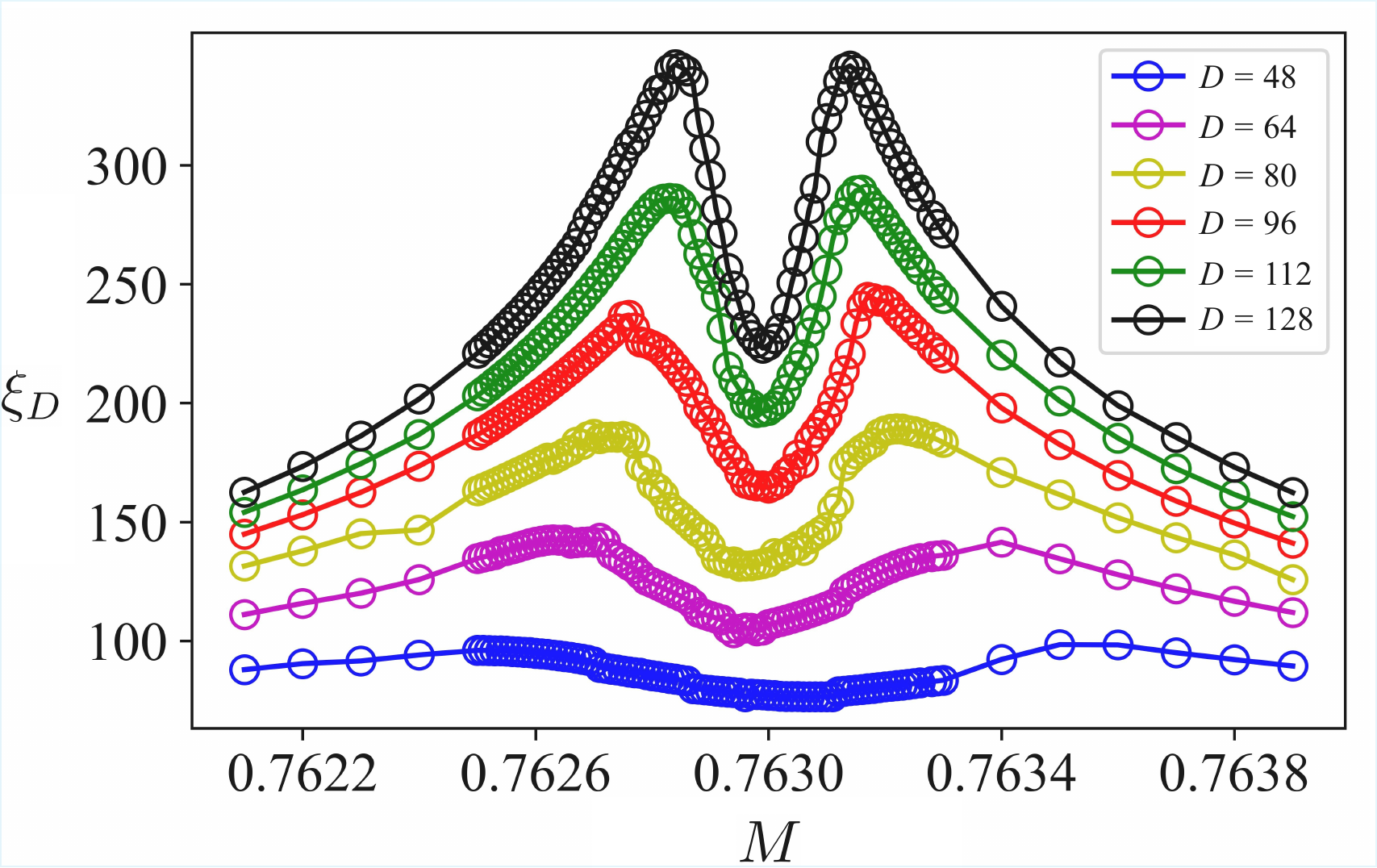}
    }
    \subfigure[]{
    \includegraphics[width=8.6cm]{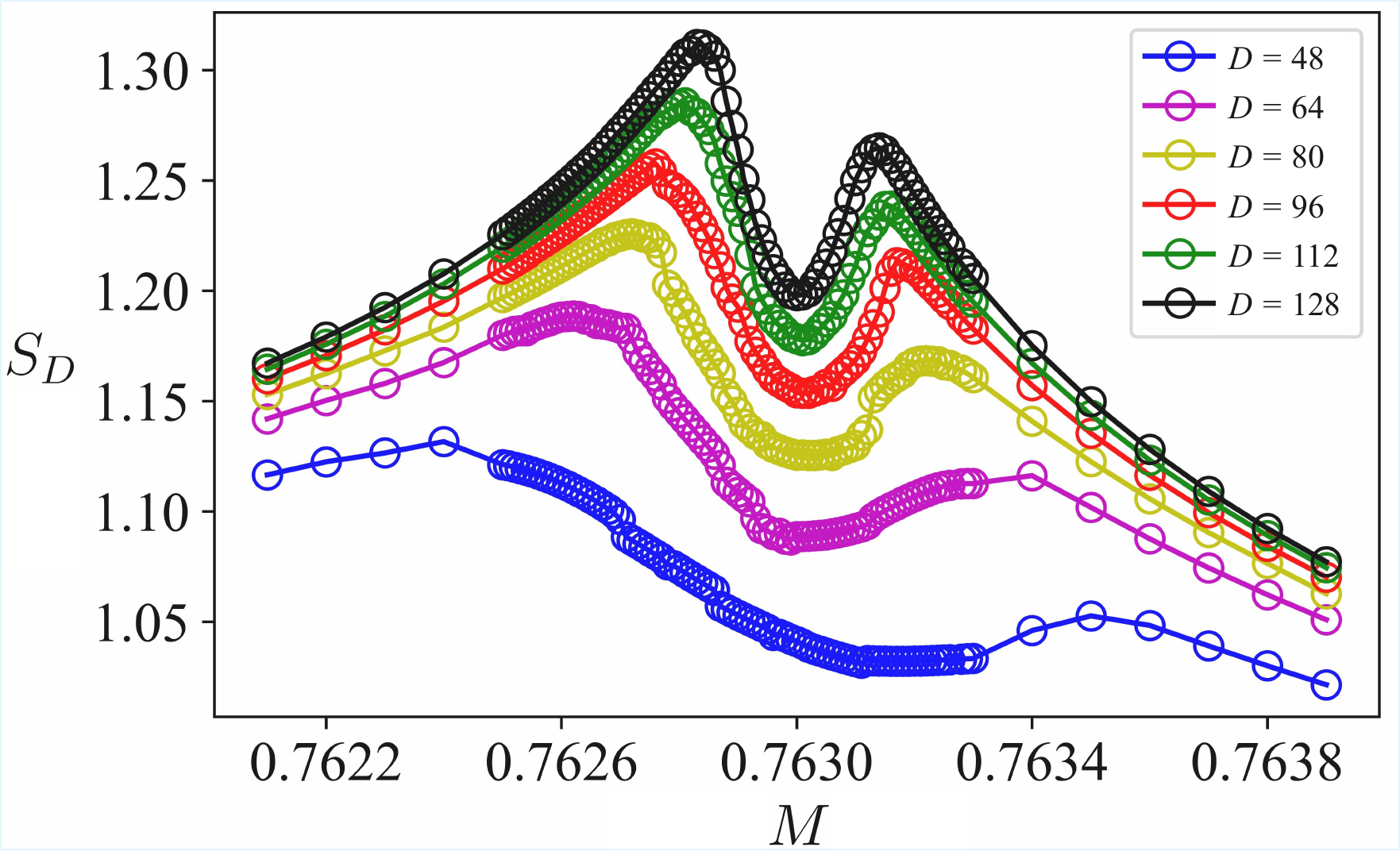}
    }
    \caption{
        Correlation length (a) and entanglement entropy (b) as a function of $M$ at $g^{2}=0.9$ varying the bond dimension $D$.
    }
    \label{fig:GNW_xi_S_g209}
\end{figure}

\begin{figure}[htbp]
	\centering
    \subfigure[]{
    \includegraphics[width=8.6cm]{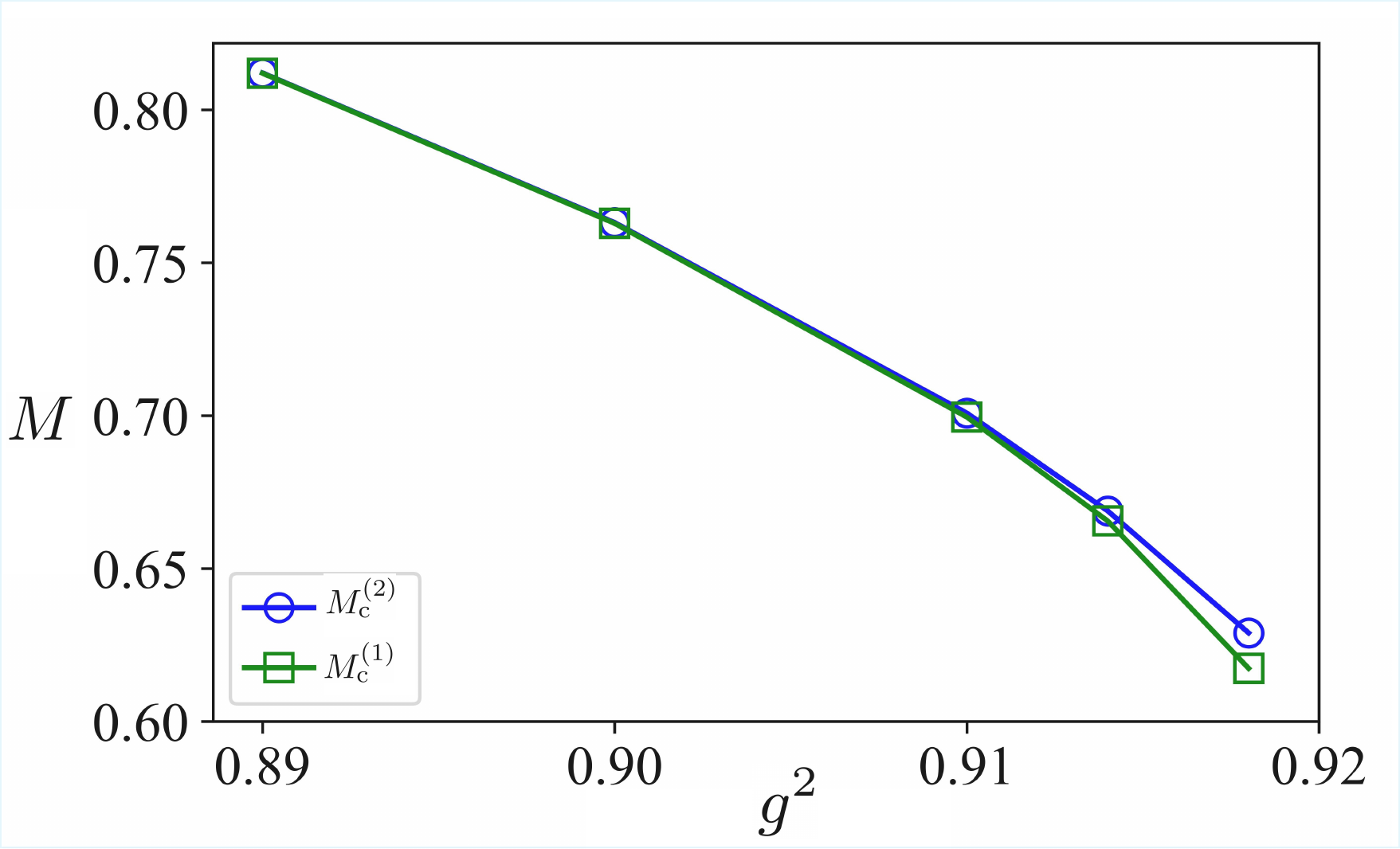}
    }
    \subfigure[]{
    \includegraphics[width=8.6cm]{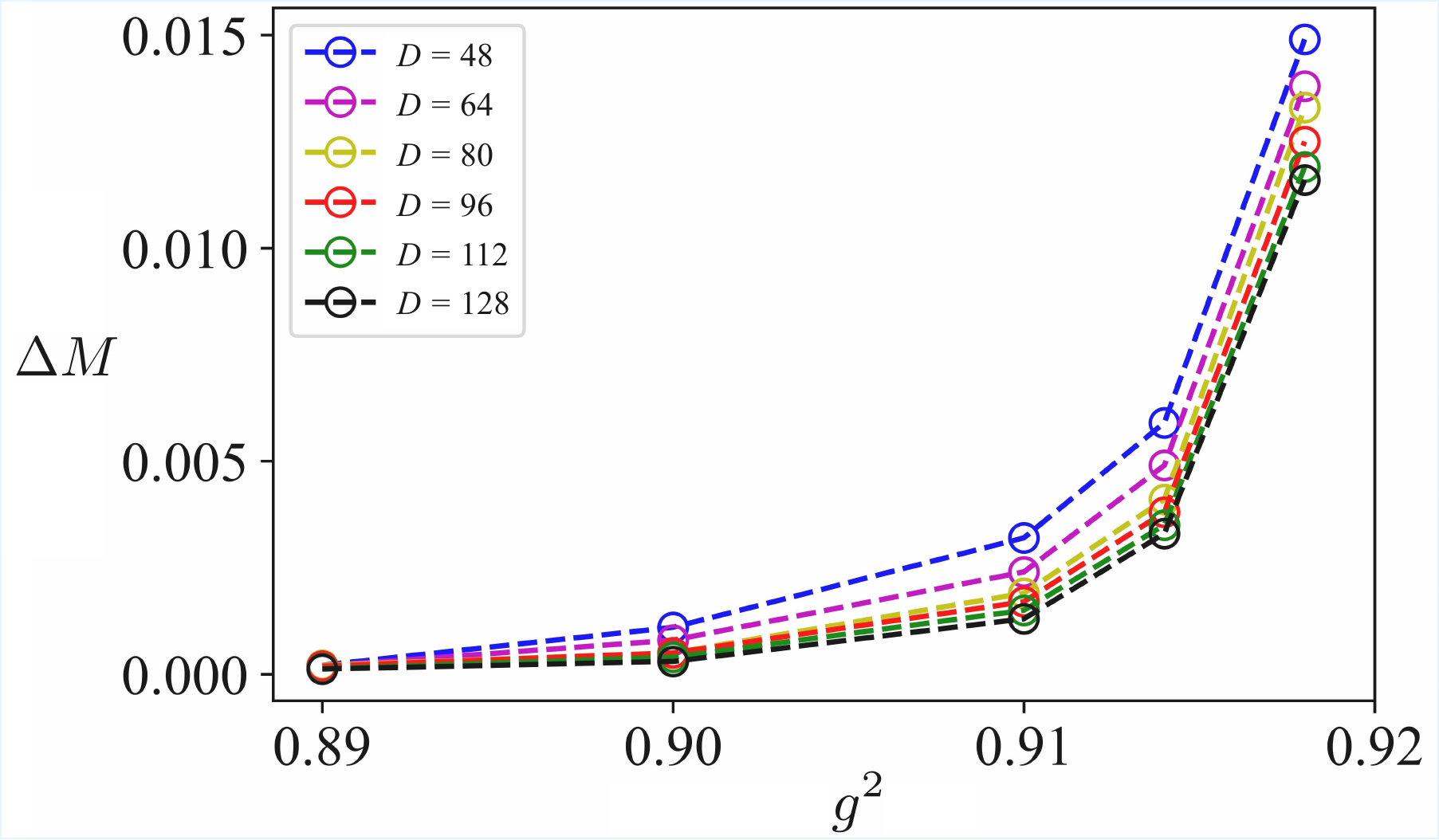}
    }
	\caption{
        (a) $M_{\rm c}^{(1)}$ and $M_{\rm c}^{(2)}$ as a function of $g^{2}$ at $D=128$.
        (b) Difference between $M_{\rm c}^{(1)}$ and $M_{\rm c}^{(2)}$ as a function of $g^{2}$ at several bond dimensions.
    }
	\label{fig:dm}
\end{figure}

As shown in Fig.~\ref{fig:GNW_xi_S_g209}, both the correlation length and the entanglement entropy at $g^{2}=0.9$ also exhibit clear double-peak structures around $M\sim0.763$.
However, since these two transition points are located very close to each other, a reliable analysis of the finite-entanglement scaling becomes difficult, and we leave a more detailed investigation for future work.
Instead, we focus here on providing an estimate for the location of the triple point.
We first estimate the two transition points simply from the positions of the peaks in the correlation length for several values of $g^{2}$.
Denoting these two points by $M_{\rm c}^{(1)}$ and $M_{\rm c}^{(2)}$, with $M_{\rm c}^{(1)} < M_{\rm c}^{(2)}$, we compute $\Delta M = M_{\rm c}^{(2)}-M_{\rm c}^{(1)}$ as a function of $g^{2}$ while varying the bond dimension.
A vanishingly small value of $\Delta M$ signals the location of the triple point.
The resulting $\Delta M$ is shown in Fig.~\ref{fig:dm}.
From this observation, the triple point is expected to be located around $(M,g^{2})\sim(0.812,0.89)$.
Assuming a phase structure symmetric with respect to $M=0$, there should be another triple point at $(M,g^{2})\sim(-0.812,0.89)$.

\section{Summary and outlook}
\label{sec:summary}

This work represents the first tensor network study to determine the full phase diagram of the (1+1)-dimensional single-flavor GNW model based on the path-integral formalism.
By developing the Grassmann CTMRG algorithm, we have successfully identified the phase diagram, which consists of three distinct phases: the Aoki phase, the topological insulating phase, and the trivial phase.
The Aoki phase is characterized by a finite pseudoscalar condensate, which is computed using the impurity tensor method within the CTMRG framework.
To determine the universality classes appearing on the phase boundaries, we analyze not only the correlation length but also the entanglement entropy, which are directly available from the reduced density matrix constructed by the CTMRG.

Our results have shown that the Aoki phase is separated from the other two phases by critical lines characterized by $c=1/2$.
The topological insulating phase is further identified through its characteristic entanglement spectrum, in which all eigenvalues appear as doubly degenerate pairs.
We have found that this topological phase exhibits a two-lobe structure, consistent with the large-$N_{f}$ prediction.
On the other hand, our CTMRG results suggest that the large-$N_{f}$ approach tends to overestimate the extent of the Aoki phase.
We have identified triple points at which the Aoki phase terminates.
At these points, the two $c=1/2$ critical lines appear to merge into a single $c=1$ critical line.
This scenario is also consistent with the recent MPS simulation of the same model based on the Hamiltonian formalism~\cite{Bermudez:2018eyh}.

A striking difference between the phase structure obtained from the CTMRG and that predicted by the large-$N_{f}$ analysis is that the Aoki phase disappears in the strong-coupling region, terminating at finite critical couplings.
However, the phase structure obtained from the Grassmann CTMRG seems to be more natural in the sense that in the strong-coupling regime, where the four-fermion interaction term dominates the action, the contribution from the kinetic term becomes negligible, and the lattice theory is expected to be trivially gapped.
The recent prediction based on ’t~Hooft anomaly matching at $M=0$ suggests that the parity symmetry is broken for any value of $g^{2}$ in the $N_{f}=1$ GNW model~\cite{Misumi:2019jrt}, which is expected to be valid only in the vicinity of the continuum limit.
Our numerical analysis instead treats the system purely as a lattice model and suggests that the large-$g^{2}$ region lies far outside the domain where the continuum description applies.

Recently, Grassmann CTMRG has also proven useful for quantitatively characterizing the ground-state phase diagram of the one-dimensional Hubbard model~\cite{Kong:2026}. 
It would therefore be interesting to apply this approach to more general interacting fermionic systems that host novel quantum states in one dimension.
As future work, it is interesting to investigate the phase structure of the multi-flavor GNW model using the Grassmann tensor networks~\cite{Akiyama:2023lvr} and to clarify how the critical coupling separating the Aoki and trivial phases evolves as the number of flavors is increased, in particular, whether it is pushed further into the strong-coupling regime.
Identifying the phase inside the two lobes for $N_{f}=2$ could also be an interesting direction for future study. 

\begin{acknowledgments}
We thank Yuya Tanizaki for valuable comments. 
Jian-Gang Kong and Z. Y. Xie are supported by the National R\&D Program of China (Grants No. 2023YFA1406500 and No. 2024YFA1408604), the National Natural Science Foundation of China (Grants No. 12274458).
SA acknowledges the support from JSPS KAKENHI Grant Numbers JP23K13096, the Center of Innovations for Sustainable Quantum AI (JST Grant Number JPMJPF2221), the Endowed Project for Quantum Software Research and Education, the University of Tokyo~\cite{qsw}, and the Top Runners in Strategy of Transborder Advanced Researches (TRiSTAR) program conducted as the Strategic Professional Development Program for Young Researchers by the MEXT. 
A part of the numerical computations for the present work was carried out with Pegasus and Miyabi, provided by the Multidisciplinary Cooperative Research Program of Center for Computational Sciences, University of Tsukuba.
\end{acknowledgments} 

\appendix

\section{Construction of the Grassmann projectors $\mathcal{P}$ and $\mathcal{Q}$} 
\label{app:construct_P_Q}

\begin{figure}
    \centering
    \subfigure[]{
        \includegraphics[width=11cm]{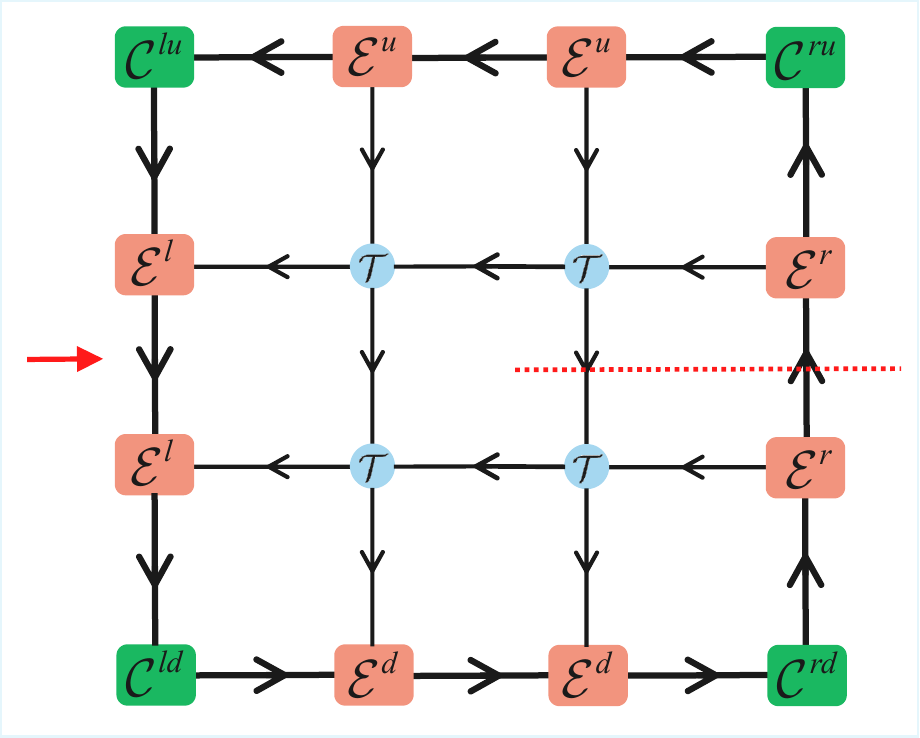}
    }\\
    \subfigure[]{
        \includegraphics[width=15cm]{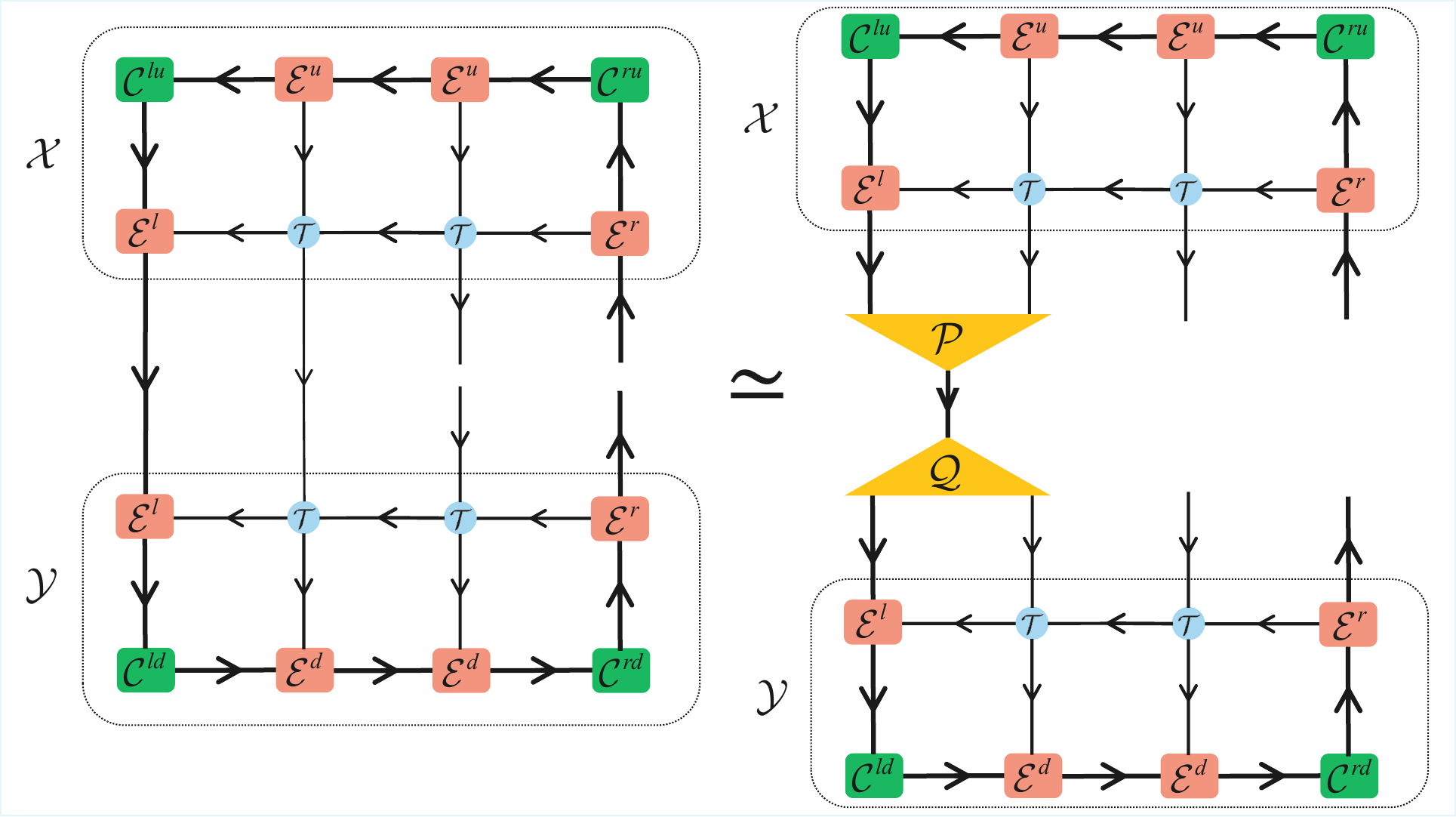}
    }
    \caption{
        Construction of the Grassmann projectors $\mathcal{P}$ and $\mathcal{Q}$. 
        (a) A $2\times2$ unit cell together with its surrounding environment. 
        (b) Graphical representation of Eq.~\eqref{eq:ctmrg_equation}.
    }
	\label{fig:ctmrg_PQ}
\end{figure}

The most essential step of the CTMRG algorithm is to properly construct the projectors $\mathcal{P}$ and $\mathcal{Q}$, whose purpose is to preserve as much information as possible while minimizing the truncation error.
Following Ref.~\cite{Corboz:2014ocg}, we consider a minimal $2\times2$ unit cell of bulk tensors with the surrounding environments as shown in Fig.~\ref{fig:ctmrg_PQ}(a), the red arrow marks the position to insert the $\mathcal{P}$ and $\mathcal{Q}$. 

We regard a $4\times4$ cluster in Fig.~\ref{fig:ctmrg_PQ}(a) as a four-leg tensor $\mathcal{W}$, by making a cut at the middle of the two columns on the right side, indicated by the red dashed line, and introducing the four-leg tensors $\mathcal{X}$ and $\mathcal{Y}$ as upper-half and lower-half clusters.

Here, we derive the Grassmann projectors $\mathcal{P}$ and $\mathcal{Q}$ in such a way that they reproduce the Grassmann singular value decomposition (SVD) of the tensor $\mathcal{W}$.
More explicitly, we need to derive $\mathcal{P}$ and $\mathcal{Q}$ as
\begin{eqnarray} 
\label{eq:ctmrg_equation}
    \mathcal{W}_{A_{1}\bar{A}_{2}\bar{A}_{1}A_{2}} 
    &=& \nonumber
    \int_{\bar{B}_{1},B_{1}} 
    \int_{\bar{B}_{2},B_{2}} 
    \mathcal{X}_{A_{1}\bar{A}_{2}\bar{B}_{2}\bar{B}_{1}}
    \mathcal{Y}_{B_{1}B_{2}\bar{A}_{1}A_{2}}
    \\ 
    &\simeq&
    \int_{\bar{B}_{1},B_{1}} 
    \int_{\bar{B}_{2},B_{2}} 
    \int_{\bar{B}_{3},B_{3}} 
    \int_{\bar{B}_{4},B_{4}} 
    \mathcal{X}_{A_{1}\bar{A}_{2}\bar{B}_{2}\bar{B}_{1}}
    \left(
        \int_{\bar{C},C}
        \mathcal{P}_{B_{1}B_{2}\bar{C}}
        \mathcal{Q}_{C\bar{B}_{4}\bar{B}_{3}}
    \right)
    \mathcal{Y}_{B_{3}B_{4}\bar{A}_{1}A_{2}},
\end{eqnarray}
which corresponds to the truncated Grassmann SVD of $\mathcal{W}$.
In the above expression, we have introduced the shorthand notation,
\begin{align}
    \int_{\eta,\xi}:=\int\int{\rm d}\eta{\rm d}\xi~{\rm e}^{-\eta\xi},
\end{align}
with the Grassmann variables $\eta$ and $\xi$.

To further describe how to derive these projectors, we focus on their coefficient tensors.
We introduce the Grassmann parity function $p(i)$, which takes $0~(1)$ when the argument $i$ describes the Grassmann even (odd) contribution.
We note that $i$ is not restricted to a single binary variable, but can represent a bit string consisting of multiple indices~\cite{Akiyama:2024ush}.
Letting $W$ be the coefficient tensor of $\mathcal{W}$, the first equality in Eq.~\eqref{eq:ctmrg_equation} results in the contraction between the coefficient tensors $X$ and $Y$ in $\mathcal{X}$ and $\mathcal{Y}$, respectively,
\begin{eqnarray} 
\label{eq:W_def}
    W_{a_{1}\bar{a}_{2}\bar{a}_{1}a_{2}}
    =
    \sum_{b_{1},b_{2}}
    X_{a_{1}\bar{a}_{2}b_{2}b_{1}}
    Y_{b_{1}b_{2}\bar{a}_{1}a_{2}}
    \times
    (-1)^{p(b_{1}) + p(b_{2})}
    \equiv
    \sum_{b_{1},b_{2}}
    X_{a_{1}\bar{a}_{2}b_{2}b_{1}}
    \tilde{Y}_{b_{1}b_{2}\bar{a}_{1}a_{2}},
\end{eqnarray}
where we have introduced $\tilde{Y}_{b_{1}b_{2}\bar{a}_{1}a_{2}}=(-1)^{p(b_{1}) + p(b_{2})}Y_{b_{1}b_{2}\bar{a}_{1}a_{2}}$.
We note that the Grassmann indices are labeled by capital letters, whereas the corresponding integer-valued indices in coefficient tensors are denoted by the corresponding lowercase letters.

Introducing the normal SVD of the $W$ tensor as $W = US V^{\dagger}$, the coefficient tensors of $\mathcal{P}$ and $\mathcal{Q}$ are now available as
\begin{align}
\label{soluion_P}
    P = \tilde{Y}VS^{-1/2} = 
    \raisebox{-0.4\height}{\includegraphics[width=0.28\textwidth, page=1]{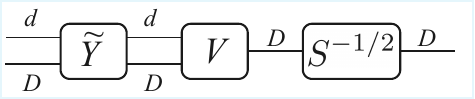}},
\end{align}
\begin{align}
\label{soluion_Q}
    Q = S^{-1/2} U^{\dagger} X
    =
    \raisebox{-0.4\height}{\includegraphics[width=0.28\textwidth, page=1]{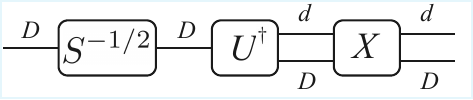}}.
\end{align}
The truncation of $P$ and $Q$ has already been performed according to the magnitude of the singular value $S$. 
The construction would be quite similar when we bring Grassmann variables back into the story. 
The coefficient tensors in $\mathcal{P}$ and $\mathcal{Q}$ are now given by
\begin{align}
    P_{b_{1}b_{2}\bar{c}} \times 
    (-1)^{p(b_{1}) + p(b_{2})},
\end{align}
\begin{align}
    Q_{c\bar{b}_{3}\bar{b}_{4}}
    \times
    (-1)^{p(c)}.
\end{align}

\section{Pseudoscalar condensate evaluated under periodic boundary conditions} 
\label{app:GTRG}

Here, we directly compute Eqs.~\eqref{eq:GNW_Z} and \eqref{eq:GNW_impurity} assuming the periodic boundary conditions.
For this purpose, we employ the HOTRG~\cite{PhysRevB.86.045139,Sakai:2017jwp} with the impurity method~\cite{Yoshimura:2017jpk,Morita:2018tpw}.

\begin{figure}[htbp]
	\centering
	\includegraphics[width=14cm]{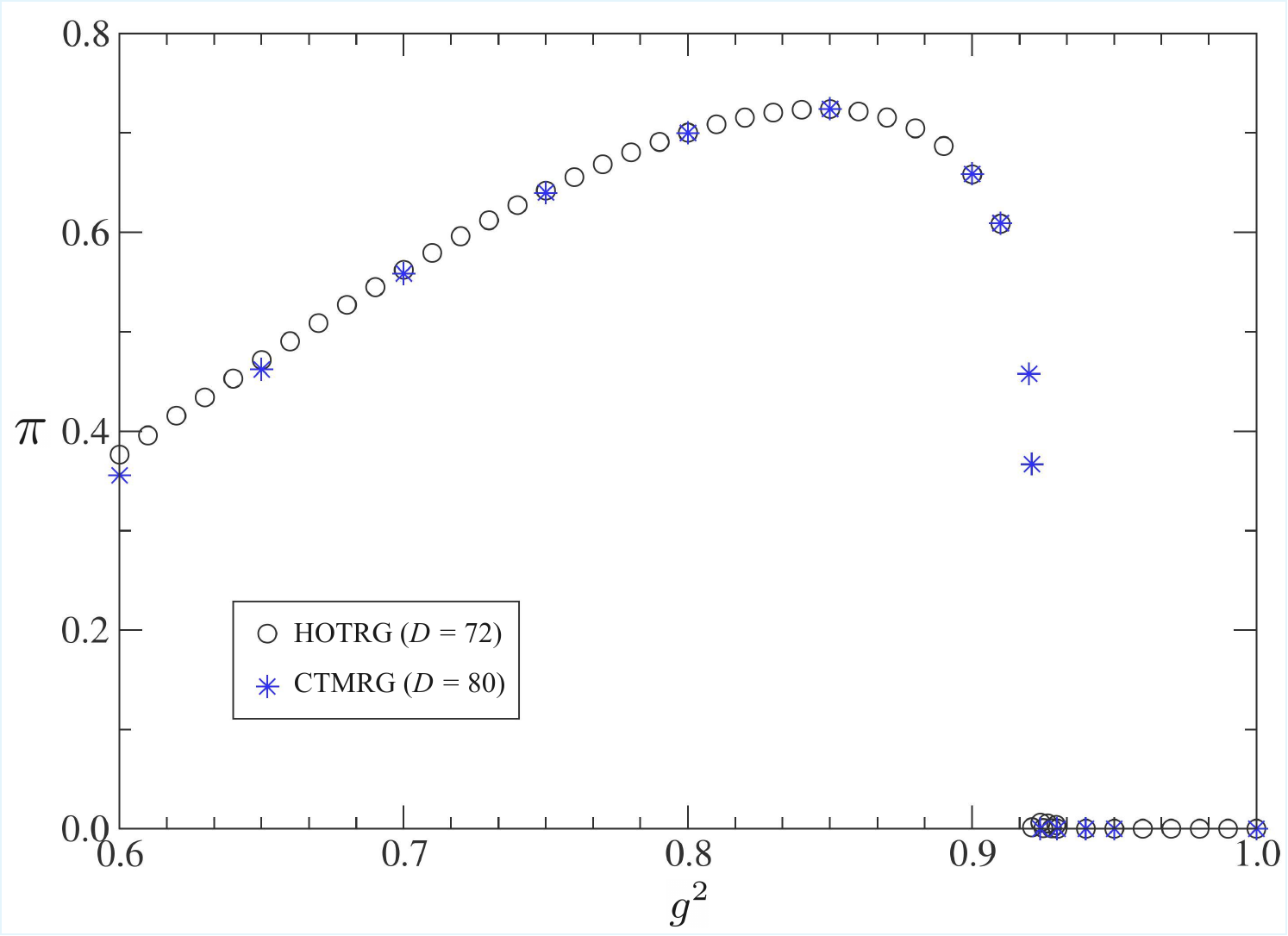}
	\caption{
        Pseudoscalar condensate at $M=0$ as a function of $g^{2}$ obtained by the HOTRG with $D=72$.
        For comparison, the resulting pseudoscalar condensate from the CTMRG at $D=80$ is also shown by star symbols.
    }
	\label{fig:GNW_m_-2_scan_HOTRG}
\end{figure}

Fig.~\ref{fig:GNW_m_-2_scan_HOTRG} shows the resulting pseudoscalar condensate in the thermodynamic limit at $M=0$ as a function of $g^{2}$, whose behavior is in good agreement with that shown in Fig.~\ref{fig:GNW_m_-2_scan}: The HOTRG also finds a critical coupling around $g^{2}\sim0.925$ at which the Aoki phase terminates.
In Fig.~\ref{fig:GNW_m_-2_scan_HOTRG}, we restrict our analysis to the region $g^{2}\in[0.6,1.0]$, since the HOTRG suffers from larger truncation errors for $g^{2}\le 0.6$, which makes it difficult to reliably take the limit $D\to\infty$, unlike in the case of the CTMRG shown in Fig.~\ref{fig:extrap_m_-2}.
We also note that the larger truncation errors in the HOTRG have already been suggested by Fig.~\ref{fig:GNW_g2_0}(b).

\section{Scaling analysis of the entanglement entropy}
\label{app:FES}

\begin{figure}
    \centering
    \subfigure[$g^{2}_{c}=0.77181$]{
        \includegraphics[width=8.6cm]{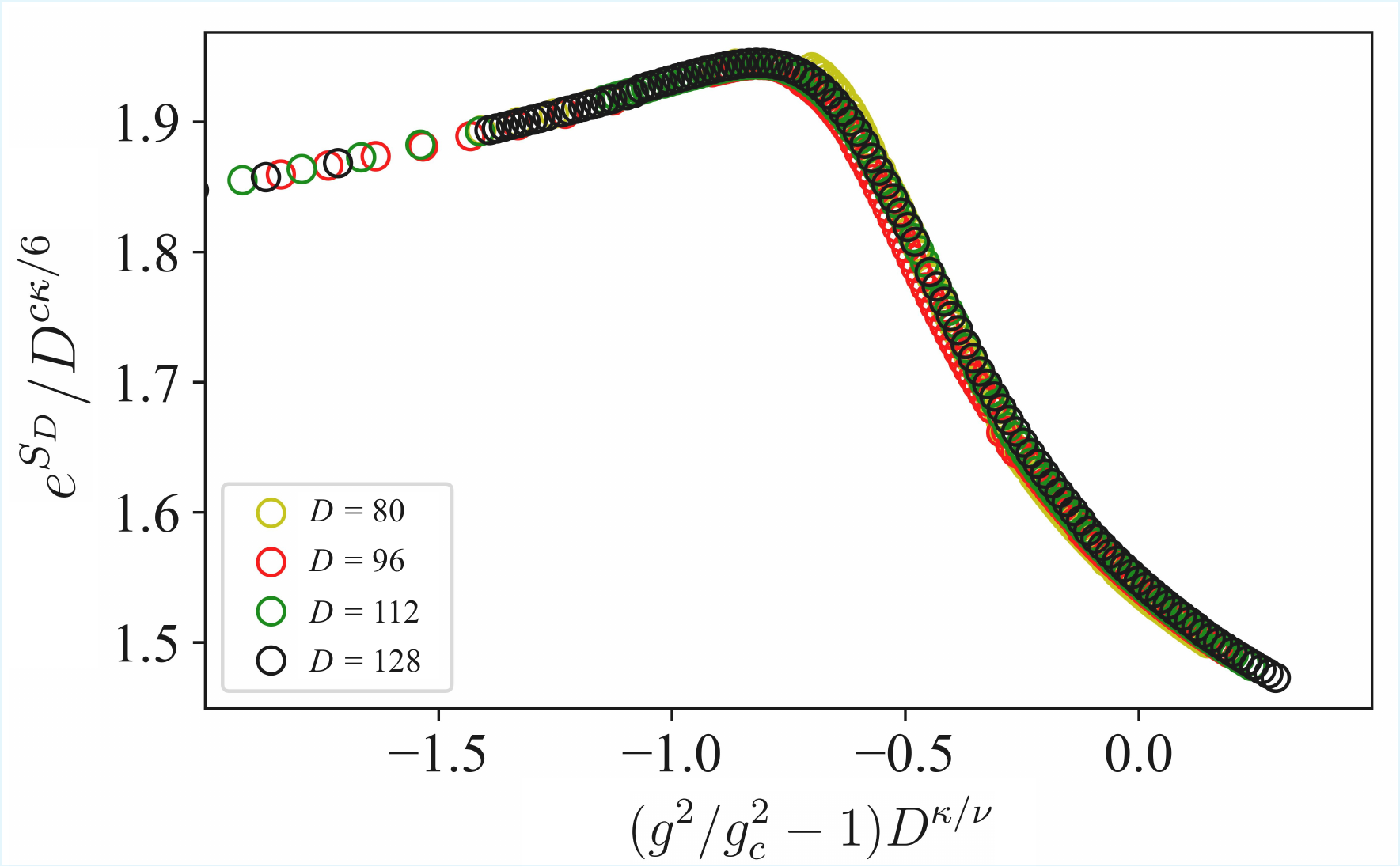}
    }
    \subfigure[$g^{2}_{c}=0.92161$]{
        \includegraphics[width=8.6cm]{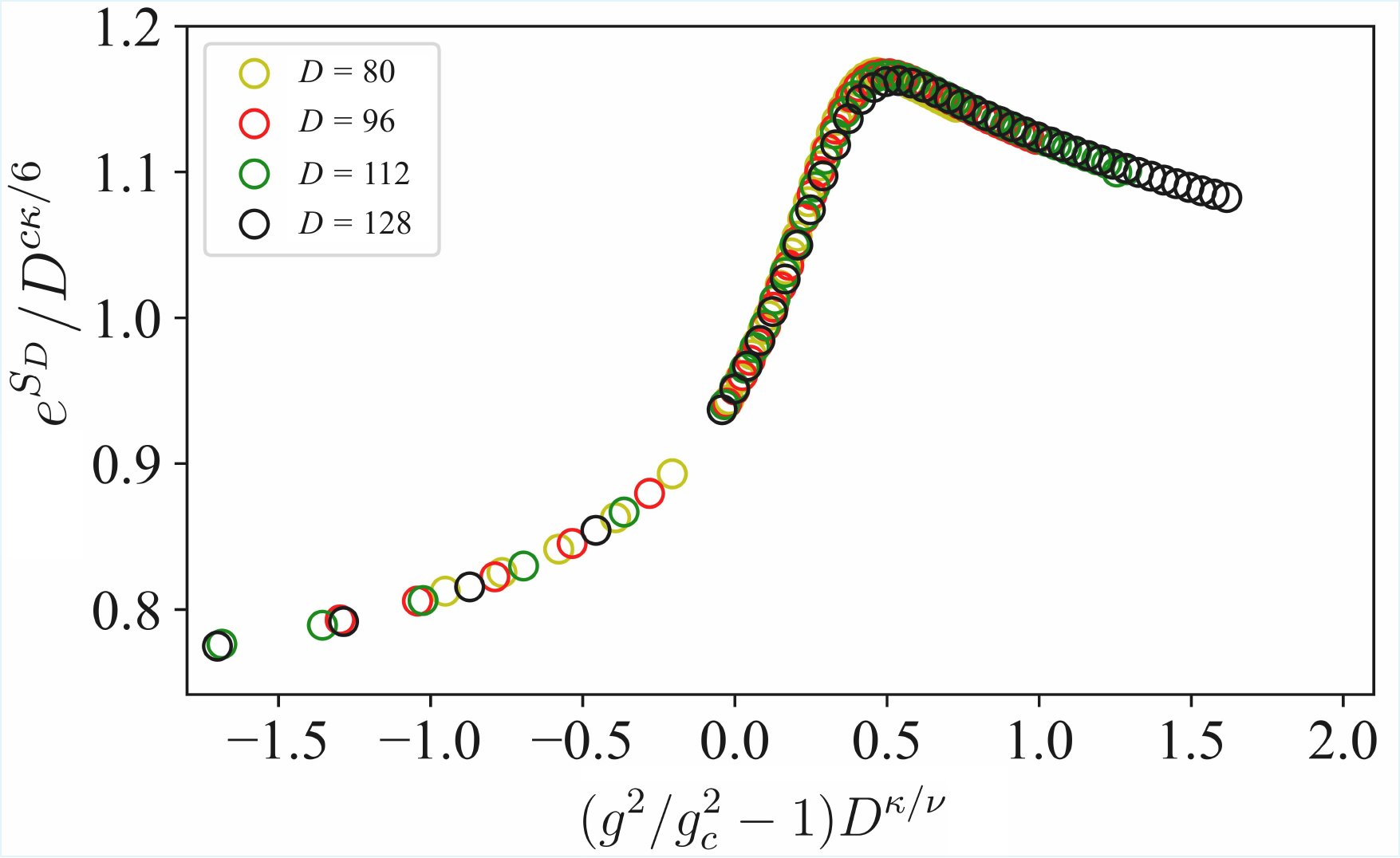}
    }
    \caption{
        Data collapse of the entanglement entropy at $M=0.1$ based on finite-entanglement scaling, assuming $c=1/2$ and $\nu=1$, with $g^{2}_{c}=0.77181$ (a) and $g^{2}_{c}=0.92161$ (b).
    }
    \label{fig:EE_m_1p9_lobes}
\end{figure}

In one-dimensional critical systems, it is known that the MPS with bond dimension $D$ induces a finite correlation length which scales as 
\begin{align}
\label{eq:finite-m_scaling}
    \xi_{D}\sim D^{\kappa},
\end{align}
where
\begin{align}
\label{eq:kappa_MPS}
    \kappa = \dfrac{6}{c(\sqrt{12/c}+1)}, 
\end{align}
with the central charge $c$~\cite{Tagliacozzo:2007rda,Pollmann:2009lnv}.
Several numerical studies have found that the finite-$D$ scaling in Eq.~\eqref{eq:finite-m_scaling} also holds in the CTMRG calculations with $\kappa$ given by Eq.~\eqref{eq:kappa_MPS}~\cite{Ueda:2017ojh,Ueda:2020etz}.
As shown in Fig.~\ref{fig:EE_m_1p9_lobes}, the curves for different values of $D$ collapse reasonably well onto a single curve described by a universal scaling function when we assume the two-dimensional Ising CFT, with $c=1/2$, the critical exponent $\nu=1$, and the critical couplings obtained in Sec.~\ref{subsubsec:phase_boundaries_Aoki} at $M=0.1$.
We find, however, that the best data collapse in Fig.~\ref{fig:EE_m_1p9_lobes} is achieved with $\kappa\sim1.48$, which differs from the value $\kappa\simeq2.03$ predicted by Eq.~\eqref{eq:kappa_MPS} at $c=1/2$.
We note that although minor discrepancies from Eq.~\eqref{eq:kappa_MPS} have been reported in previous numerical studies~\cite{PhysRevB.99.205121,Huang:2019moo,Ueda:2020igb}, the deviation observed in this study is significantly larger, suggesting that it may not be identical to those found earlier.
Further investigation of the finite-$D$ scaling is needed to understand the origin of this deviation, which we leave for future work.

\bibliographystyle{JHEP}
\bibliography{bib/ref,bib/review,bib/formalism,bib/algorithm,bib/fermi_dof}

@article{Levin:2006jai,
      author         = "Levin, Michael and Nave, Cody P.",
      title          = "{Tensor renormalization group approach to two-dimensional classical
                        lattice models}",
      journal        = "Phys. Rev. Lett.",
      volume         = "99",
      year           = "2007",
      number         = "12",
      pages          = "120601",
      doi            = "10.1103/PhysRevLett.99.120601",
      eprint         = "cond-mat/0611687",
      archivePrefix  = "arXiv",
      primaryClass   = "cond-mat.stat-mech",
      SLACcitation   = "%%CITATION = COND-MAT/0611687;%%"
}

@article{Gu:2010yh,
      author         = "Gu, Zheng-Cheng and Verstraete, Frank and Wen, Xiao-Gang",
      title          = "{Grassmann tensor network states and its renormalization
                        for strongly correlated fermionic and bosonic states}",
      year           = "2010",
      eprint         = "1004.2563",
      archivePrefix  = "arXiv",
      primaryClass   = "cond-mat.str-el",
      SLACcitation   = "%%CITATION = ARXIV:1004.2563;%%"
}

@article{Gu:2013gba,
      author         = "Gu, Zheng-Cheng",
      title          = "{Efficient simulation of Grassmann tensor product
                        states}",
      journal        = "Phys. Rev.",
      volume         = "B88",
      year           = "2013",
      number         = "11",
      pages          = "115139",
      doi            = "10.1103/PhysRevB.88.115139",
      eprint         = "1109.4470",
      archivePrefix  = "arXiv",
      primaryClass   = "cond-mat.str-el",
      SLACcitation   = "%%CITATION = 1109.4470;%%"
}

@article{PhysRevB.86.045139,
  title = {Coarse-graining renormalization by higher-order singular value decomposition},
  author = {Xie, Z. Y. and Chen, J. and Qin, M. P. and Zhu, J. W. and Yang, L. P. and Xiang, T.},
  journal = {Phys. Rev. B},
  volume = {86},
  issue = {4},
  pages = {045139},
  numpages = {9},
  year = {2012},
  month = {Jul},
  publisher = {American Physical Society},
  doi = {10.1103/PhysRevB.86.045139},
  eprint = "1201.1144",
  archivePrefix = "arXiv",
  primaryClass = "cond-mat.stat-mech"
}

@article{Sakai:2017jwp,
      author         = "Sakai, Ryo and Takeda, Shinji and Yoshimura, Yusuke",
      title          = "{Higher order tensor renormalization group for
                        relativistic fermion systems}",
      journal        = "PTEP",
      volume         = "2017",
      year           = "2017",
      number         = "6",
      pages          = "063B07",
      doi            = "10.1093/ptep/ptx080",
      eprint         = "1705.07764",
      archivePrefix  = "arXiv",
      primaryClass   = "hep-lat",
      reportNumber   = "KANAZAWA-17-03",
      SLACcitation   = "%%CITATION = ARXIV:1705.07764;%%"
}

@article{Yoshimura:2017jpk,
      author         = "Yoshimura, Yusuke and Kuramashi, Yoshinobu and Nakamura,
                        Yoshifumi and Takeda, Shinji and Sakai, Ryo",
      title          = "{Calculation of fermionic Green functions with Grassmann
                        higher-order tensor renormalization group}",
      journal        = "Phys. Rev.",
      volume         = "D97",
      year           = "2018",
      number         = "5",
      pages          = "054511",
      doi            = "10.1103/PhysRevD.97.054511",
      eprint         = "1711.08121",
      archivePrefix  = "arXiv",
      primaryClass   = "hep-lat",
      SLACcitation   = "%%CITATION = ARXIV:1711.08121;%%"
}

@article{Morita:2018tpw,
    author = "Morita, Satoshi and Kawashima, Naoki",
    title = "{Calculation of higher-order moments by higher-order tensor renormalization group}",
    eprint = "1806.10275",
    archivePrefix = "arXiv",
    primaryClass = "cond-mat.stat-mech",
    doi = "10.1016/j.cpc.2018.10.014",
    journal = "Comput. Phys. Commun.",
    volume = "236",
    pages = "65--71",
    year = "2019"
}

@article{PhysRevB.105.L060402,
  title = {Bond-weighted tensor renormalization group},
  author = {Adachi, Daiki and Okubo, Tsuyoshi and Todo, Synge},
  journal = {Phys. Rev. B},
  volume = {105},
  issue = {6},
  pages = {L060402},
  numpages = {6},
  year = {2022},
  month = {Feb},
  publisher = {American Physical Society},
  doi = {10.1103/PhysRevB.105.L060402},
  eprint         = "2011.01679",
  archivePrefix  = "arXiv",
  primaryClass   = "cond-mat.stat-mech"
}

@article{Akiyama:2022pse,
    author = "Akiyama, Shinichiro",
    title = "{Bond-weighting method for the Grassmann tensor renormalization group}",
    eprint = "2208.03227",
    archivePrefix = "arXiv",
    primaryClass = "hep-lat",
    doi = "10.1007/JHEP11(2022)030",
    journal = "JHEP",
    volume = "11",
    pages = "030",
    year = "2022"
}

@article{Akiyama:2023rih,
    author = "Akiyama, Shinichiro",
    title = "{Implementation of bond weighting method for the Grassmann tensor renormalization group}",
    eprint = "2311.17691",
    archivePrefix = "arXiv",
    primaryClass = "hep-lat",
    reportNumber = "UTCCS-P-150",
    doi = "10.22323/1.453.0370",
    journal = "PoS",
    volume = "LATTICE2023",
    pages = "370",
    year = "2024"
}

@article{Asaduzzaman:2022pnw,
    author = "Asaduzzaman, Muhammad and Catterall, Simon and Meurice, Yannick and Sakai, Ryo and Toga, Goksu Can",
    title = "{Improved coarse-graining methods for two dimensional tensor networks including fermions}",
    eprint = "2210.03834",
    archivePrefix = "arXiv",
    primaryClass = "hep-lat",
    doi = "10.1007/JHEP01(2023)024",
    journal = "JHEP",
    volume = "01",
    pages = "024",
    year = "2023"
}

@article{Akiyama:2023lvr,
    author = "Akiyama, Shinichiro",
    title = "{Matrix product decomposition for two- and three-flavor Wilson fermions: Benchmark results in the lattice Gross-Neveu model at finite density}",
    eprint = "2304.01473",
    archivePrefix = "arXiv",
    primaryClass = "hep-lat",
    doi = "10.1103/PhysRevD.108.034514",
    journal = "Phys. Rev. D",
    volume = "108",
    number = "3",
    pages = "034514",
    year = "2023"
}

@article{Yosprakob:2023tyr,
    author = "Yosprakob, Atis and Nishimura, Jun and Okunishi, Kouichi",
    title = "{A new technique to incorporate multiple fermion flavors in tensor renormalization group method for lattice gauge theories}",
    eprint = "2309.01422",
    archivePrefix = "arXiv",
    primaryClass = "hep-lat",
    reportNumber = "KEK-TH-2549",
    doi = "10.1007/JHEP11(2023)187",
    journal = "JHEP",
    volume = "11",
    pages = "187",
    year = "2023"
}

@article{Yosprakob:2023flr,
	title={{GrassmannTN: A Python package for Grassmann tensor network computations}},
	author={Atis Yosprakob},
    eprint = "2309.07557",
    archivePrefix = "arXiv",
    primaryClass = "hep-lat",
	journal={SciPost Phys. Codebases},
	pages={20},
	year={2023},
	publisher={SciPost},
	doi={10.21468/SciPostPhysCodeb.20},
}

@article{Shimizu:2014uva,
      author         = "Shimizu, Yuya and Kuramashi, Yoshinobu",
      title          = "{Grassmann tensor renormalization group approach to
                        one-flavor lattice Schwinger model}",
      journal        = "Phys. Rev. D",
      volume         = "90",
      year           = "2014",
      number         = "1",
      pages          = "014508",
      doi            = "10.1103/PhysRevD.90.014508",
      eprint         = "1403.0642",
      archivePrefix  = "arXiv",
      primaryClass   = "hep-lat",
      SLACcitation   = "%%CITATION = ARXIV:1403.0642;%%"
}

@article{Shimizu:2014fsa,
      author         = "Shimizu, Yuya and Kuramashi, Yoshinobu",
      title          = "{Critical behavior of the lattice Schwinger model with a
                        topological term at $\theta=\pi$ using the Grassmann
                        tensor renormalization group}",
      journal        = "Phys. Rev. D",
      volume         = "90",
      year           = "2014",
      number         = "7",
      pages          = "074503",
      doi            = "10.1103/PhysRevD.90.074503",
      eprint         = "1408.0897",
      archivePrefix  = "arXiv",
      primaryClass   = "hep-lat",
      SLACcitation   = "%%CITATION = ARXIV:1408.0897;%%"
}

@article{Shimizu:2017onf,
      author         = "Shimizu, Yuya and Kuramashi, Yoshinobu",
      title          = "{Berezinskii-Kosterlitz-Thouless transition in lattice
                        Schwinger model with one flavor of Wilson fermion}",
      journal        = "Phys. Rev. D",
      volume         = "97",
      year           = "2018",
      number         = "3",
      pages          = "034502",
      doi            = "10.1103/PhysRevD.97.034502",
      eprint         = "1712.07808",
      archivePrefix  = "arXiv",
      primaryClass   = "hep-lat",
      reportNumber   = "UTHEP-711, UTCCS-P-109",
      SLACcitation   = "%%CITATION = ARXIV:1712.07808;%%"
}

@article{Kanno:2024elz,
    author = "Kanno, Hayato and Akiyama, Shinichiro and Murakami, Kotaro and Takeda, Shinji",
    title = "{Grassmann tensor renormalization group for the massive Schwinger model with a {\ensuremath{\theta}} term using staggered fermions}",
    eprint = "2412.08959",
    archivePrefix = "arXiv",
    primaryClass = "hep-lat",
    reportNumber = "UTCCS-P-156, KANAZAWA 24-07, RIKEN-iTHEMS-Report-24",
    doi = "10.1007/JHEP11(2025)036",
    journal = "JHEP",
    volume = "11",
    pages = "036",
    year = "2025"
}

@article{Takeda:2014vwa,
      author         = "Takeda, Shinji and Yoshimura, Yusuke",
      title          = "{Grassmann tensor renormalization group for the
                        one-flavor lattice Gross-Neveu model with finite
                        chemical potential}",
      journal        = "PTEP",
      volume         = "2015",
      year           = "2015",
      number         = "4",
      pages          = "043B01",
      doi            = "10.1093/ptep/ptv022",
      eprint         = "1412.7855",
      archivePrefix  = "arXiv",
      primaryClass   = "hep-lat",
      reportNumber   = "KANAZAWA-14-10",
      SLACcitation   = "%%CITATION = ARXIV:1412.7855;%%"
}

@article{Kadoh:2018hqq,
      author         = "Kadoh, Daisuke and Kuramashi, Yoshinobu and Nakamura,
                        Yoshifumi and Sakai, Ryo and Takeda, Shinji and Yoshimura,
                        Yusuke",
      title          = "{Tensor network formulation for two-dimensional lattice $
                        \mathcal{N} $ = 1 Wess-Zumino model}",
      journal        = "JHEP",
      volume         = "03",
      year           = "2018",
      pages          = "141",
      doi            = "10.1007/JHEP03(2018)141",
      eprint         = "1801.04183",
      archivePrefix  = "arXiv",
      primaryClass   = "hep-lat",
      reportNumber   = "UTHEP-708, UTCCS-P-107, KANAZAWA-17-11",
      SLACcitation   = "%%CITATION = ARXIV:1801.04183;%%"
}

@article{Akiyama:2021xxr,
    author = "Akiyama, Shinichiro and Kuramashi, Yoshinobu",
    title = "{Tensor renormalization group approach to (1+1)-dimensional Hubbard model}",
    eprint = "2105.00372",
    archivePrefix = "arXiv",
    primaryClass = "hep-lat",
    reportNumber = "UTHEP-756, UTCCS-P-137",
    doi = "10.1103/PhysRevD.104.014504",
    journal = "Phys. Rev. D",
    volume = "104",
    number = "1",
    pages = "014504",
    year = "2021"
}

@article{Bloch:2022vqz,
    author = "Bloch, Jacques and Lohmayer, Robert",
    title = "{Grassmann higher-order tensor renormalization group approach for two-dimensional strong-coupling QCD}",
    eprint = "2206.00545",
    archivePrefix = "arXiv",
    primaryClass = "hep-lat",
    doi = "10.1016/j.nuclphysb.2022.116032",
    journal = "Nucl. Phys. B",
    volume = "986",
    pages = "116032",
    year = "2023"
}

@article{Asaduzzaman:2023pyz,
    author = "Asaduzzaman, Muhammad and Catterall, Simon and Meurice, Yannick and Sakai, Ryo and Toga, Goksu Can",
    title = "{Tensor network representation of non-abelian gauge theory coupled to reduced staggered fermions}",
    eprint = "2312.16167",
    archivePrefix = "arXiv",
    primaryClass = "hep-lat",
    doi = "10.1007/JHEP05(2024)195",
    journal = "JHEP",
    volume = "05",
    pages = "195",
    year = "2024"
}

@article{Pai:2024tip,
    author = "Pai, Kwok Ho and Akiyama, Shinichiro and Todo, Synge",
    title = "{Grassmann tensor renormalization group approach to (1+1)-dimensional two-color lattice QCD at finite density}",
    eprint = "2410.09485",
    archivePrefix = "arXiv",
    primaryClass = "hep-lat",
    doi = "10.1007/JHEP03(2025)027",
    journal = "JHEP",
    volume = "03",
    pages = "027",
    year = "2025"
}

@article{Pai:2025eia,
    author = "Pai, Kwok Ho and Akiyama, Shinichiro and Todo, Synge",
    title = "{Two-color lattice QCD in (1 + 1) dimensions with Grassmann tensor renormalization group}",
    eprint = "2501.18918",
    archivePrefix = "arXiv",
    primaryClass = "hep-lat",
    doi = "10.22323/1.466.0364",
    journal = "PoS",
    volume = "LATTICE2024",
    pages = "364",
    year = "2025"
}

@article{Akiyama:2021glo,
    author = "Akiyama, Shinichiro and Kuramashi, Yoshinobu and Yamashita, Takumi",
    title = "{Metal-insulator transition in (2+1)-dimensional Hubbard model with tensor renormalization group}",
    eprint = "2109.14149",
    archivePrefix = "arXiv",
    primaryClass = "cond-mat.str-el",
    reportNumber = "UTHEP-760, UTCCS-P-139",
    doi = "10.1093/ptep/ptac014",
    journal = "PTEP",
    volume = "2022",
    pages = "023",
    month = "9",
    year = "2021"
}

@article{Akiyama:2020soe,
    author = "Akiyama, Shinichiro and Kuramashi, Yoshinobu and Yamashita, Takumi and Yoshimura, Yusuke",
    title = "{Restoration of chiral symmetry in cold and dense Nambu--Jona-Lasinio model with tensor renormalization group}",
    eprint = "2009.11583",
    archivePrefix = "arXiv",
    primaryClass = "hep-lat",
    reportNumber = "UTHEP-752, UTCCS-P-134",
    doi = "10.1007/JHEP01(2021)121",
    journal = "JHEP",
    volume = "01",
    pages = "121",
    year = "2021"
}

@article{Sugimoto:2025vui,
    author = "Sugimoto, Yuto and Akiyama, Shinichiro and Kuramashi, Yoshinobu",
    title = "{Phase structure of (3+1)-dimensional dense two-color QCD at T=0 in the strong coupling limit with the tensor renormalization group}",
    eprint = "2509.23637",
    archivePrefix = "arXiv",
    primaryClass = "hep-lat",
    doi = "10.1103/jd1r-cqrc",
    journal = "Phys. Rev. D",
    volume = "113",
    number = "3",
    pages = "034503",
    year = "2026"
}

@article{Sugimoto:2026wnw,
    author = "Sugimoto, Yuto and Akiyama, Shinichiro and Kuramashi, Yoshinobu",
    title = "{Tensor renormalization group study of cold and dense QCD in the strong coupling limit}",
    eprint = "2601.20690",
    archivePrefix = "arXiv",
    primaryClass = "hep-lat",
    month = "1",
    year = "2026"
}

@article{Akiyama:2020sfo,
    author = "Akiyama, Shinichiro and Kadoh, Daisuke",
    title = "{More about the Grassmann tensor renormalization group}",
    eprint = "2005.07570",
    archivePrefix = "arXiv",
    primaryClass = "hep-lat",
    reportNumber = "UTHEP-751",
    doi = "10.1007/JHEP10(2021)188",
    journal = "JHEP",
    volume = "10",
    pages = "188",
    year = "2021"
}

@article{Nambu:1961tp,
    author = "Nambu, Yoichiro and Jona-Lasinio, G.",
    editor = "Eguchi, T.",
    title = "{Dynamical Model of Elementary Particles Based on an Analogy with Superconductivity. 1.}",
    doi = "10.1103/PhysRev.122.345",
    journal = "Phys. Rev.",
    volume = "122",
    pages = "345--358",
    year = "1961"
}

@article{Nambu:1961fr,
    author = "Nambu, Yoichiro and Jona-Lasinio, G.",
    editor = "Eguchi, T.",
    title = "{Dynamical model of elementary particles based on an analogy with superconductivity. II.}",
    doi = "10.1103/PhysRev.124.246",
    journal = "Phys. Rev.",
    volume = "124",
    pages = "246--254",
    year = "1961"
}

@article{Gross:1974jv,
    author = "Gross, David J. and Neveu, Andre",
    title = "{Dynamical Symmetry Breaking in Asymptotically Free Field Theories}",
    reportNumber = "COO-2220-19",
    doi = "10.1103/PhysRevD.10.3235",
    journal = "Phys. Rev. D",
    volume = "10",
    pages = "3235",
    year = "1974"
}

@article{Mermin:1966fe,
    author = "Mermin, N. D. and Wagner, H.",
    title = "{Absence of ferromagnetism or antiferromagnetism in one-dimensional or two-dimensional isotropic Heisenberg models}",
    doi = "10.1103/PhysRevLett.17.1133",
    journal = "Phys. Rev. Lett.",
    volume = "17",
    pages = "1133--1136",
    year = "1966"
}

@article{Coleman:1973ci,
    author = "Coleman, Sidney R.",
    title = "{There are no Goldstone bosons in two-dimensions}",
    doi = "10.1007/BF01646487",
    journal = "Commun. Math. Phys.",
    volume = "31",
    pages = "259--264",
    year = "1973"
}

@article{Nielsen:1981hk,
    author = "Nielsen, Holger Bech and Ninomiya, M.",
    title = "{No Go Theorem for Regularizing Chiral Fermions}",
    reportNumber = "RL-81-052",
    doi = "10.1016/0370-2693(81)91026-1",
    journal = "Phys. Lett. B",
    volume = "105",
    pages = "219--223",
    year = "1981"
}

@article{Wilson:1974sk,
    author = "Wilson, Kenneth G.",
    editor = "Taylor, J. C.",
    title = "{Confinement of Quarks}",
    reportNumber = "CLNS-262",
    doi = "10.1103/PhysRevD.10.2445",
    journal = "Phys. Rev. D",
    volume = "10",
    pages = "2445--2459",
    year = "1974"
}

@inproceedings{Wilson:1975id,
    author = "Wilson, Kenneth G.",
    title = "{Quarks and Strings on a Lattice}",
    booktitle = "{13th International School of Subnuclear Physics: New Phenomena in Subnuclear Physics}",
    reportNumber = "CLNS-321",
    month = "11",
    year = "1975"
}

@article{Kawamoto:1980fd,
    author = "Kawamoto, Noboru",
    title = "{Towards the Phase Structure of Euclidean Lattice Gauge Theories with Fermions}",
    reportNumber = "Print-80-0488 (AMSTERDAM)",
    doi = "10.1016/0550-3213(81)90450-8",
    journal = "Nucl. Phys. B",
    volume = "190",
    pages = "617--669",
    year = "1981"
}

@article{Aoki:1983qi,
    author = "Aoki, Sinya",
    title = "{New Phase Structure for Lattice QCD with Wilson Fermions}",
    reportNumber = "UT-421-TOKYO",
    doi = "10.1103/PhysRevD.30.2653",
    journal = "Phys. Rev. D",
    volume = "30",
    pages = "2653",
    year = "1984"
}

@article{Aoki:1986xr,
    author = "Aoki, Sinya",
    title = "{A Solution to the U(1) Problem on a Lattice}",
    reportNumber = "UT-488-TOKYO",
    doi = "10.1103/PhysRevLett.57.3136",
    journal = "Phys. Rev. Lett.",
    volume = "57",
    pages = "3136",
    year = "1986"
}

@article{Aoki:1987us,
    author = "Aoki, Sinya",
    title = "{U(1) Problem and Lattice {QCD}}",
    reportNumber = "UT-506-TOKYO",
    doi = "10.1016/0550-3213(89)90113-2",
    journal = "Nucl. Phys. B",
    volume = "314",
    pages = "79--111",
    year = "1989"
}

@article{Vafa:1984xg,
    author = "Vafa, Cumrun and Witten, Edward",
    title = "{Parity Conservation in QCD}",
    reportNumber = "Print-84-0549 (PRINCETON)",
    doi = "10.1103/PhysRevLett.53.535",
    journal = "Phys. Rev. Lett.",
    volume = "53",
    pages = "535",
    year = "1984"
}

@article{Aoki:1989rw,
    author = "Aoki, Sinya and Gocksch, Andreas",
    title = "{Spontaneous Breaking of Parity in Quenched Lattice {QCD} With Wilson Fermions}",
    reportNumber = "BNL-43054, HET-001/89",
    doi = "10.1016/0370-2693(89)90692-8",
    journal = "Phys. Lett. B",
    volume = "231",
    pages = "449--452",
    year = "1989"
}

@article{Aoki:1995yf,
    author = "Aoki, S. and Ukawa, A. and Umemura, T.",
    title = "{Finite temperature phase structure of lattice QCD with Wilson quark action}",
    eprint = "hep-lat/9508008",
    archivePrefix = "arXiv",
    reportNumber = "UTHEP-313",
    doi = "10.1103/PhysRevLett.76.873",
    journal = "Phys. Rev. Lett.",
    volume = "76",
    pages = "873--876",
    year = "1996"
}

@article{Sharpe:1998xm,
    author = "Sharpe, Stephen R. and Singleton, Jr, Robert L.",
    title = "{Spontaneous flavor and parity breaking with Wilson fermions}",
    eprint = "hep-lat/9804028",
    archivePrefix = "arXiv",
    reportNumber = "UW-PT-98-2",
    doi = "10.1103/PhysRevD.58.074501",
    journal = "Phys. Rev. D",
    volume = "58",
    pages = "074501",
    year = "1998"
}

@article{Azcoiti:2012ns,
    author = "Azcoiti, Vicente and Di Carlo, Giuseppe and Follana, Eduardo and Vaquero, Alejandro",
    title = "{Elucidating the Vacuum Structure of the Aoki Phase}",
    eprint = "1208.0761",
    archivePrefix = "arXiv",
    primaryClass = "hep-lat",
    doi = "10.1016/j.nuclphysb.2013.01.008",
    journal = "Nucl. Phys. B",
    volume = "870",
    pages = "138--158",
    year = "2013"
}

@article{Misumi:2019jrt,
    author = "Misumi, Tatsuhiro and Tanizaki, Yuya",
    title = "{Lattice gauge theory for the Haldane conjecture and central-branch Wilson fermion}",
    eprint = "1910.09604",
    archivePrefix = "arXiv",
    primaryClass = "hep-lat",
    doi = "10.1093/ptep/ptaa003",
    journal = "PTEP",
    volume = "2020",
    number = "3",
    pages = "033B03",
    year = "2020"
}

@article{Kenna:2001fs,
    author = "Kenna, R. and Sexton, J. C.",
    title = "{The Weakly coupled Gross-Neveu model with Wilson fermions}",
    eprint = "hep-lat/0103014",
    archivePrefix = "arXiv",
    doi = "10.1103/PhysRevD.65.014507",
    journal = "Phys. Rev. D",
    volume = "65",
    pages = "014507",
    year = "2002"
}

@article{Kuno:2018pcp,
    author = "Kuno, Yoshihito",
    title = "{Phase structure of the interacting Su-Schrieffer-Heeger model and the relationship with the Gross-Neveu model on lattice}",
    eprint = "1811.01487",
    archivePrefix = "arXiv",
    primaryClass = "cond-mat.quant-gas",
    doi = "10.1103/PhysRevB.99.064105",
    journal = "Phys. Rev. B",
    volume = "99",
    number = "6",
    pages = "064105",
    year = "2019"
}

@article{Junemann:2016fxu,
    author = {J{\"u}nemann, J. and Piga, A. and Ran, S. -J. and Lewenstein, M. and Rizzi, M. and Bermudez, A.},
    title = "{Exploring Interacting Topological Insulators with Ultracold Atoms: the Synthetic Creutz-Hubbard Model}",
    eprint = "1612.02996",
    archivePrefix = "arXiv",
    primaryClass = "cond-mat.quant-gas",
    doi = "10.1103/PhysRevX.7.031057",
    journal = "Phys. Rev. X",
    volume = "7",
    number = "3",
    pages = "031057",
    year = "2017"
}

@article{Bermudez:2018eyh,
    author = "Bermudez, A. and Tirrito, E. and Rizzi, M. and Lewenstein, M. and Hands, S.",
    title = "{Gross{\textendash}Neveu{\textendash}Wilson model and correlated symmetry-protected topological phases}",
    eprint = "1807.03202",
    archivePrefix = "arXiv",
    primaryClass = "cond-mat.quant-gas",
    doi = "10.1016/j.aop.2018.10.007",
    journal = "Annals Phys.",
    volume = "399",
    pages = "149--180",
    year = "2018"
}

@article{Schnyder:2008tya,
    author = "Schnyder, Andreas and Ryu, Shinsei and Furusaki, Akira and Ludwig, Andreas",
    title = "{Classification of topological insulators and superconductors in three spatial dimensions}",
    eprint = "0803.2786",
    archivePrefix = "arXiv",
    primaryClass = "cond-mat.mes-hall",
    doi = "10.1103/PhysRevB.78.195125",
    journal = "Phys. Rev. B",
    volume = "78",
    number = "19",
    pages = "195125",
    year = "2008"
}

@article{Ryu:2010zza,
    author = "Ryu, Shinsei and Schnyder, Andreas P. and Furusaki, Akira and Ludwig, Andreas W. W.",
    title = "{Topological insulators and superconductors: Tenfold way and dimensional hierarchy}",
    eprint = "0912.2157",
    archivePrefix = "arXiv",
    primaryClass = "cond-mat.mes-hall",
    doi = "10.1088/1367-2630/12/6/065010",
    journal = "New J. Phys.",
    volume = "12",
    pages = "065010",
    year = "2010"
}

@article{Roose:2021pba,
    author = "Roose, Gertian and Haegeman, Jutho and Van Acoleyen, Karel and Vanderstraeten, Laurens and Bultinck, Nick",
    title = "{The chiral Gross-Neveu model on the lattice via a Landau-forbidden phase transition}",
    eprint = "2111.14652",
    archivePrefix = "arXiv",
    primaryClass = "hep-th",
    doi = "10.1007/JHEP06(2022)019",
    journal = "JHEP",
    volume = "06",
    pages = "019",
    year = "2022"
}

@article{Asaduzzaman:2022bpi,
    author = "Asaduzzaman, Muhammad and Catterall, Simon and Toga, Goksu Can and Meurice, Yannick and Sakai, Ryo",
    title = "{Quantum simulation of the N-flavor Gross-Neveu model}",
    eprint = "2208.05906",
    archivePrefix = "arXiv",
    primaryClass = "hep-lat",
    doi = "10.1103/PhysRevD.106.114515",
    journal = "Phys. Rev. D",
    volume = "106",
    number = "11",
    pages = "114515",
    year = "2022"
}

@article{Nishino_1996,
   title={Corner Transfer Matrix Renormalization Group Method},
   volume={65},
   ISSN={1347-4073},
   DOI={10.1143/jpsj.65.891},
   number={4},
   journal={Journal of the Physical Society of Japan},
   publisher={Physical Society of Japan},
   author={Nishino, Tomotoshi and Okunishi, Kouichi},
   year={1996},
   month=apr, pages={891–894} }

@article{Baxter:1968krk,
    author = "Baxter, R. J.",
    title = "{Dimers on a Rectangular Lattice}",
    doi = "10.1063/1.1664623",
    journal = "J. Math. Phys.",
    volume = "9",
    number = "4",
    pages = "650",
    year = "1968"
}

@article{baxter1978variational,
  title={Variational approximations for square lattice models in statistical mechanics},
  doi = "https://doi.org/10.1007/BF01011693

",
  author={Baxter, Rodney J},
  journal={Journal of Statistical Physics},
  volume={19},
  number={5},
  pages={461--478},
  year={1978},
  publisher={Springer}
}

@article{Nishino_1997,
   title={Corner Transfer Matrix Algorithm for Classical    Renormalization Group},
   volume={66},
   ISSN={1347-4073},
   DOI={10.1143/jpsj.66.3040},
   number={10},
   journal={Journal of the Physical Society of Japan},
   publisher={Physical Society of Japan},
   author={Nishino, Tomotoshi and Okunishi, Kouichi},
   year={1997},
   month=oct, pages={3040–3047} }

@book{Baxter:1982zz,
    author = "Baxter, R. J.",
    title = "{Exactly solved models in statistical mechanics}",
    isbn = "978-0-486-46271-4",
    year = "1982"
}

@article{White:1992zz,
    author = "White, Steven R.",
    title = "{Density matrix formulation for quantum renormalization groups}",
    doi = "10.1103/PhysRevLett.69.2863",
    journal = "Phys. Rev. Lett.",
    volume = "69",
    pages = "2863--2866",
    year = "1992"
}

@article{White:1993zza,
    author = "White, Steven R.",
    title = "{Density-matrix algorithms for quantum renormalization groups}",
    doi = "10.1103/PhysRevB.48.10345",
    journal = "Phys. Rev. B",
    volume = "48",
    pages = "10345--10356",
    year = "1993"
}

@article{Verstraete:2004cf,
    author = "Verstraete, F. and Cirac, J. I.",
    title = "{Renormalization algorithms for quantum-many body systems in two and higher dimensions}",
    eprint = "cond-mat/0407066",
    archivePrefix = "arXiv",
    month = "7",
    year = "2004"
}

@article{Nishino:2000ygc,
    author = "Nishino, Tomotoshi and Hieida, Yasuhiro and Okunishi, Kouichi and Maeshima, Nobuya and Akutsu, Yasuhiro and Gendiar, Andrej",
    title = "{Two-Dimensional Tensor Product Variational Formulation}",
    eprint = "cond-mat/0011103",
    archivePrefix = "arXiv",
    doi = "10.1143/ptp.105.409",
    journal = "Prog. Theor. Phys.",
    volume = "105",
    number = "3",
    pages = "409--417",
    year = "2001"
}

@article{Li2022,
  title = {Magnetization of the spin-$\frac{1}{2}$ Heisenberg antiferromagnet on the triangular lattice},
  author = {Li, Qian and Li, Hong and Zhao, Jize and Luo, Hong-Gang and Xie, Z. Y.},
  journal = {Phys. Rev. B},
  volume = {105},
  issue = {18},
  pages = {184418},
  numpages = {7},
  year = {2022},
  month = {May},
  publisher = {American Physical Society},
  doi = {10.1103/PhysRevB.105.184418},
  eprint = "2009.03765",
  archivePrefix = "arXiv",
  primaryClass = "cond-mat.str-el",
}

@article{Orus:2009wuu,
    author = "Or{\'u}s, Rom{\'a}n and Vidal, Guifr{\'e}",
    title = "{Simulation of two-dimensional quantum systems on an infinite lattice revisited: Corner transfer matrix for tensor contraction}",
    eprint = "0905.3225",
    archivePrefix = "arXiv",
    primaryClass = "cond-mat.str-el",
    doi = "10.1103/PhysRevB.80.094403",
    journal = "Phys. Rev. B",
    volume = "80",
    number = "9",
    pages = "094403",
    year = "2009"
}

@article{PhysRevB.82.245119,
  title = {Simulation of fermionic lattice models in two dimensions with projected entangled-pair states: Next-nearest neighbor Hamiltonians},
  author = {Corboz, Philippe and Jordan, Jacob and Vidal, Guifr\'e},
  eprint = "1008.3937",
  archivePrefix = "arXiv",
  primaryClass = "cond-mat.str-el",
  journal = {Phys. Rev. B},
  volume = {82},
  issue = {24},
  pages = {245119},
  numpages = {9},
  year = {2010},
  month = {Dec},
  publisher = {American Physical Society},
  doi = {10.1103/PhysRevB.82.245119}
}

@article{Corboz:2014ocg,
    author = "Corboz, Philippe and Rice, T. {\,}M. and Troyer, Matthias",
    title = "{Competing States in the t-J Model: Uniform d-Wave State versus Stripe State}",
    eprint = "1402.2859",
    archivePrefix = "arXiv",
    primaryClass = "cond-mat.str-el",
    doi = "10.1103/PhysRevLett.113.046402",
    journal = "Phys. Rev. Lett.",
    volume = "113",
    number = "4",
    pages = "046402",
    year = "2014"
}

@article{Fishman:2018lnr,
    author = "Fishman, M. T. and Vanderstraeten, L. and Zauner-Stauber, V. and Haegeman, J. and Verstraete, F.",
    title = "{Faster methods for contracting infinite two-dimensional tensor networks}",
    eprint = "1711.05881",
    archivePrefix = "arXiv",
    primaryClass = "cond-mat.str-el",
    doi = "10.1103/PhysRevB.98.235148",
    journal = "Phys. Rev. B",
    volume = "98",
    number = "23",
    pages = "235148",
    year = "2018"
}

@article{Liu:2022dht,
    author = "Liu, X. F. and Fu, Y. F. and Yu, W. Q. and Yu, J. F. and Xie, Z. Y.",
    title = "{Variational Corner Transfer Matrix Renormalization Group Method for Classical Statistical Models}",
    eprint = "2203.17098",
    archivePrefix = "arXiv",
    primaryClass = "cond-mat.str-el",
    doi = "10.1088/0256-307X/39/6/067502",
    journal = "Chin. Phys. Lett.",
    volume = "39",
    number = "6",
    pages = "067502",
    year = "2022"
}

@article{NISHINO199669,
   title={Numerical renormalization group at criticality},
   volume={213},
   ISSN={0375-9601},
   DOI={10.1016/0375-9601(96)00128-4},
   number={1–2},
   journal={Physics Letters A},
   publisher={Elsevier BV},
   author={Nishino, T. and Okunishi, K. and Kikuchi, M.},
   year={1996},
   month=apr, pages={69–72} }

@article{Calabrese:2004eu,
    author = "Calabrese, Pasquale and Cardy, John L.",
    title = "{Entanglement entropy and quantum field theory}",
    eprint = "hep-th/0405152",
    archivePrefix = "arXiv",
    doi = "10.1088/1742-5468/2004/06/P06002",
    journal = "J. Stat. Mech.",
    volume = "0406",
    pages = "P06002",
    year = "2004"
}

@article{Ueda:2014jxq,
    author = "Ueda, Hiroshi and Okunishi, Kouichi and Nishino, Tomotoshi",
    title = "{Doubling of Entanglement Spectrum in Tensor Renormalization Group}",
    eprint = "1306.6829",
    archivePrefix = "arXiv",
    primaryClass = "cond-mat.stat-mech",
    doi = "10.1103/PhysRevB.89.075116",
    journal = "Phys. Rev. B",
    volume = "89",
    pages = "075116",
    year = "2014"
}

@article{Ueda:2017ojh,
    author = "Ueda, Hiroshi and Okunishi, Kouichi and Kr{\v{c}}m{\'a}r, Roman and Gendiar, Andrej and Yunoki, Seiji and Nishino, Tomotoshi",
    title = "{Critical behavior of the two-dimensional icosahedron model}",
    eprint = "1709.01275",
    archivePrefix = "arXiv",
    primaryClass = "cond-mat.stat-mech",
    doi = "10.1103/PhysRevE.96.062112",
    journal = "Phys. Rev. E",
    volume = "96",
    number = "6",
    pages = "062112",
    year = "2017"
}

@article{Ueda:2020igb,
    author = "Ueda, Hiroshi and Okunishi, Kouichi and Harada, Kenji and Kr{\v{c}}m{\'a}r, Roman and Gendiar, Andrej and Yunoki, Seiji and Nishino, Tomotoshi",
    title = "{Finite-$m$ scaling analysis of Berezinskii-Kosterlitz-Thouless phase transitions and entanglement spectrum for the six-state clock model}",
    eprint = "2001.10176",
    archivePrefix = "arXiv",
    primaryClass = "cond-mat.stat-mech",
    doi = "10.1103/PhysRevE.101.062111",
    journal = "Phys. Rev. E",
    volume = "101",
    number = "6",
    pages = "062111",
    year = "2020"
}

@article{Tagliacozzo:2007rda,
    author = "Tagliacozzo, L. and de Oliveira, Thiago R. and Iblisdir, S. and Latorre, J. I.",
    title = "{Scaling of entanglement support for Matrix Product States}",
    eprint = "0712.1976",
    archivePrefix = "arXiv",
    primaryClass = "cond-mat.stat-mech",
    doi = "10.1103/PhysRevB.78.024410",
    journal = "Phys. Rev. B",
    volume = "78",
    pages = "024410",
    year = "2008"
}

@article{Pollmann:2009lnv,
    author = "Pollmann, Frank and Mukerjee, Subroto and Turner, Ari M. and Moore, Joel E.",
    title = "{Theory of Finite-Entanglement Scaling at One-Dimensional Quantum Critical Points}",
    eprint = "0812.2903",
    archivePrefix = "arXiv",
    primaryClass = "cond-mat.str-el",
    doi = "10.1103/PhysRevLett.102.255701",
    journal = "Phys. Rev. Lett.",
    volume = "102",
    number = "25",
    pages = "255701",
    year = "2009"
}

@article{Li:2008kda,
    author = "Li, Hui and Haldane, F.",
    title = "{Entanglement Spectrum as a Generalization of Entanglement Entropy: Identification of Topological Order in Non-Abelian Fractional Quantum Hall Effect States}",
    eprint = "0805.0332",
    archivePrefix = "arXiv",
    primaryClass = "cond-mat.mes-hall",
    doi = "10.1103/PhysRevLett.101.010504",
    journal = "Phys. Rev. Lett.",
    volume = "101",
    number = "1",
    pages = "010504",
    year = "2008"
}

@article{Pollmann:2009ryx,
    author = "Pollmann, Frank and Turner, Ari M. and Berg, Erez and Oshikawa, Masaki",
    title = "{Entanglement spectrum of a topological phase in one dimension}",
    eprint = "0910.1811",
    archivePrefix = "arXiv",
    primaryClass = "cond-mat.str-el",
    doi = "10.1103/PhysRevB.81.064439",
    journal = "Phys. Rev. B",
    volume = "81",
    number = "6",
    pages = "064439",
    year = "2010"
}

@article{Creutz:2011cd,
    author = "Creutz, Michael and Kimura, Taro and Misumi, Tatsuhiro",
    title = "{Aoki Phases in the Lattice Gross-Neveu Model with Flavored Mass terms}",
    eprint = "1101.4239",
    archivePrefix = "arXiv",
    primaryClass = "hep-lat",
    reportNumber = "BNL-94608-2011-JA, YITP-11-6",
    doi = "10.1103/PhysRevD.83.094506",
    journal = "Phys. Rev. D",
    volume = "83",
    pages = "094506",
    year = "2011"
}

@article{Kimura:2011ik,
    author = "Kimura, Taro and Komatsu, Shota and Misumi, Tatsuhiro and Noumi, Toshifumi and Torii, Shingo and Aoki, Sinya",
    title = "{Revisiting symmetries of lattice fermions via spin-flavor representation}",
    eprint = "1111.0402",
    archivePrefix = "arXiv",
    primaryClass = "hep-lat",
    reportNumber = "RIKEN-MP-36, UTHEP-635, UT-KOMABA-11-11, YITP-11-91",
    doi = "10.1007/JHEP01(2012)048",
    journal = "JHEP",
    volume = "01",
    pages = "048",
    year = "2012"
}

@Misc{qsw,
  title        = "",
  howpublished = "\url{https://qsw.phys.s.u-tokyo.ac.jp/}"
}

@article{Ueda:2020etz,
    author = "Ueda, Hiroshi and Okunishi, Kouichi and Yunoki, Seiji and Nishino, Tomotoshi",
    title = "{Corner transfer matrix renormalization group analysis of the two-dimensional dodecahedron model}",
    eprint = "2004.08669",
    archivePrefix = "arXiv",
    primaryClass = "cond-mat.stat-mech",
    doi = "10.1103/PhysRevE.102.032130",
    journal = "Phys. Rev. E",
    volume = "102",
    number = "3",
    pages = "032130",
    year = "2020"
}

@article{Huang:2019moo,
    author = "Huang, Rui-Zhen and Lu, Da-Chuan and You, Yi-Zhuang and Meng, Zi Yang and Xiang, Tao",
    title = "{Emergent Symmetry and Conserved Current at a One Dimensional Incarnation of Deconfined Quantum Critical Point}",
    eprint = "1904.00021",
    archivePrefix = "arXiv",
    primaryClass = "cond-mat.str-el",
    doi = "10.1103/PhysRevB.100.125137",
    journal = "Phys. Rev. B",
    volume = "100",
    number = "12",
    pages = "125137",
    year = "2019"
}

@article{PhysRevB.99.205121,
  title = {Quantum criticality on a chiral ladder: An SU(2) infinite density matrix renormalization group study},
  author = {Schmoll, Philipp and Haller, Andreas and Rizzi, Matteo and Or\'us, Rom\'an},
  eprint = "1812.01311",
  archivePrefix = "arXiv",
  primaryClass = "cond-mat.str-el",
  journal = {Phys. Rev. B},
  volume = {99},
  issue = {20},
  pages = {205121},
  numpages = {17},
  year = {2019},
  month = {May},
  publisher = {American Physical Society},
  doi = {10.1103/PhysRevB.99.205121}
}

@article{Kong:2026,
    author = "Kong, Jian-Gang and Xie, Zhi Yuan",
    title = "{Grassmann corner transfer-matrix renormalization group approach to one-dimensional fermionic models}",
    eprint = "2604.05582",
    archivePrefix = "arXiv",
    primaryClass = "cond-mat.str-el",
    year = "2026"
}

@article{Okunishi:2021but,
    author = "Okunishi, Kouichi and Nishino, Tomotoshi and Ueda, Hiroshi",
    title = "{Developments in the Tensor Network -- from Statistical Mechanics to Quantum Entanglement}",
    eprint = "2111.12223",
    archivePrefix = "arXiv",
    primaryClass = "cond-mat.stat-mech",
    doi = "10.7566/JPSJ.91.062001",
    journal = "J. Phys. Soc. Jap.",
    volume = "91",
    pages = "062001",
    year = "2022"
}

@article{Akiyama:2024ush,
    author = "Akiyama, Shinichiro and Meurice, Yannick and Sakai, Ryo",
    title = "{Tensor renormalization group for fermions}",
    eprint = "2401.08542",
    archivePrefix = "arXiv",
    primaryClass = "hep-lat",
    reportNumber = "UTCCS-P-152",
    doi = "10.1088/1361-648X/ad4760",
    journal = "J. Phys. Condens. Matter",
    volume = "36",
    number = "34",
    pages = "343002",
    year = "2024"
}

\end{document}